\documentclass[12pt,a4paper]{article}
\usepackage{subcaption}
\usepackage{multirow}

\usepackage{ifthen} % for conditional statements
\newboolean{pdflatex}
\setboolean{pdflatex}{true} % False for eps figures 

\newboolean{articletitles}
\setboolean{articletitles}{true} % False removes titles in references

\newboolean{uprightparticles}
\setboolean{uprightparticles}{false} %True for upright particle symbols

\def\paperauthors{LHCb collaboration} % Leave as is for PAPER, CONF and FIGURE
\def\paperasciititle{Study of Ds1(2460)+ -> Ds+pi+pi- in B -> Dbar(*)Ds+pi+pi- decays} % Set ASCII title here !! MAKE sure it's only ASCII characters !! 
\def\papertitle{Study of $D_{s1}(2460)^{+}\to D_{s}^{+}\pi^{+}\pi^{-}$ in $B\to {\offsetoverline{D\xspace}}\xspace{}^{(*)}D_{s}^{+}\pi^{+}\pi^{-}$ decays} % Latex formatted title
\def\paperkeywords{{High Energy Physics}, {LHCb}} % Comma separated list
\def\papercopyright{\the\year\ CERN for the benefit of the LHCb collaboration} % new since 9/Apr/2018
\def\paperlicence{CC BY 4.0 licence}
\def\paperlicenceurl{https://creativecommons.org/licenses/by/4.0/}

\newif\ifEnableSectionTOCLinks
\EnableSectionTOCLinksfalse % deactivated

\usepackage[top=1in, bottom=1.25in, left=1in, right=1in]{geometry}

\usepackage{microtype}
\usepackage{lineno}  % for line numbering during review
\usepackage{xspace} % To avoid problems with missing or double spaces after
\usepackage{caption} %these three command get the figure and table captions automatically small

\usepackage{graphicx}  % to include figures (can also use other packages)
\usepackage{color}
\usepackage{colortbl}
\graphicspath{{./figs/}} % Make Latex search fig subdir for figures
\usepackage{amsmath} % Adds a large collection of math symbols
\usepackage{amssymb}
\usepackage{amsfonts}
\usepackage{upgreek} % Adds in support for greek letters in roman typeset

\newcommand*\patchAmsMathEnvironmentForLineno[1]{%
\expandafter\let\csname old#1\expandafter\endcsname\csname #1\endcsname
\expandafter\let\csname oldend#1\expandafter\endcsname\csname
end#1\endcsname
 \renewenvironment{#1}%
   {\linenomath\csname old#1\endcsname}%
   {\csname oldend#1\endcsname\endlinenomath}%
}
\newcommand*\patchBothAmsMathEnvironmentsForLineno[1]{%
  \patchAmsMathEnvironmentForLineno{#1}%
  \patchAmsMathEnvironmentForLineno{#1*}%
}
\AtBeginDocument{%
\patchBothAmsMathEnvironmentsForLineno{equation}%
\patchBothAmsMathEnvironmentsForLineno{align}%
\patchBothAmsMathEnvironmentsForLineno{flalign}%
\patchBothAmsMathEnvironmentsForLineno{alignat}%
\patchBothAmsMathEnvironmentsForLineno{gather}%
\patchBothAmsMathEnvironmentsForLineno{multline}%
\patchBothAmsMathEnvironmentsForLineno{eqnarray}%
}

\usepackage{hyperxmp}

\usepackage[pdftex,
            pdfauthor={\paperauthors},
            pdftitle={\paperasciititle},
            pdfkeywords={\paperkeywords},
            pdfcopyright={Copyright (C) \papercopyright},
            pdflicenseurl={\paperlicenceurl}]{hyperref}
\usepackage[bottom,flushmargin,hang,multiple]{footmisc}

\usepackage[all]{hypcap} % Internal hyperlinks to floats.

\usepackage{xspace} 
\usepackage{upgreek}

\def\lhcb   {\mbox{LHCb}\xspace}

\def\MagUp {\mbox{\em Mag\kern -0.05em Up}\xspace}

\ifthenelse{\boolean{uprightparticles}}%
{

 \def\Ppi         {\ensuremath{\uppi}\xspace}

 \def\PDelta      {\ensuremath{\Delta}\xspace}                 
 \def\PXi         {\ensuremath{\Xi}\xspace}                 
 \def\PLambda     {\ensuremath{\Lambda}\xspace}                 
 \def\PSigma      {\ensuremath{\Sigma}\xspace}                 
 \def\POmega      {\ensuremath{\Omega}\xspace}                 
 \def\PUpsilon    {\ensuremath{\Upsilon}\xspace}
 \let\oldPi\Pi
 \def\PPi         {\ensuremath{\oldPi}\xspace}

 \def\PB      {\ensuremath{\mathrm{B}}\xspace}                 
 \def\PD      {\ensuremath{\mathrm{D}}\xspace}                 

 \def\PK      {\ensuremath{\mathrm{K}}\xspace}                 
 \def\Pb      {\ensuremath{\mathrm{b}}\xspace}                 
 \def\Pc      {\ensuremath{\mathrm{c}}\xspace}

 \def\Ps      {\ensuremath{\mathrm{s}}\xspace}

 \def\thebaroffset{0.0em}
}
{

 \def\Ppi         {\ensuremath{\pi}\xspace}

 \mathchardef\PDelta="7101
 \mathchardef\PXi="7104
 \mathchardef\PLambda="7103
 \mathchardef\PSigma="7106
 \mathchardef\POmega="710A
 \mathchardef\PUpsilon="7107
 \mathchardef\PPi="7105
 \def\PB      {\ensuremath{B}\xspace}                 
 \def\PD      {\ensuremath{D}\xspace}                 

 \def\PK      {\ensuremath{K}\xspace}                 
 \def\Pb      {\ensuremath{b}\xspace}                 
 \def\Pc      {\ensuremath{c}\xspace}

 \def\Ps      {\ensuremath{s}\xspace}

 \def\thebaroffset{0.18em}
}
\newcommand{\offsetoverline}[2][\thebaroffset]{\kern #1\overline{\kern -#1 #2}}%

\makeatletter
\ifcase \@ptsize \relax% 10pt
  \newcommand{\miniscule}{\@setfontsize\miniscule{4}{5}}% \tiny: 5/6
\or% 11pt
  \newcommand{\miniscule}{\@setfontsize\miniscule{5}{6}}% \tiny: 6/7
\or% 12pt
  \newcommand{\miniscule}{\@setfontsize\miniscule{5}{6}}% \tiny: 6/7
\fi
\makeatother

\DeclareRobustCommand{\optbar}[1]{\shortstack{{\miniscule (\rule[.5ex]{1.25em}{.18mm})}
  \\ [-.7ex] $#1$}}

\def\squark    {{\ensuremath{\Ps}}\xspace}

\def\cquark    {{\ensuremath{\Pc}}\xspace}

\def\bquark    {{\ensuremath{\Pb}}\xspace}

\def\pion   {{\ensuremath{\Ppi}}\xspace}
\def\piz    {{\ensuremath{\pion^0}}\xspace}
\def\pip    {{\ensuremath{\pion^+}}\xspace}
\def\pim    {{\ensuremath{\pion^-}}\xspace}
\def\pipm   {{\ensuremath{\pion^\pm}}\xspace}

\def\kaon    {{\ensuremath{\PK}}\xspace}
\def\KorKbar {\kern \thebaroffset\optbar{\kern -\thebaroffset \PK}{}\xspace}

\def\Kp      {{\ensuremath{\kaon^+}}\xspace}
\def\Km      {{\ensuremath{\kaon^-}}\xspace}

\def\Dbar    {{\ensuremath{\offsetoverline{\PD}}}\xspace}
\def\D       {{\ensuremath{\PD}}\xspace}

\def\DorDbar {\kern \thebaroffset\optbar{\kern -\thebaroffset \PD}\xspace}

\def\Dzb     {{\ensuremath{\Dbar{}^0}}\xspace}
\def\Dp      {{\ensuremath{\D^+}}\xspace}
\def\Dm      {{\ensuremath{\D^-}}\xspace}

\def\DpDm    {\ensuremath{\Dp {\kern -0.16em \Dm}}\xspace}
\def\Dstar   {{\ensuremath{\D^*}}\xspace}

\def\Dstarm  {{\ensuremath{\D^{*-}}}\xspace}

\def\Ds      {{\ensuremath{\D^+_\squark}}\xspace}
\def\Dsp     {{\ensuremath{\D^+_\squark}}\xspace}

\def\B       {{\ensuremath{\PB}}\xspace}

\def\BorBbar {\kern \thebaroffset\optbar{\kern -\thebaroffset \PB}\xspace}
\def\Bz      {{\ensuremath{\B^0}}\xspace}

\def\Bd      {{\ensuremath{\B^0}}\xspace}

\def\BdorBdbar {\kern \thebaroffset\optbar{\kern -\thebaroffset \Bd}\xspace}
\def\Bu      {{\ensuremath{\B^+}}\xspace}
\def\Bub     {{\ensuremath{\B^-}}\xspace}
\def\Bp      {{\ensuremath{\Bu}}\xspace}
\def\Bm      {{\ensuremath{\Bub}}\xspace}

\def\Bs      {{\ensuremath{\B^0_\squark}}\xspace}

\def\BsorBsbar {\kern \thebaroffset\optbar{\kern -\thebaroffset \Bs}\xspace}

\def\Y#1S{\ensuremath{\PUpsilon{(#1S)}}\xspace}

\def\LorLbar     {\kern \thebaroffset\optbar{\kern -\thebaroffset \PLambda}\xspace}

\newcommand{\decay}[2]{\mbox{\ensuremath{#1\!\to #2}}\xspace} 

\def\to                 {\ensuremath{\rightarrow}\xspace}

\def\AT#1     {\ensuremath{A_{\mathrm{T}}^{#1}}\xspace}           % 2

\def\C#1      {\ensuremath{\mathcal{C}_{#1}}\xspace}                       % 9
\def\Cp#1     {\ensuremath{\mathcal{C}_{#1}^{'}}\xspace}                    % 7
\def\Ceff#1   {\ensuremath{\mathcal{C}_{#1}^{\mathrm{(eff)}}}\xspace}        % 9  
\def\Cpeff#1  {\ensuremath{\mathcal{C}_{#1}^{'\mathrm{(eff)}}}\xspace}       % 7
\def\Ope#1    {\ensuremath{\mathcal{O}_{#1}}\xspace}                       % 2
\def\Opep#1   {\ensuremath{\mathcal{O}_{#1}^{'}}\xspace}                    % 7

\newcommand{\nospaceunit}[1]{\ensuremath{\text{#1}}}       
\newcommand{\aunit}[1]{\ensuremath{\text{\,#1}}}       
\newcommand{\tev}{\aunit{Te\kern -0.1em V}\xspace}
\newcommand{\gev}{\aunit{Ge\kern -0.1em V}\xspace}
\newcommand{\mev}{\aunit{Me\kern -0.1em V}\xspace}
\newcommand{\kev}{\aunit{ke\kern -0.1em V}\xspace}
\newcommand{\ev}{\aunit{e\kern -0.1em V}\xspace}
 
\newcommand{\mevc}{\ensuremath{\aunit{Me\kern -0.1em V\!/}c}\xspace}
\newcommand{\gevc}{\ensuremath{\aunit{Ge\kern -0.1em V\!/}c}\xspace}
\newcommand{\mevcc}{\ensuremath{\aunit{Me\kern -0.1em V\!/}c^2}\xspace}
\newcommand{\gevcc}{\ensuremath{\aunit{Ge\kern -0.1em V\!/}c^2}\xspace}
\def\mum  {\ensuremath{\,\upmu\nospaceunit{m}}\xspace}

\def\fm   {\aunit{fm}\xspace}

\def\fb   {\ensuremath{\aunit{fb}}\xspace}
\def\invfb   {\ensuremath{\fb^{-1}}\xspace}

\def\deriv {\ensuremath{\mathrm{d}}}

\def\gsim{{~\raise.15em\hbox{$>$}\kern-.85em
          \lower.35em\hbox{$\sim$}~}\xspace}
\def\lsim{{~\raise.15em\hbox{$<$}\kern-.85em
          \lower.35em\hbox{$\sim$}~}\xspace}

\def\pt         {\ensuremath{p_{\mathrm{T}}}\xspace}

\def\ptot       {\ensuremath{p}\xspace}

\def\evtgen     {\mbox{\textsc{EvtGen}}\xspace}

\def\geant      {\mbox{\textsc{Geant4}}\xspace}

\def\photos     {\mbox{\textsc{Photos}}\xspace}

\def\pythia     {\mbox{\textsc{Pythia}}\xspace}

\def\tell1  {TELL1\xspace}
\def\ukl1   {UKL1\xspace}

\newcommand{\lhcborcid}[1]{\href{https://orcid.org/#1}{\hspace*{0.1em}\raisebox{-0.45ex}{\includegraphics[width=1em]{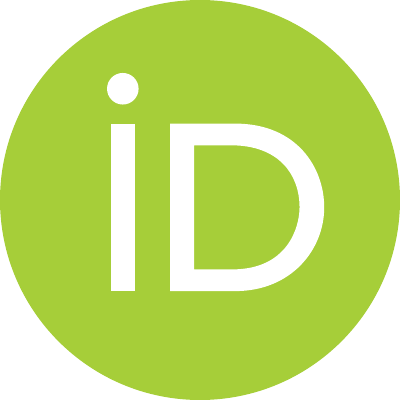}}}}

\hypersetup{
  colorlinks   = true, %Colours links instead of ugly boxes
  urlcolor     = blue, %Colour for external hyperlinks
  linkcolor    = blue, %Colour of internal links
  citecolor    = red   %Colour of citations
}

\ifEnableSectionTOCLinks
    \usepackage[explicit]{titlesec} % to change headings

    \let\oldcontentsline\contentsline
    \renewcommand

    \titleformat{\section}{\normalfont\Large\bf}{\hyperlink{tocsection.\thesection}{{\thesection} \parbox[t]{\dimexpr\textwidth-1pc}{#1}}}{1pc}{}

    \titleformat{\subsection}{\normalfont\bf}{\hyperlink{tocsubsection.\thesubsection}{{\thesubsection} \parbox[t]{\dimexpr\textwidth-1pc}{#1}}}{1pc}{}

    \titleformat{name=\section,numberless}[display]{}{}{0pt}{\normalfont\Huge\bfseries #1}
\fi

\usepackage{cite} % Allows for ranges in citations
\usepackage{mciteplus}
\usepackage{longtable} % only for template; not usually to be used in PAPERs

\begin{document}

%%%%%%%%%%%%%%%%%%%%%%%%%
%%%%% Title     %%%%%%%%%
%%%%%%%%%%%%%%%%%%%%%%%%%
\renewcommand{\thefootnote}{\fnsymbol{footnote}}
\setcounter{footnote}{1}

% %%%%%%% CHOOSE TITLE PAGE--------
%\onecolumn
%\input{title-LHCb-INT}
%\input{title-LHCb-ANA}
%\input{title-LHCb-CONF}
%\input{title-LHCb-FIGURE}
% ===============================================================================
% Purpose: LHCb-PAPER journal paper title page template
% Author: 
% Created on: 2010-09-25
% ===============================================================================

%%%%%%%%%%%%%%%%%%%%%%%%%
%%%%%  TITLE PAGE  %%%%%%
%%%%%%%%%%%%%%%%%%%%%%%%%
\begin{titlepage}
\pagenumbering{roman}

% Header ---------------------------------------------------
\vspace*{-1.5cm}
\centerline{\large EUROPEAN ORGANIZATION FOR NUCLEAR RESEARCH (CERN)}
\vspace*{1.5cm}
\noindent
\begin{tabular*}{\linewidth}{lc@{\extracolsep{\fill}}r@{\extracolsep{0pt}}}
\ifthenelse{\boolean{pdflatex}}% Logo format choice
{\vspace*{-1.5cm}\mbox{\!\!\!\includegraphics[width=.14\textwidth]{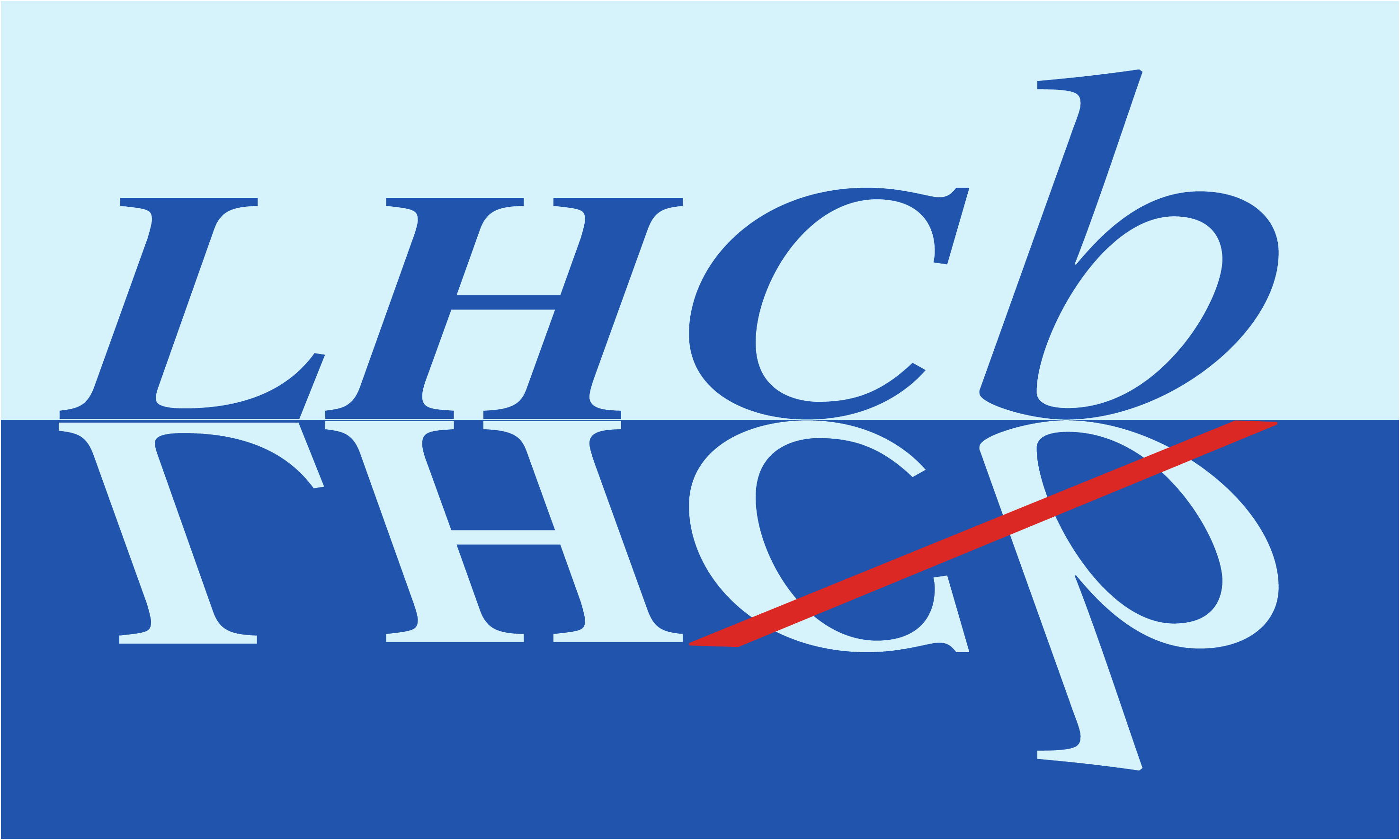}} & &}%
{\vspace*{-1.2cm}\mbox{\!\!\!\includegraphics[width=.12\textwidth]{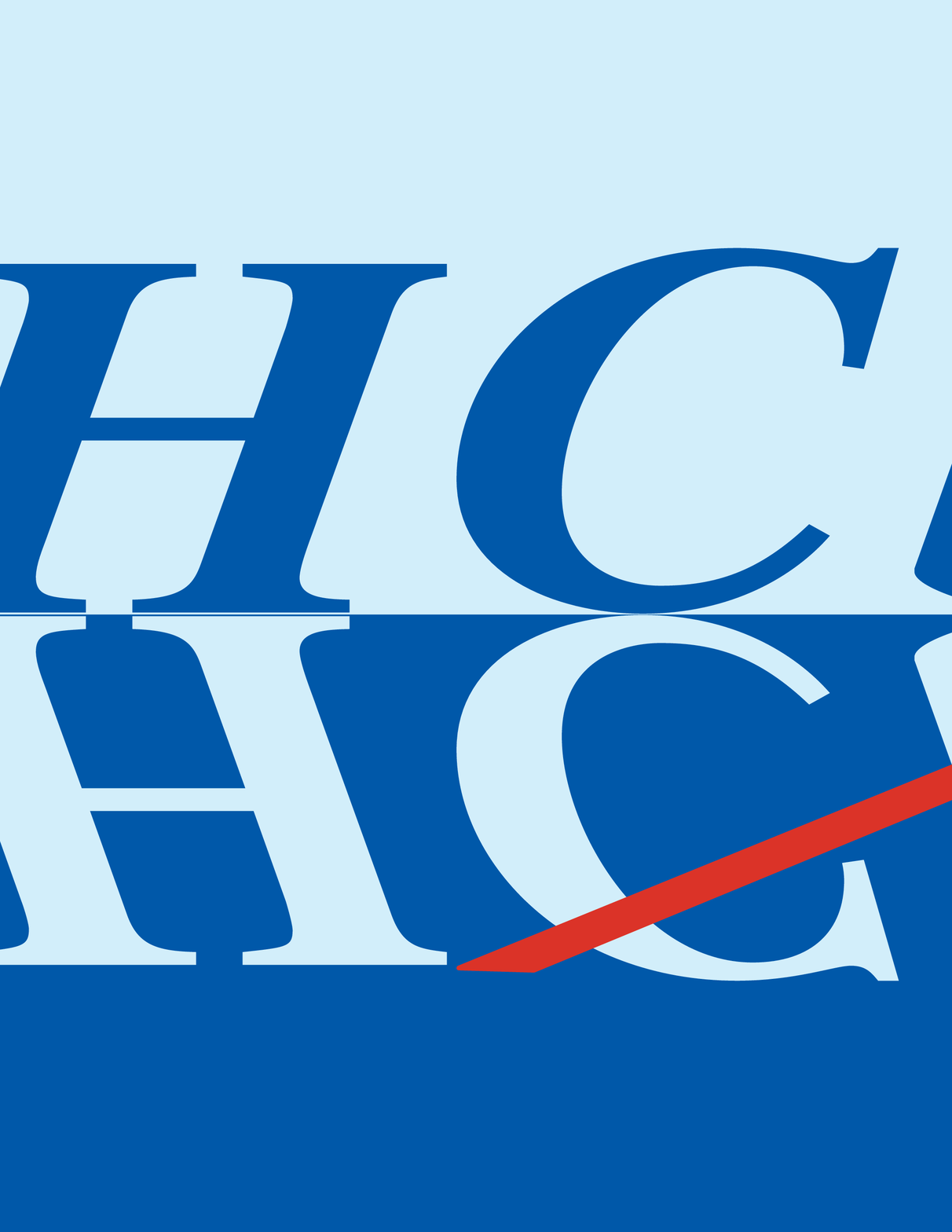}} & &}%
\\
 & & CERN-EP-2024-264 \\  % ID 
 & & LHCb-PAPER-2024-033 \\  % ID 
 & & 19 Jan 2025 \\ % & & \today \\ % Date - Can also hardwire e.g.: 23 March 2010
 & & \\
% not in paper \hline
\end{tabular*}

\vspace*{2.0cm}

% Title --------------------------------------------------
{\normalfont\bfseries\boldmath\huge
\begin{center}
% DO NOT EDIT HERE. Instead edit macro in main.tex to keep metadata correct
  \papertitle 
\end{center}
}

\vspace*{1.0cm}

% Authors -------------------------------------------------
\begin{center}
%In the footnote, replace 'paper' by 'Letter' in case of submission to PRL or PLB 
% Edit macro in main.tex to keep metadata correct
\paperauthors\footnote{Authors are listed at the end of this paper.}
\end{center}

\vspace{\fill}

% Abstract -----------------------------------------------
\begin{abstract}
\noindent
An amplitude analysis of the \mbox{$D_{s1}(2460)^+\to D_{s}^{+}\pi^{+}\pi^{-}$} transition is performed simultaneously in \mbox{$B^{0}\to D^{-}D_{s}^{+}\pi^{+}\pi^{-}$}, \mbox{$\B^{+}\to{{\offsetoverline{D}}\xspace}{}^{0} D_{s}^{+}\pi^{+}\pi^{-}$}, and \mbox{$B^{0}\to\D^{*-}D_{s}^{+}\pi^{+}\pi^{-}$} decays. 
The study is based on a data sample of proton-proton collisions recorded with the LHCb detector at centre-of-mass energies of $\sqrt{s}=7,8,$ and $13\,$TeV, corresponding to a total integrated luminosity of $9\,\rm{fb}^{-1}$. 
A clear double-peak structure is observed in the $m(\pi^{+}\pi^{-})$ spectrum of the \mbox{$D_{s1}(2460)^{+}\to D_{s}^{+}\pi^{+}\pi^{-}$} decay. 
The data can be described either with a model including $f_0(500)$, $f_0(980)$ and $f_2(1270)$ resonances, in which the contributions of $f_0(980)$ and $f_2(1270)$ are unexpectedly large, or with a model including $f_0(500)$, a doubly charged open-charm tetraquark state $T_{c\bar{s}}^{++}$ and its isospin partner $T_{c\bar{s}}^{0}$. 
If the former is considered implausible, the $T_{c\bar{s}}$ states are observed with high significance, and the data are consistent with isospin symmetry.
When imposing isospin constraints between the two $T_{c\bar{s}}$ states, their mass and width are determined to be $2327\pm13\pm13\,$MeV and $96\pm16\,^{+170}_{-23}\,$MeV, respectively, where the first uncertainty is statistical and the second is systematic. 
The mass is slightly below the $DK$ threshold, and a spin-parity of $0^+$ is favoured with high significance. 
\end{abstract}

\vspace*{1.0cm}

\begin{center}
    Published in \href{https://doi.org/10.1016/j.scib.2025.02.025}{%
  \textit{Sci. Bull.} \textbf{70}, 1432--1444 (2025)%
}
\end{center}

\vspace{\fill}

{\footnotesize 
% Edit macro in main.tex to keep metadata correct
\centerline{\copyright~\papercopyright. \href{\paperlicenceurl}{\paperlicence}.}}
\vspace*{2mm}

\end{titlepage}

%%%%%%%%%%%%%%%%%%%%%%%%%%%%%%%%
%%%%%  EOD OF TITLE PAGE  %%%%%%
%%%%%%%%%%%%%%%%%%%%%%%%%%%%%%%%

%  empty page follows the title page ----
\newpage
\setcounter{page}{2}
\mbox{~}
%\newpage
%
%% Author List ----------------------------
%%  You need to get a new author list!
%\input{LHCb_authorlist.tex}
%
%The author list for journal publications is provided by the Membership Committee shortly after 'approval to go to paper' has been given.
%%It will be made available on the page
%%\verb!http://www.physik.uzh.ch/~strauman/forMemCo/LHCb-PAPER-XXXX-XXX/! .
%It will be sent to you by email shortly after a paper number has beens assigned.
%The author list should be included already at first circulation, 
%to allow new members of the collaboration to verify whether they have been included correctly.
%Occasionally a misspelled name is corrected or associated institutions become full members.
%In that case, a new author list will be sent to you.
%In case line numbering doesn't work well after including the authorlist, try moving the \verb!\bigskip! after the last author to a separate line.
%
%
%The authorship for Conference Reports should be ``The LHCb
%  collaboration'', with a footnote giving the name(s) of the contact
%  author(s), but without the full list of collaboration names.

%\twocolumn
% %%%%%%%%%%%%% ---------

\renewcommand{\thefootnote}{\arabic{footnote}}
\setcounter{footnote}{0}

%%%%%%%%%%%%%%%%%%%%%%%%%%%%%%%%
%%%%%  Table of Content   %%%%%%
%%%%%%%%%%%%%%%%%%%%%%%%%%%%%%%%
%%%% Uncomment if desired
%\tableofcontents
\cleardoublepage

%%%%%%%%%%%%%%%%%%%%%%%%%
%%%%% Main text %%%%%%%%%
%%%%%%%%%%%%%%%%%%%%%%%%%

\pagestyle{plain} % restore page numbers for the main text
\setcounter{page}{1}
\pagenumbering{arabic}

%% Uncomment during review phase. 
%% Comment before a final submission.
%\linenumbers

%% This is the main body
%% It is useful to have a single file so comments are not missed in overleaf.
\section{Introduction}
\label{sec:Introduction}
Since the observation of the $D_{s0}^{*}(2317)^{+}$ and $D_{s1}(2460)^+$ mesons in 2003~\cite{BaBar:2003oey,CLEO:2003ggt}, their nature has been discussed extensively but without a firm conclusion~\cite{Chen:2004dy, Guo:2006rp, Lutz:2007sk, Rosner:2006vc, Feng:2012zze, Ortega:2016mms, Zhang:2009pn, Mehen:2004uj, PhysRevD.91.092011, Yeo:2024chk}.
The $D_{s0}^{*}(2317)^{+}$ and $D_{s1}(2460)^{+}$ masses are much lower than the expectation in the quark model~\cite{Godfrey:1985xj, PhysRevD.43.1679, PhysRevD.64.114004}. 
The observed degeneracy between the masses of the charmed mesons ($D_{0}^{*}(2300)^{0(\pm)}$ and $D_{1}(2430)^{0}$) and the charmed-strange mesons ($D_{s0}^{*}(2317)^{\pm}$ and $D_{s1}(2460)^{\pm}$) in the $(0^+, 1^+)$ doublet~\cite{PDG2024}, indicates that the $D_{s0}^{*}(2317)^{+}$ and $D_{s1}(2460)^+$ states probably have nontrivial internal structure.\footnote{The form $J^P$ denotes the total spin $J$ and parity $P$.} Due to their relatively small masses, their decays to $D^{(*)}K$ states are forbidden, resulting in total widths of a few \mev\ or less~\cite{BaBar:2006eep} and substantial branching fractions for the isospin-violating decays to $D_s^{(*)+}\piz$ final states.\footnote{Natural units in which $c = \hbar = 1$ are used throughout the article.}
%This motivates the study of three-body decays of $D_{s1}(2460)^+$, which could shed new light on this question.
The isospin-conserving decay, $\decay{D_{s1}(2460)^{+}}{\Dsp\pip\pim}$, also occurs at a sizeable rate~\cite{Belle:2003kup,BaBar:2006eep}.  
Theoretical calculations predict a double-bump lineshape in the $\pip\pim$ invariant-mass spectrum in this decay if the $D_{s1}(2460)^{+}$ meson is a $D^{(*)}K$ hadronic molecule~\cite{Tang:2023yls}. 

The LHCb collaboration recently reported the observation of two neutral tetraquark states, labelled $T_{cs0}(2900)^{0}$ and $T_{cs1}(2900)^{0}$, in $\decay{\Bm}{\Dp\Dm\Km}$ decays~\cite{LHCb-PAPER-2020-024,LHCb-PAPER-2020-025}.\footnote{The inclusion of charge-conjugate processes is implied throughout the article.} Later, LHCb also observed a doubly charged tetraquark and its neutral partner, labelled $T_{c\bar{s}}(2900)^{++}$ and $T_{c\bar{s}}(2900)^{0}$ with $I(J^P)=1(0^+)$ where $I$ denotes the isospin of the particle, in $\decay{B}{\Dbar \Dsp\pi}$ decays~\cite{LHCb-PAPER-2022-026,LHCb-PAPER-2022-027}. 
The proximity of the masses of these states to the $\Dstar\!\KorKbar\!^*$ threshold suggests that they might be $\Dstar\!\KorKbar\!^*$ bound states~\cite{Ke:2022ocs, Agaev:2022eyk, Duan:2023lcj}.
Furthermore, recent theoretical work suggests that the multiplet including $T_{c\bar{s}}(2900)^{++}$, $T_{c\bar{s}}(2900)^{0}$, and $T_{cs0}(2900)^{0}$ tetraquarks could be the radial excitation of a lighter multiplet containing the $D_{s0}^{*}(2317)$ state~\cite{Maiani:2024quj}. 
If so, scalar $DK$ bound states with isospin 1 near the $DK$ threshold are also expected and the relationship between this triplet and the $D_{s0}^{*}(2317)^{+}$ state needs further clarification~\cite{Terasaki:2003qa}.  
This motivates the study of three-body $D_{s1}(2460)^+$ decays to investigate the potential existence of $D_{s}\pi$ structures that may couple to the $DK$ channel. 
Such research could shed new light on the internal structures of the $D_{s0}^{*}(2317)^{+}$ and $D_{s1}(2460)^+$ mesons.

In this paper, the results from a combined amplitude analysis of the $\decay{D_{s1}(2460)^+}{\Dsp \pip\pim}$ transition in $\decay{\Bz}{\Dm\Dsp\pip\pim}$, $\decay{\Bp}{\Dzb\Dsp\pip\pim}$, and $\decay{\Bz}{\Dstarm\Dsp\pip\pim}$ decays are presented.
The study is based on a data sample of proton-proton ($pp$) collisions recorded with the LHCb detector at centre-of-mass energies of $\sqrt{s}=7,8,$ and $13\tev$, corresponding to a total integrated luminosity of $9\invfb$.
The use of fully reconstructed $B$-meson decays allows kinematic constraints on the decay chain to be applied, which improves the resolution, suppresses background contributions and enables the determination of the quantum numbers that affect the decay amplitudes.

\section{Detector and simulation}

The \lhcb detector~\cite{LHCb-DP-2008-001,LHCb-DP-2014-002} is a single-arm forward spectrometer covering the \mbox{pseudorapidity} range $2<\eta <5$,
designed for the study of particles containing \bquark or \cquark quarks. The detector includes a high-precision tracking system consisting of a silicon-strip vertex detector surrounding the $pp$
interaction region, a large-area silicon-strip detector located upstream of a dipole magnet with a bending power of about $4{\mathrm{\,T\,m}}$, and three stations of silicon-strip detectors and straw
drift tubes placed downstream of the magnet.
The tracking system provides a measurement of the momentum, \ptot, of charged particles with a relative uncertainty that varies from 0.5\% at low momentum to 1.0\% at $200\gev$.
The minimum distance of a track to a primary $pp$ collision vertex (PV), the impact parameter (IP), is measured with a resolution of $(15+29/\pt)\mum$,
where \pt is the component of the momentum transverse to the beam, in \gev.
Different types of charged hadrons are distinguished using information from two ring-imaging Cherenkov detectors. 

In this analysis, the online selections include hardware and software triggers. 
The hardware trigger criteria are satisfied by energy deposits in the calorimeter associated with the signal candidate decay. 
The software trigger requires a two-, three- or four-track secondary vertex with significant displacement from any primary $pp$ interaction vertex. 
In general, at least one charged particle must have a transverse momentum $\pt > 1.6\gev$ and be inconsistent with originating from a PV.

Simulation is required to determine the detector efficiency (which includes the detector acceptance  and selection requirements). 
In the simulation, $pp$ collisions are generated using \pythia~\cite{Sjostrand:2007gs, Sjostrand:2006za} with a specific \lhcb configuration~\cite{LHCb-PROC-2010-056}. 
Decays of unstable particles are described by \evtgen~\cite{Lange:2001uf}, in which final-state radiation is generated using \photos~\cite{davidson2015photos}.
The interaction of the generated particles with the detector, and its response, are implemented using the \geant toolkit~\cite{Allison:2006ve, Agostinelli:2002hh} as described in Ref.~\cite{LHCb-PROC-2011-006}. 
The underlying $pp$ interaction is reused multiple times using \textsc{ReDecay}~\cite{LHCb-DP-2018-004}, with an independently generated signal decay for each interaction.

\section{Selection}

The intermediate $D_{s1}(2460)^+$, $\Dstarm$, $\Dsp$, $\Dzb$, and $\Dm$ mesons are reconstructed through the following decays: $\decay{D_{s1}(2460)^+}{\Dsp\pip\pim}$, $\decay{\Dstarm}{\Dzb\pim}$, $\decay{\Dsp}{\Km\Kp\pip}$, $\decay{\Dzb}{\Kp\pim}$, and $\decay{\Dm}{\Kp\pim\pim}$. 
The charged $K$ and $\pi$ candidates are formed from well-reconstructed tracks that are inconsistent with originating from any PV, with particle identification information consistent with the corresponding mass hypothesis. 
The $\Dsp$, $\Dzb$, and $\Dm$ candidates  
%reconstructed from combinations of $K$ and $\pi$ candidates, and 
are required to have good vertex quality and significant displacement with respect to all PVs. 
Combinatorial background is suppressed with requirements on the outputs of trained Boosted Decision Tree (BDT) classifiers~\cite{Breiman, AdaBoost, Hocker:2007ht, TMVA4}, which take as input transverse momentum, tracking, vertexing and particle identification variables. 
Each BDT classifier is trained with simulated $D$ mesons from $B$ decays as signal and combinatorial background from mass sideband regions in data. 
The mass of the reconstructed candidates must be within $\pm15\mev$ of the corresponding known mass.\footnote{Unless otherwise specified, known values of particle properties are taken from Ref.~\cite{PDG2024}.}  
The $D_{s1}(2460)^+$ and $\Dstarm$ candidates are formed from combinations of charged pions with $\Dsp$ and $\Dzb$ candidates, respectively, where the combined vertices must be of good quality and displaced from all PVs. 
The mass difference between the $\Dstarm$ candidate and its $\Dzb$ decay product is required to be less than $150\mev$. 
The reconstructed $D_{s1}(2460)^+$ candidate mass must be less than $2700\mev$ for the fit to the $m(\Dsp\pip\pim)$ invariant-mass distribution used to determine the signal and background yields, and must be within $\pm10\mev$ of its known value for the subsequent amplitude fit. 

The $B$ candidates are formed by combining a $\Dbar{}^{(*)}$ and a $D_{s1}(2460)^{+}$ candidate, and requiring a well-reconstructed vertex which is displaced from all PVs. 
The momentum vector of each $B$ candidate is required to point back to the PV where it is hypothesised to have been produced, referred to hereafter as the associated PV\@. 
The reconstructed $B$-meson mass is required to be within $\pm20\mev$ of its known value. 
For the $\decay{\Bp}{\Dzb D_{s1}(2460)^+}$ signal channel, an additional requirement that the $\Dzb\pim$ invariant-mass be larger than $2020\mev$ is applied to veto potential $\Dstarm$ background contamination. 
After applying the selection criteria, around 5\% of the remaining events contain more than one $B$ candidate; in these cases only one is kept randomly. 

Kinematic fits~\cite{Hulsbergen:2005pu} are used at different stages of the data analysis. 
By default, the $B$ candidate is constrained to have originated from the associated PV\@. 
When considering  the $B$-candidate mass distribution, further constraints on the masses of the $\Dbar{}^{(*)}$ and $\Dsp$ candidates to their known values are applied. 
For the invariant-mass fit described in the next section, in addition to the above constraints, the $B$-candidate mass is fixed in the kinematic fit to its known value to improve the $D_{s1}(2460)^+$ mass resolution. 
Finally, an additional $D_{s1}(2460)^+$ mass constraint is applied in the fit used to obtain the four-momenta of the final-state particles for the amplitude analysis. 
This is valid since the small $D_{s1}(2460)^+$ width~\cite{BaBar:2006eep} has negligible impact on the analysis. 

\section{Invariant-mass fit}
\begin{figure}[tb]
    \centering
    \begin{minipage}{0.48\linewidth}
        \centering
        \includegraphics[width=\linewidth]{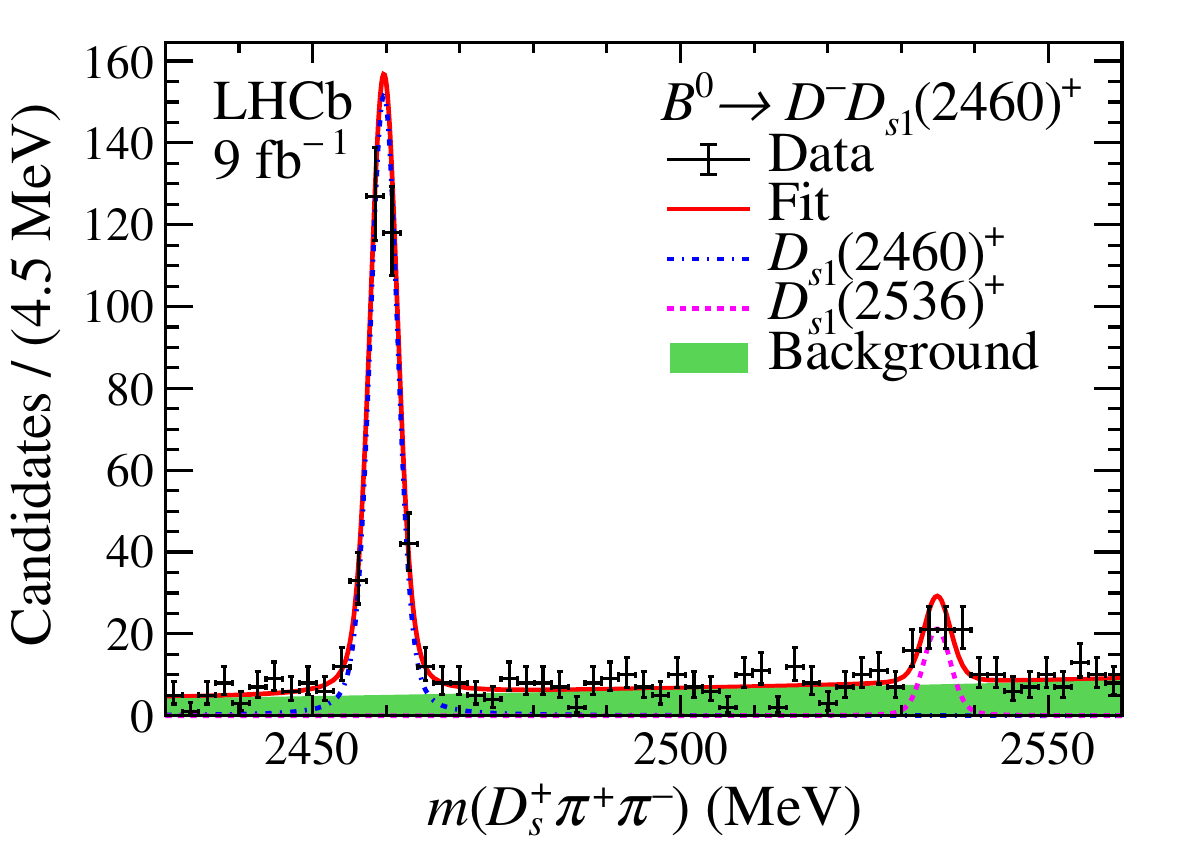}
    \end{minipage}
    \begin{minipage}{0.48\linewidth}
        \centering
        \includegraphics[width=\linewidth]{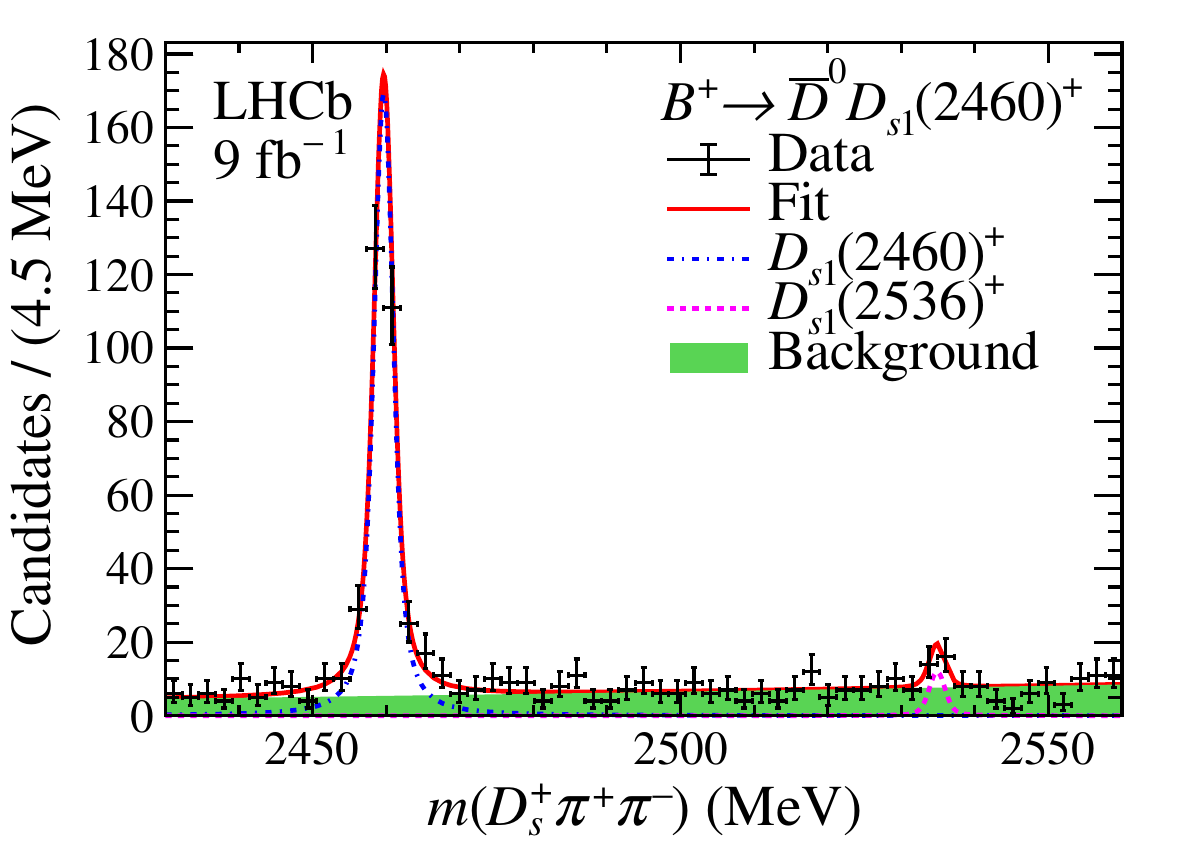}
    \end{minipage}
    \begin{minipage}{0.48\linewidth}
        \centering
        \includegraphics[width=\linewidth]{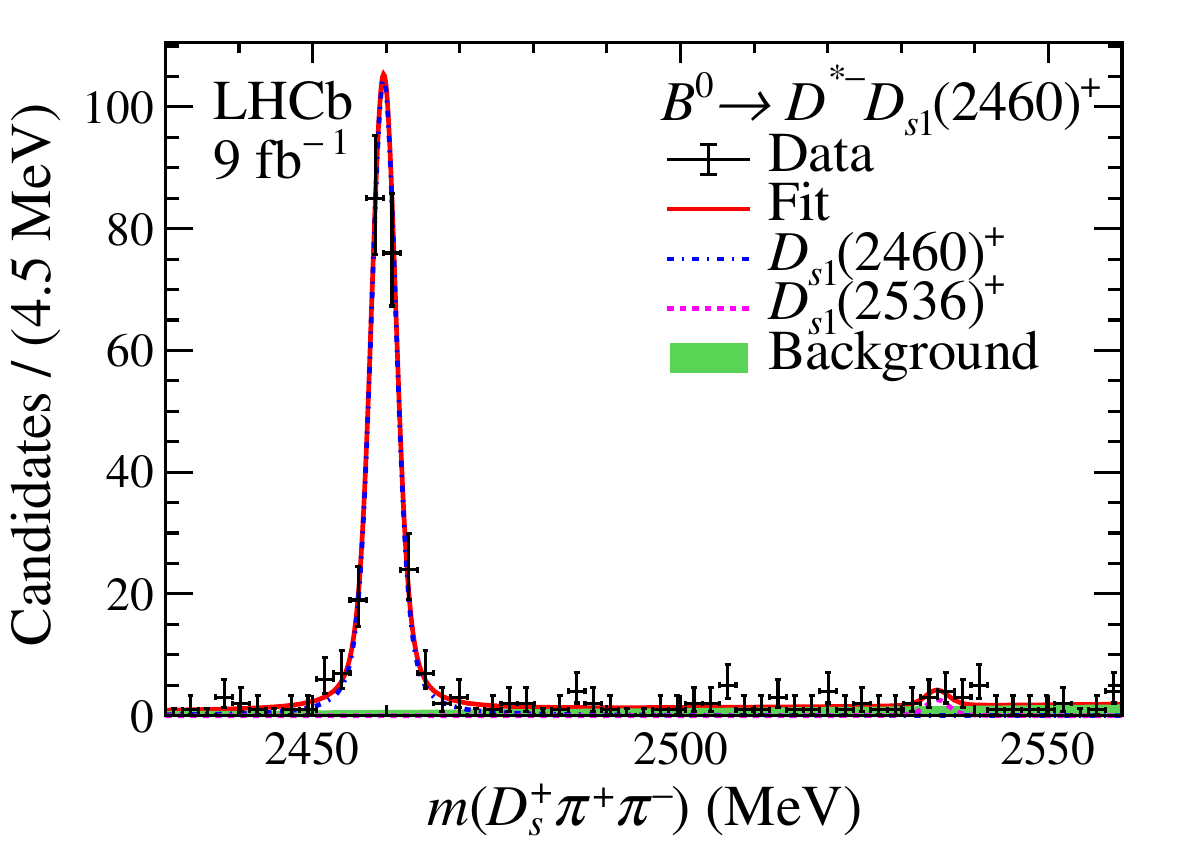}
    \end{minipage}   
   \caption{Invariant-mass distributions for the $D_{s1}(2460)^+$ candidates in the three signal channels (black dots with error bars) shown with the fit model (solid lines). The coloured region shows the combinatorial background. }
    \label{fig:mass_fit}
\end{figure}

The $m(\Dsp\pip\pim)$ invariant-mass spectra for the three signal channels after all selection criteria are shown in Fig.~\ref{fig:mass_fit}. 
Clear $D_{s1}(2460)^+$ signals are observed in all three channels, together with a smoothly varying combinatorial background and a small contribution from $\decay{D_{s1}(2536)^+}{\Dsp\pip\pim}$ decays.
Extended unbinned maximum-likelihood fits to the data, where the lower bound is the kinematic threshold $m_{\Dsp} + 2m_{\pip}$, are performed to extract signal and background yields for the subsequent amplitude fit. 
The $D_{s1}(2460)^+$ and $D_{s1}(2536)^+$ components are modelled by relativistic Breit--Wigner functions~\cite{PDG2024} convolved with a common Gaussian function to account for experimental resolution. 
The $D_{s1}(2460)^+$ Breit--Wigner mass and width are free to vary in the fit while the $D_{s1}(2536)^+$ parameters are fixed to their known values. 
The width of the Gaussian resolution function is shared between the $D_{s1}(2460)^+$ and $D_{s1}(2536)^+$ components and is allowed to vary in the fit.
The combinatorial background is modelled by an ARGUS function~\cite{ARGUS:1990hfq} with fixed kinematic threshold of $2247\mev$ and the shape parameter governing the slope is free to vary in the fit.

The fit results for the three signal channels are shown together with the data distributions in Fig.~\ref{fig:mass_fit}. 
The signal and background yields inside the signal region, defined to be $\pm 5\mev$ around the $D_{s1}(2460)^+$ known mass, are summarised in \tablename~\ref{tab:fit_yields}. In total around 800 signal events are obtained.

\begin{table}[tb]
\centering
\caption{Estimated signal and background yields inside the $D_{s1}(2460)^+$ mass window, together with the signal fraction. Note that the extrapolation of the background yield into the signal window allows uncertainties on the yields ($N$) to be smaller than $\sqrt{N}$.
}
\label{tab:fit_yields}
\begin{tabular}{lccc}
\hline
Channel & Signal yield & Background yield & Signal fraction (\%) \\
\hline
$\decay{\Bz}{\Dm \Dsp \pip \pim}$        & $305\pm20$ & $22\pm1$           & $93.2\pm0.4 $ \\
$\decay{\Bp}{\Dzb \Dsp \pip \pim}$       &$279\pm18$  & $24\pm1$           & $92.2\pm0.5$ \\
$\decay{\Bz}{\Dstarm \Dsp \pip \pim}$    & $205\pm14$ & $\phantom{2}4\pm1$ & $98.0\pm0.2$ \\
\hline
\end{tabular}
\end{table}

\section{Amplitude analysis formalism}

The amplitudes of the signal decays are expressed using the helicity formalism with an isobar approach~\cite{PhysRev.135.B551, PhysRev.166.1731, PhysRevD.11.3165}, where the total amplitude is a coherent sum of quasi-two-body amplitudes. The Blatt--Weisskopf factor in the amplitudes is fixed to $3.0\gev^{-1}$ in the amplitude fit. 
Each resonant lineshape is modelled by a relativistic Breit--Wigner (RBW) function, if not specified otherwise. 
The $f_0(980)$ state is modelled by a modified Flatt\'{e} function~\cite{FLATTE1976224,Bugg:2008ig}, with its parameters fixed according to Refs.~\cite{Bugg:2008ig,LHCb-PAPER-2013-069}. 
The $K$-matrix model suggested in Ref.~\cite{Aitchison:1972ay, Anisovich:2002ij} is used as an alternative $\pi\pi$ S-wave lineshape. Due to the small phase space available in $\decay{D_{s1}(2460)^+}{\Dsp\pip\pim}$ decays, the accessible $m(\pi\pi)$ range is limited and the analysis has little sensitivity to the parameters related to the $K\bar{K}$, $4\pi$, $\eta\eta$, and $\eta\eta^{\prime}$ coupled channels; such parameters are fixed to zero in the fit. 

An alternative model for $\pi\pi$ lineshapes, 
%described in Ref.~\cite{Tang:2023yls} 
based on the assumption of the $D_{s1}(2460)^+$ meson being a compact or a molecular state~\cite{Tang:2023yls} and hereafter referred to as the chiral dynamics model, is also tested. This model includes separate compact and molecular components, each obtained using a two-dimensional interpolation of the $\cos\theta_1$ (defined below) and $m(\pip\pim)$ distributions from the data points provided in Ref.~\cite{Tang:2023yls}, with relative fractions set by a parameter that is determined in the fit.

Another $K$-matrix model based on the scattering length approximation, considering $DK$ and $D_s\pi$ coupled-channel effects, is used as a possible lineshape to describe $T_{c\bar{s}}$ states. Details of this model can be found in the review of resonances in Ref.~\cite{PDG2024} and also in  Ref.~\cite{Fernandez-Ramirez:2019koa}.
The scattering $K$-matrix is parameterised as
\begin{equation}
    \label{eq:kscatk}
    K=
    \begin{pmatrix}
        \gamma & \beta \\
        \beta & \gamma_{2}
    \end{pmatrix}\,,
\end{equation}
where $\gamma$ is proportional to the scattering length in the elastic $DK$ channel, $\beta$ describes the coupling to the inelastic $D_s\pi$ channel, and $\gamma_2$ includes the possible interaction in the $D_{s}\pi$ channel. The lineshape for the $D_s\pi$ decay is  
\begin{equation}
    \label{eq:kscatm2}
    f^{K\text{-matrix}}=
    \frac{\beta^{2}\rho_{DK}+i\gamma_{2}(i\gamma\rho_{DK}-1)}{\beta^{2}\rho_{DK}\rho_{D_s\pi}+(i\gamma\rho_{DK}-1)(i\gamma_{2}\rho_{D_s\pi}-1)}\,,
\end{equation}
and the scattering length is 
\begin{equation}
    \label{eq:kscata}
    a = \frac{1}{8\pi\sqrt{s_{\rm thr}}}\left(\gamma+i\beta^2\rho_{D_{s}\pi}(s_{\rm thr})\right)\,,
\end{equation}
where $s_{\rm thr} =(m_D+m_K)^2$, 
and $\rho_{DK/D_{s}\pi}$ denotes the dimensionless phase-space term. 
The parameter $\gamma_2$ is fixed to zero in the amplitude fit, since there is little sensitivity to it in the channels under study. 

An unbinned maximum-likelihood fit is performed simultaneously to the three signal channels. 
The negative log-likelihood function for each channel is defined as 
\begin{equation}
    \label{eq:nllc}
    -\ln{\cal{L}} = -\sum_{i\in \text{data}}\ln{\left[f_{\rm sig}{\cal P}_{\rm sig}(\xi_{i};\Lambda)+(1 - f_{\rm sig}){\cal P}_{\rm bkg}(\xi_{i};\Lambda)\right]}\,,
\end{equation}
where $f_{\rm sig}$ denotes the signal fraction in the signal region, determined from the $m(\Dsp\pip\pim)$ fit described previously. %, and $f_{\rm bkg}=1-f_{\rm sig}$. 
The term ${\cal P}_{\rm sig}$ stands for the signal probability density function (PDF) for candidate $i$ at position $\xi_i$ in phase space, 
\begin{equation}
    \label{eq:probampsig}
    {\cal P}_{\rm sig}(\xi_{i};\Lambda) = \frac{|A(\xi_{i};\Lambda)|^{2}}{\int|A(\xi;\Lambda)|^{2}\varepsilon(\xi)\mathrm{d}\xi}\,,
\end{equation}
where $\Lambda$ denotes the set of parameters to be determined in the fit to data.
Here, $A$ is the total amplitude and $\varepsilon(\xi)$ denotes the efficiency variation over the phase space, which is determined from simulated samples after applying simulation corrections on the tracking and trigger efficiencies, obtained using control samples~\cite{LHCb-DP-2013-002, LHCb-DP-2012-004}. The masses and widths of the considered resonances and their coupling constants are shared between the three channels. The background PDF ${\cal P}_{\rm bkg}(\xi_{i};\Lambda)$ is estimated using events in the $m(\Dsp \pip\pim)$ sidebands $[2247, 2440]\mev$ and $[2560, 2660]\mev$, and is described using kernel density estimation~\cite{Poluektov:2014rxa}. 

The results of the amplitude analysis are expressed in terms of fit fractions. 
The fit fraction $F_{i}$ for resonance $i$ is calculated based on the fitted values of the parameters $\hat{\Lambda}$, and is defined as
\begin{equation}
    \label{eq:ff}
    F_{i} = \frac{\int|A_{i}(\xi;\hat{\Lambda})|^{2}\mathrm{d}\xi}{\int|\sum_{k}A_{k}(\xi;\hat{\Lambda})|^{2}\deriv \xi}\,,
\end{equation}
where $A_i(\xi)$ is the contribution to the amplitude from resonance $i$.
The interference between any two components $i$ and $j$, $F_{ij}$, is quantified as
\begin{equation}
    \label{eq:ffitf}
    F_{ij} = \frac{\int2\,\rm{Re}\left\{A_{i} A_{j}^{*}(\xi;\hat{\Lambda})\right\}\mathrm{d}\xi}{\int|\sum_{k}A_{k}(\xi;\hat{\Lambda})|^{2}\deriv \xi}\,.
\end{equation}

\section{Amplitude fit}
\begin{figure}[tb]
    \centering
    \begin{minipage}{0.48\linewidth}
        \centering
        \includegraphics[width=\linewidth]{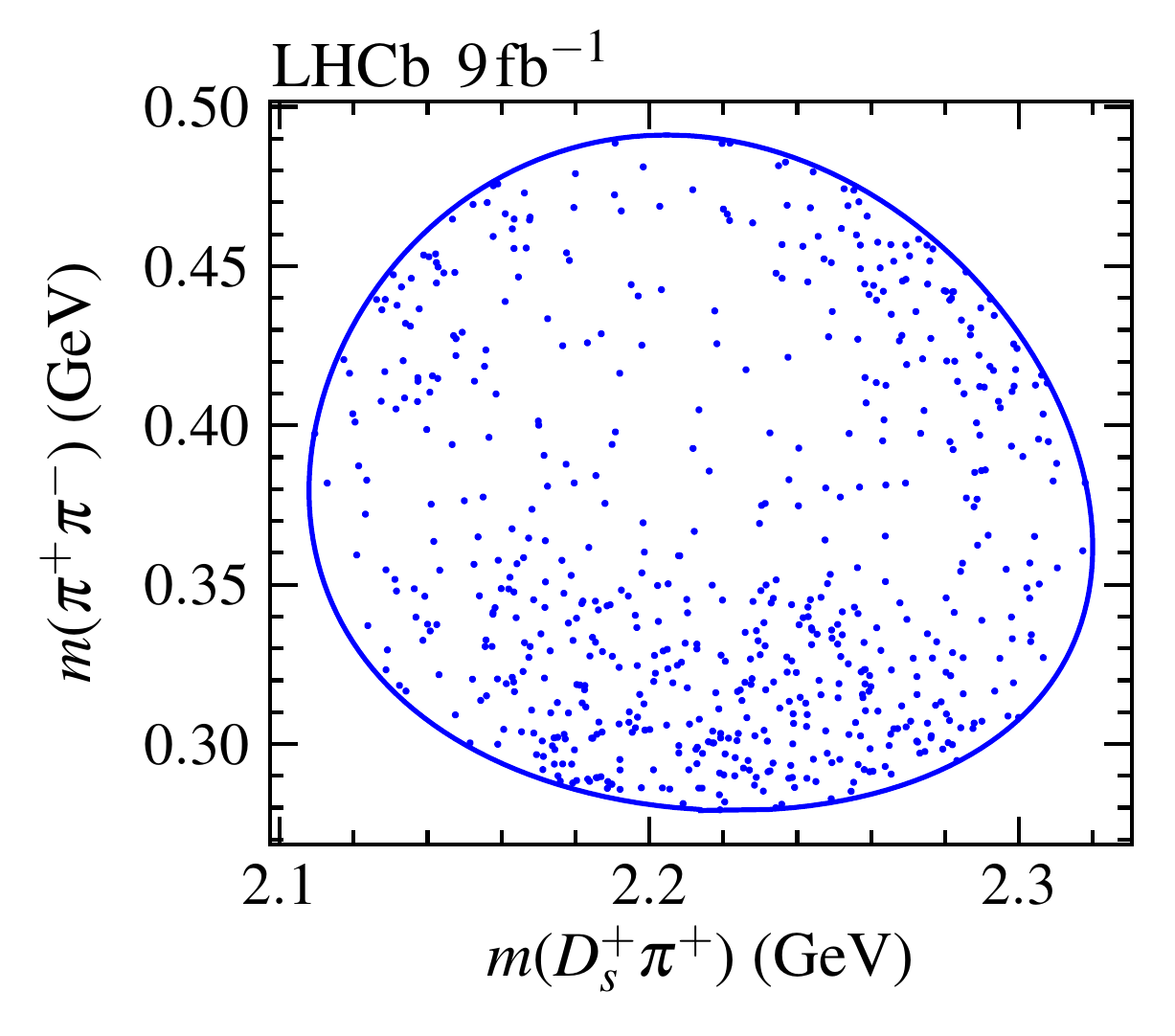}
        \begin{minipage}{\linewidth}
            \vspace{-33.0em}\hspace{16em}
            (a)
        \end{minipage}
    \end{minipage}
    \begin{minipage}{0.48\linewidth}
        \centering
        \includegraphics[width=\linewidth]{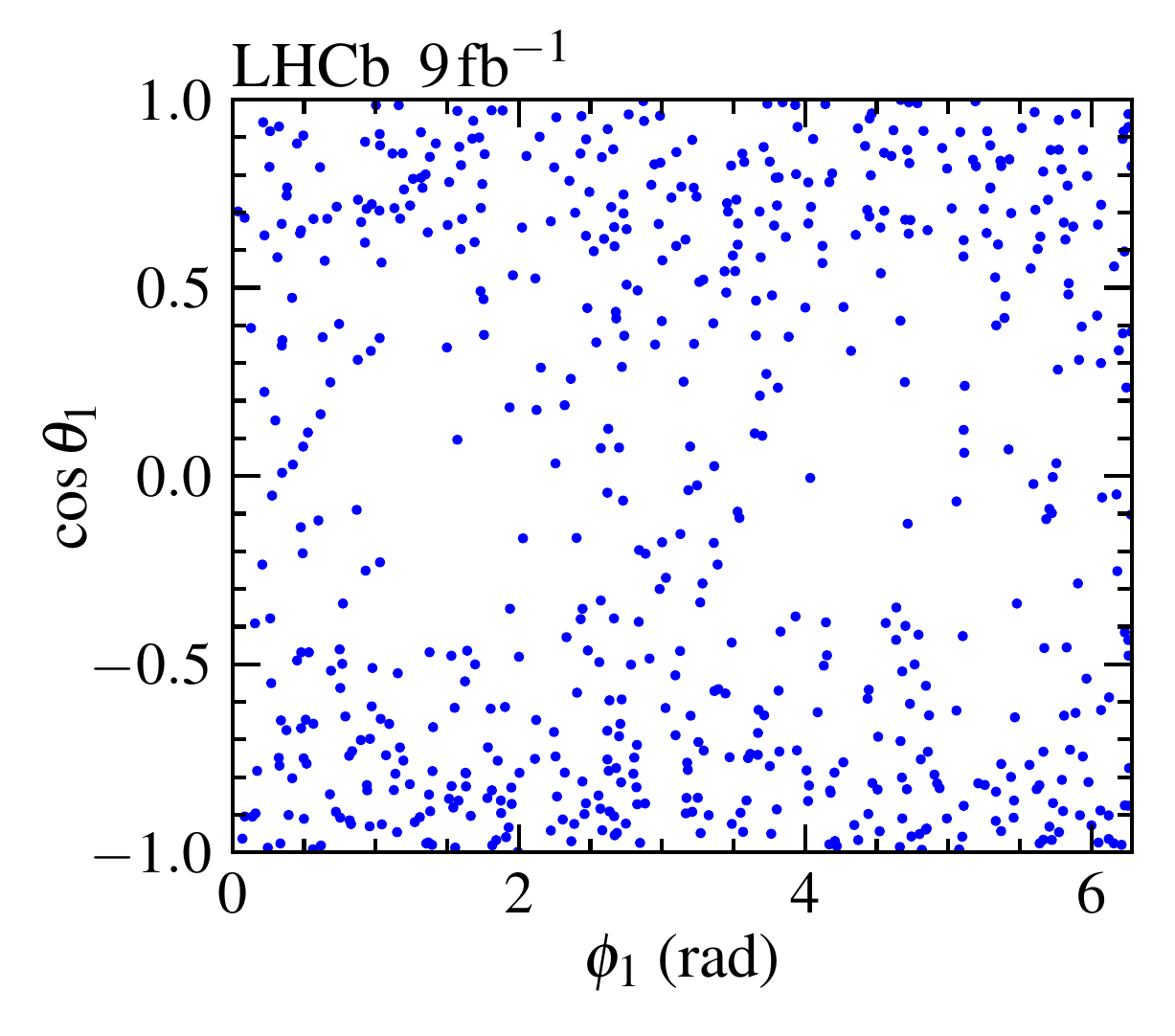}
        \begin{minipage}{\linewidth}
            \vspace{-33.0em}\hspace{16em}
            (b)
        \end{minipage}
    \end{minipage}
    \vspace{-4ex}
    \caption{Distributions of selected candidates in the (a)~$m(D_{s}^{+}\pi^{+})-m(\pi^{+}\pi^{-})$ plane and (b)~$\phi_1-\cos\theta_1$ plane, combining the \mbox{$B^{0}\to D^{-} D_{s1}(2460)^+$} and \mbox{$B^{+}\to {{\ensuremath{\offsetoverline{\PD}}}\xspace}{}^{0} D_{s1}(2460)^+$} channels. The blue solid line on the left plot denotes the boundary of the \mbox{$D_{s1}(2460)^+\to D_{s}^{+}\pi^{+}\pi^{-}$} Dalitz plot. Background contributions are not subtracted and no efficiency corrections are applied.}
    \label{fig:mpipi_mdspi}
\end{figure}

\begin{figure}[tb]
    \centering
    \begin{minipage}{0.45\linewidth}
        \centering
        \includegraphics[width=\linewidth]{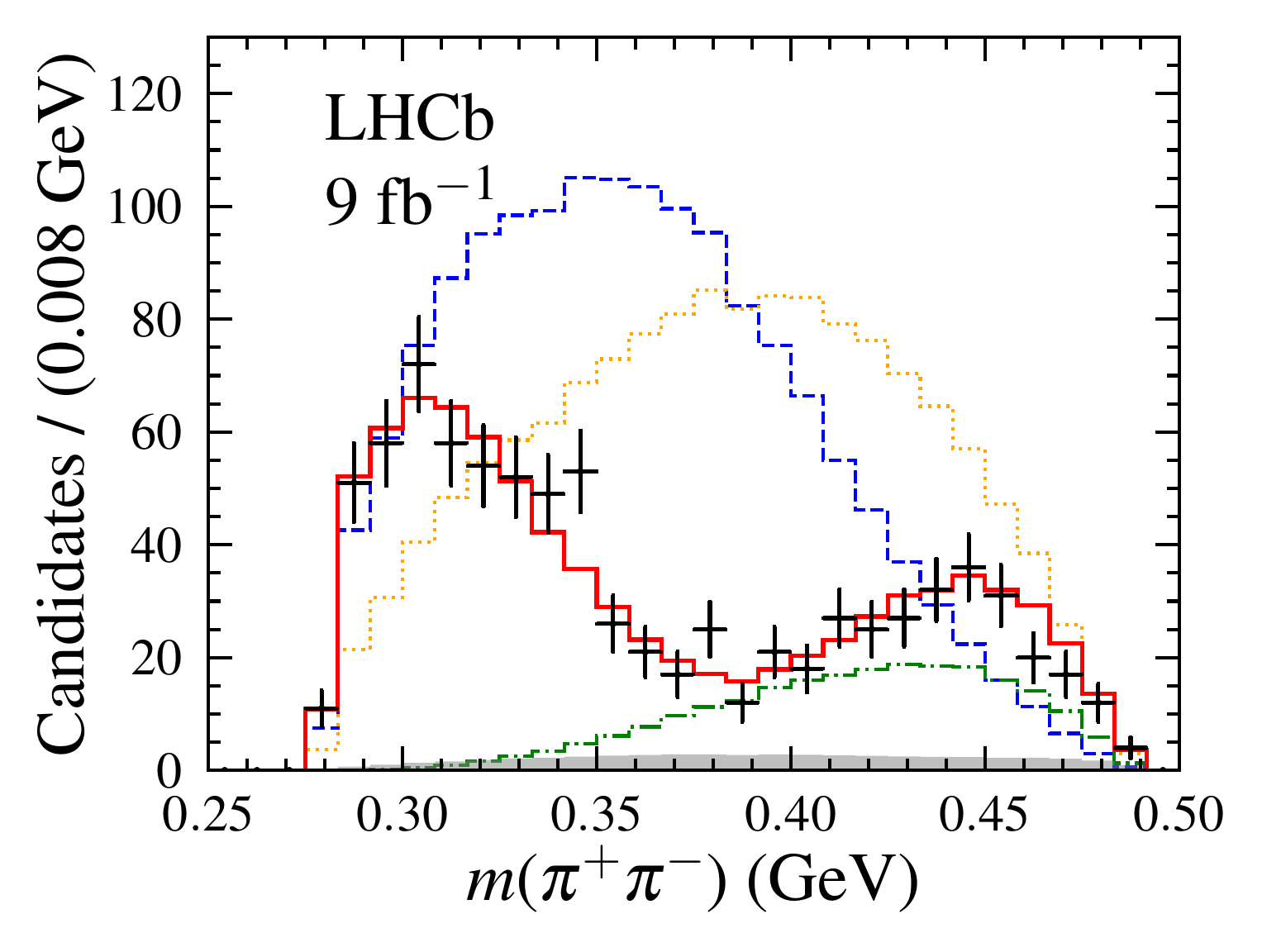}
        \begin{minipage}{\linewidth}
            \vspace{-22.5em}\hspace{14.0em}
            (a)
        \end{minipage}
        \vspace{-2em}
    \end{minipage}
    \begin{minipage}{0.45\linewidth}
        \centering
        \includegraphics[width=\linewidth]{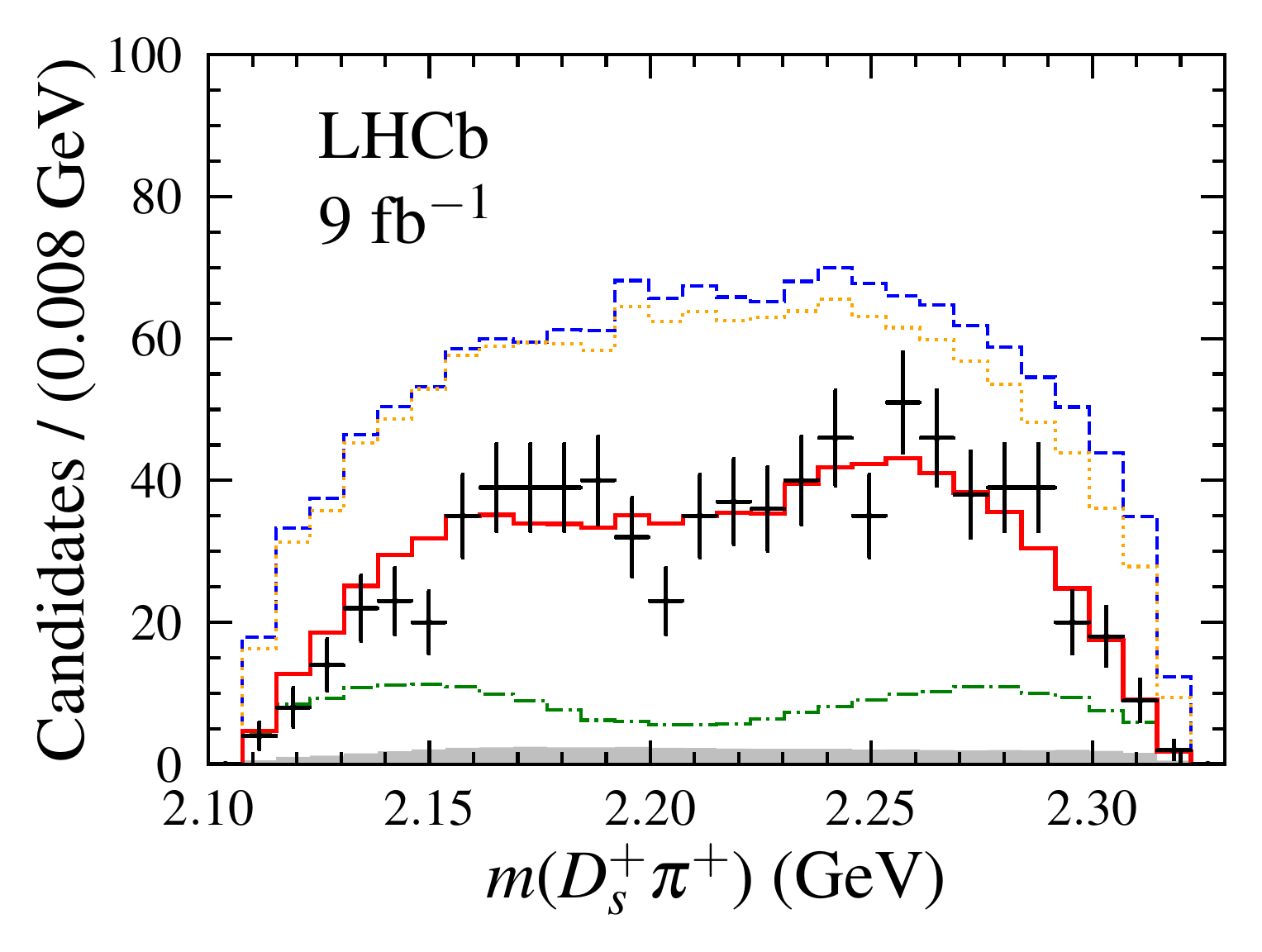}
        \begin{minipage}{\linewidth}
            \vspace{-22.5em}\hspace{14.0em}
            (b)
        \end{minipage}
        \vspace{-2em}
    \end{minipage}
    
        \hspace{3em}
        \includegraphics[width=0.40\linewidth]{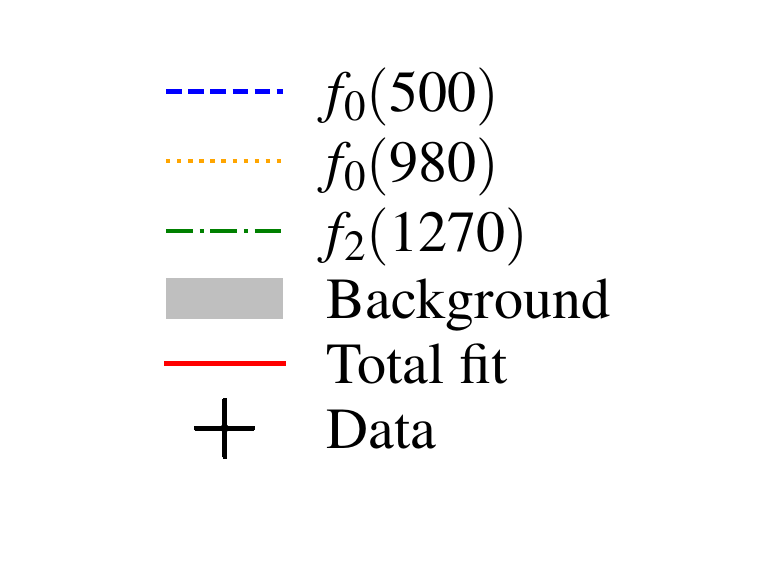}
        \hspace{-1.2em}
    \begin{minipage}{0.45\linewidth}
        \vspace{-9em}
        \centering
        \includegraphics[width=\linewidth]{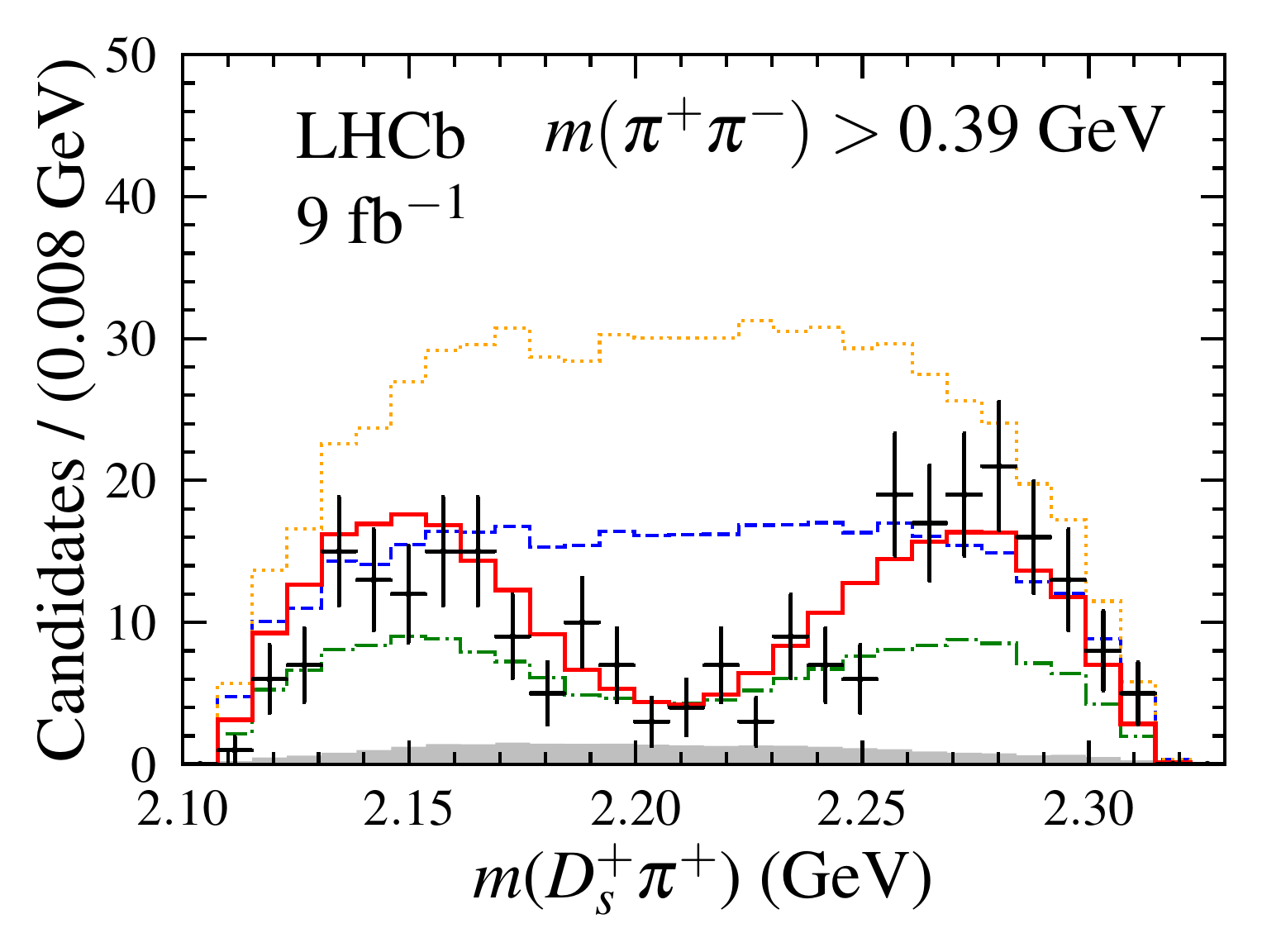}
        \begin{minipage}{\linewidth}
            \vspace{-22.5em}\hspace{14.0em}
            (c)
        \end{minipage}
    \end{minipage}
    \vspace{-2em}
    \caption{Comparison between data (black dots with error bars) and results of the fit with the $f_0(500)+f_0(980)+f_2(1270)$ model (red solid line). The distributions are for the three channels combined in (a)~$m(\pi^{+}\pi^{-})$, (b)~$m(D_{s}^{+}\pi^{+})$, and (c)~$m(D_{s}^{+}\pi^{+})$ requiring $m(\pi^{+}\pi^{-})>0.39\,\mathrm{GeV}$. Individual components, corresponding to the background contribution estimated from $m(D_{s}^{+}\pi^{+}\pi^{-})$ sideband regions (gray-filled) and the different resonant contributions (coloured dashed lines), are also shown as indicated in the legend. }
    \label{fig:model_sgm_f0980_f21270}
\end{figure}

\figurename~\ref{fig:mpipi_mdspi} shows the distributions of selected candidates in the $m(\Dsp\pip)-m(\pip\pim)$ and $\phi_{1}-\cos\theta_{1}$ planes, combining the $\decay{\Bz}{\Dm D_{s1}(2460)^+}$ and $\decay{\Bp}{\Dzb D_{s1}(2460)^+}$ channels. 
These four variables fully describe the dynamics of the two included decays, while in the \decay{\Bz}{\Dstarm D_{s1}(2460)^{+}} case two additional angles related to \Dstarm\ decays are necessary, making it inappropriate to combine the three distributions.
The variable $\theta_1$ is the helicity angle of the $R(\pi\pi)$ resonance in $\decay{D_{s1}(2460)^+}{\Dsp R(\pi\pi)}$ decays and $\phi_1$ is the angle between the decay planes of $\decay{D_{s1}(2460)^+}{\Dsp R(\pi\pi)}$ and $\decay{R(\pi\pi)}{\pip \pim}$ decays. 
Complete definitions of the angles are shown in Fig.~\ref{fig:helangle} in the supplemental material. 
Efficiency-corrected one-dimensional projections onto each of the phase-space variables, including $m(\Dsp\pim)$ which is expected to be consistent with $m(\Dsp\pip)$ due to isospin symmetry, are shown in Figs.~\ref{fig:noeffD} and~\ref{fig:noeffDst} in the supplemental material. 
The data cluster in three phase-space regions, two of which are seen as a double bump in the $m(\Dsp\pip)$ distribution when requiring \mbox{$m(\pip\pip)>0.39\gev$}, as shown in Fig.~\hyperref[fig:model_sgm_f0980_f21270]{3(c)} for example. 
The corresponding distributions for $\decay{\Bz}{\Dstarm D_{s1}(2460)^{+}}$ decays and projections on the efficiency-corrected $m(\pip\pim)$, $m(\Dsp\pip)$, $\cos\theta_1$ and $\phi_1$ distributions for the three channels are shown in the supplemental material.  

In the $D_{s1}(2460)^+\to \Dsp\pip\pim$ decays, conventional quark-antiquark resonances are only possible in the $\pi\pi$ channel. 
Therefore, models with only $\pi\pi$ resonances are attempted first. 

\begin{table}[tb]
\centering
\caption{Relative negative log likelihoods ($\Delta$NLL) and numbers of fit parameters for all tested models. The $\Delta$NLL value is calculated with the model $f_0(500)+K\text{-matrix}~T_{c\bar{s}}(0^+)$ as reference. Smaller values of $\Delta$NLL correspond to better fits. The upper section is for models containing only $\pi\pi$ resonances, while the lower section is for models with $T_{c\bar{s}}$ contributions.}
\label{tab:fit_res_sum}
\renewcommand{\arraystretch}{1.2}
\begin{tabular}{lcc}
\hline
Model                                   & $\Delta$NLL & Number of fit parameters \\ 
\hline
Chiral dynamics                         & $177.8$     & 5 \\    
$K$-matrix $\pi\pi$ S-wave                & $249.0$     & 6 \\
$f_0(500)+f_0(980)$                     & $245.2$     & 8 \\
$f_0(500)+f_0(980)+\rho(770)^0$           & $148.0$     & 12 \\
%\hline
$f_0(500)+f_0(980)+f_2(1270)$           & $3.7$       & 12 \\
$f_0(500)+f_0(980)+f_2(1270)+\rho(770)^0$ & $-2.8$      & 16 \\
$K$-matrix $\pi\pi$ S-wave $+f_2(1270)$   & $5.9$       & 10 \\
\hline
$f_0(500)+\text{RBW}~T_{c\bar{s}}(0^+)$           & $3.5$    & 10 \\
$f_0(500)+K\text{-matrix}~T_{c\bar{s}}(0^+)$      & 0.0      & 10 \\
$f_0(500)+f_0(980)+\text{RBW}~T_{c\bar{s}}(0^+)$  & $-3.0$   & 12 \\
$f_0(500)+\rho(770)^0+\text{RBW}~T_{c\bar{s}}(0^+)$ & $-1.1$   & 14 \\
$f_0(500)+f_2(1270)+\text{RBW}~T_{c\bar{s}}(0^+)$ & $-4.3$   & 14 \\
$f_0(500)+\text{RBW}~T_{c\bar{s}}(1^-)$           & $62.9$   & 12 \\

\hline
\end{tabular}
\end{table}

\begin{table}[tb]
\centering
\caption{Summary of fit results for different models described in detail. Values quoted without uncertainties are taken from previous measurements~\cite{PDG2024,LHCb-PAPER-2013-069} and are fixed in the fits. The two sources of uncertainty are statistical and systematic. For the models containing $T_{c\bar{s}}$ states the quoted fit fraction is the value for each of the isospin partners, and the quoted $T_{c\bar{s}}$ mass and width parameters are the pole mass and width.}
\label{tab:fit_res_sum_2}
\resizebox{\linewidth}{!}{
\renewcommand{\arraystretch}{1.2}
\begin{tabular}{lcccc}
\hline
Model  & Resonance & Mass (MeV) & Width (MeV) & Fractions (\%) \\
\hline
 \multirow{3}{*}{$f_0(500)+f_0(980)+f_2(1270)$}       
  & $f_0(500)$  & $376\pm9\pm16$ & $175\pm23\pm16$ & $197\pm35\pm23$ \\
  & $f_0(980)$  &  945.5         &  167            & $187\pm38\pm43$ \\
  & $f_2(1270)$ & 1275.4         & 186.6           & $29\pm2\pm1$ \\
\hline
\multirow{2}{*}{$f_0(500)+\text{RBW}~T_{c\bar{s}}(0^+)$}       & $f_0(500)$           & $464\pm23\pm14$ & $214\pm28\pm8$ & $199\,^{+42}_{-47}\pm39$ \\
  & $T_{c\bar{s}}^{++}/T_{c\bar{s}}^{0}$ & $2312\pm27\pm11$ & $264\pm46\pm21$ & $126\,^{+27}_{-17}\pm20$ \\
\hline
\multirow{2}{*}{$f_0(500)+K\text{-matrix}~T_{c\bar{s}}(0^+)$}   &   $f_0(500)$  & $474\pm30\pm18$   & $224\pm23\pm16$& $248\,^{+40}_{-54}\pm39$ \\
  & $T_{c\bar{s}}^{++}/T_{c\bar{s}}^{0}$ & $2327\pm13\pm13$  & $96\pm16\pm23$ & $156\,^{+27}_{-38}\pm25$ \\
\hline
\end{tabular}
}
\end{table}

A summary of the relative negative log likelihoods ($\Delta$NLLs) for different models containing only $\pi\pi$ resonances is given in the upper section of Table~\ref{tab:fit_res_sum}. 
When considering only $\pi\pi$ resonance contributions, two models give the best description of the data without adding nonsignificant resonant contributions. 
One contains the $f_0(500)$, $f_0(980)$ and $f_2(1270)$ states, and the other describes the $\pi\pi$ S-wave with a $K$-matrix component and includes an additional $f_2(1270)$ resonance. 
The projections onto $m(\pip \pim)$, $m(\Dsp\pip)$ and $m(\Dsp\pip)$ requiring $m(\pip\pim)>0.39\gev$ for the first model are shown in Fig.~\ref{fig:model_sgm_f0980_f21270}.
The corresponding $m(\pip \pim)-m(\Dsp\pip)$ and $\phi_1-\cos \theta_1$ distributions are shown in Fig.~\ref{fig:model_sgm_f0980_f21270_chisq} in the supplemental material. 
The inclusion of the $f_2(1270)$ component is necessary to obtain good agreement with the data. The fits with models excluding this component have much higher $\Delta$NLL values, as seen in Table~\ref{tab:fit_res_sum}.
The inclusion of a $\rho(770)^0$ component leads to a small improvement in $\Delta$NLL, but this is insignificant bearing in mind the change in the number of free parameters of the fit.

Although these models give reasonable descriptions of the data across the \mbox{$D_{s1}(2460)^+ \to \Dsp \pip\pim$} phase space, there are several reasons to doubt their credibility as physical descriptions of the decay amplitude.  
First, there is a large contribution from the $f_2(1270)$ resonance, despite the fact that the kinematic upper limit of $m(\pip\pim)$ is around \mbox{$m_{f_{2}(1270)} - 4\cdot\Gamma_{f_{2}(1270)}$}, where $m_{f_2(1270)}$ and $\Gamma_{f_2(1270)}$ are the known $f_2(1270)$ mass and width. 
Such a large contribution from the tail of a lineshape is barely plausible. 
A similar argument applies to the $f_0(980)$ contribution. 
Secondly, the model including both $f_0(500)$ and $f_0(980)$ components requires large destructive interference to generate the observed $m(\pip\pim)$ structures, with the total fit fraction summing to $(413\pm66)\%$. 
%This destructive interference between different components of the $\pi\pi$ S-wave is also present in the $K$-matrix description. The lineshape caused by destructive interference between the $f_{0}(500)$ and $f_{0}(980)$ poles is also present in the $K$-matrix description. 
While large interference effects are inevitable in an amplitude analysis with broad components overlapping in a small phase space, the dramatic effects seen here are markedly different from what is seen in $\pi\pi$ S-waves in other processes~\cite{LHCb-PAPER-2013-069,LHCb-PAPER-2014-012,LHCb-PAPER-2014-070,LHCb-PAPER-2019-017,LHCb-PAPER-2022-016,LHCb-PAPER-2022-030,ABLIKIM2005243,PhysRevLett.118.012001}.
Furthermore, as shown in Table~\ref{tab:fit_res_sum_2}, the fitted value of the $f_0(500)$ mass, and to a lesser extent also that of its width, are different from what is seen in other processes~\cite{PDG2024}. 

In addition to the results shown above, some fits with the $f_0(500)$, $f_0(980)$ and $f_2(1270)$ states converge to another solution with a similar $\Delta$NLL value. This solution, however, has a very large interference between the $f_0(500)$ and $f_0(980)$ resonances leading to unstable results.
This solution also finds the $f_0(500)$ mass to be even smaller (around 190\mev) and the $f_0(500)$ width larger than 700\mev. It is not discussed further. 

Due to these unsatisfactory aspects of the fit results for models containing only $\pip\pim$ resonances, models with additional exotic states decaying to $\Dsp\pipm$, referred to as $T_{c\bar{s}}^{++}$ and $T_{c\bar{s}}^{0}$ states, are considered. 
In all cases, both the $T_{c\bar{s}}^{++}$ and $T_{c\bar{s}}^{0}$ isospin partners are included, and by default their coupling constants, masses and widths (or parameters $\beta$ and $\gamma$ for the $K$-matrix model) are constrained to be the same following isospin symmetry. 

\begin{figure}[tb]
    \centering
    \begin{minipage}{0.45\linewidth}
        \centering
        \includegraphics[width=\linewidth]{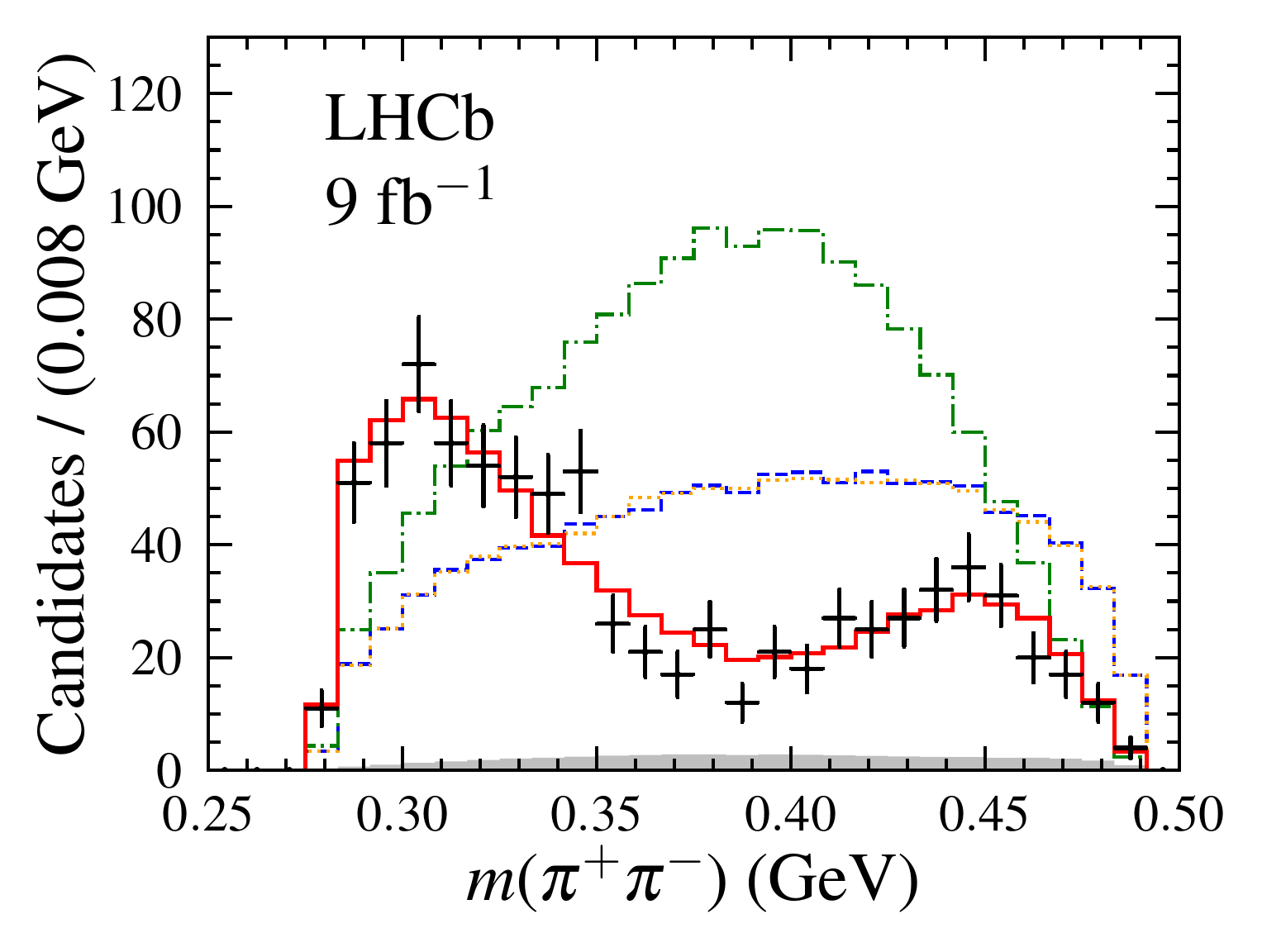}
        \begin{minipage}{\linewidth}
            \vspace{-22.5em}\hspace{14.0em}
            (a)
        \end{minipage}
        \vspace{-2em}
    \end{minipage}
    \begin{minipage}{0.45\linewidth}
        \centering
        \includegraphics[width=\linewidth]{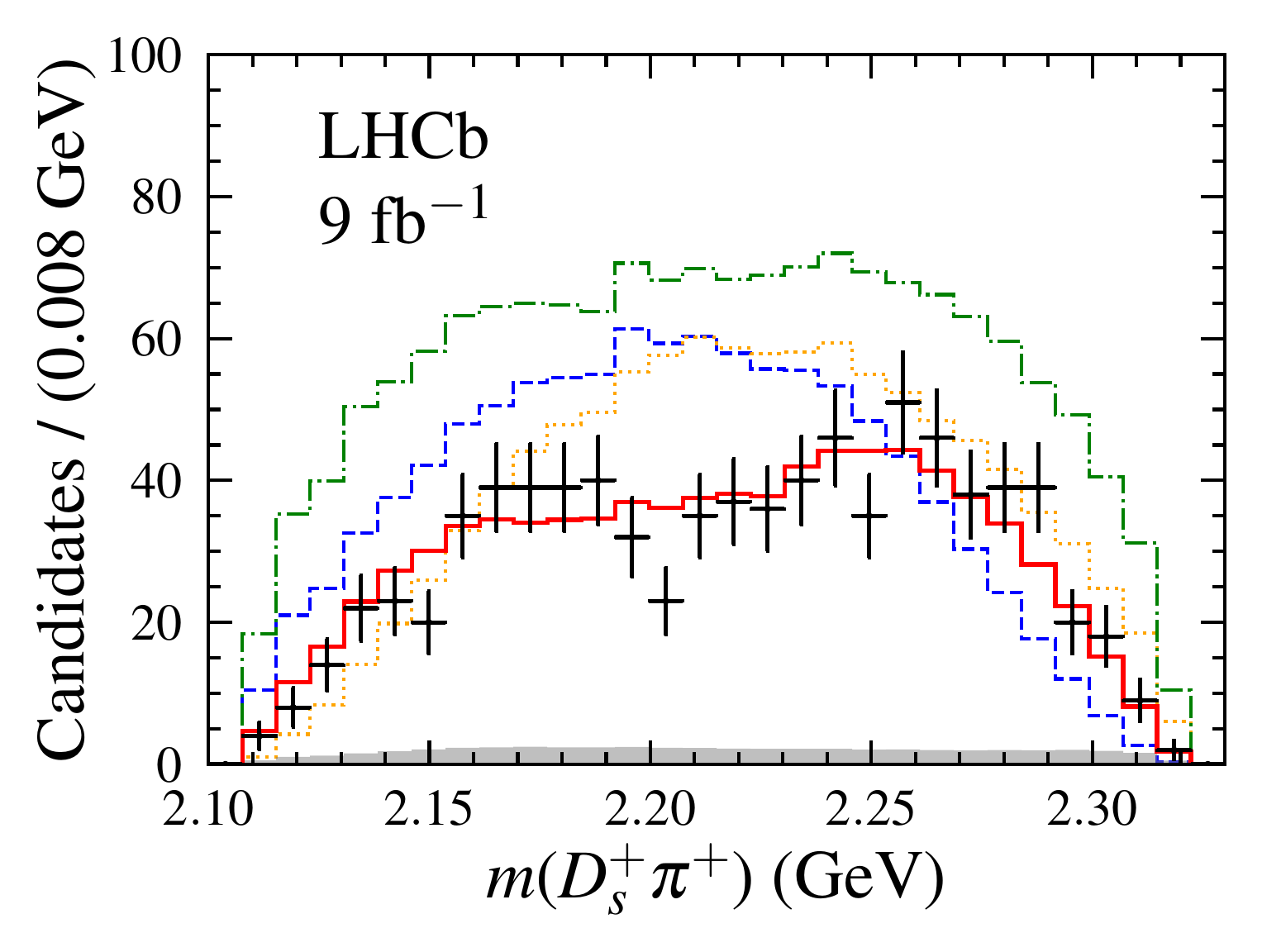}
        \begin{minipage}{\linewidth}
            \vspace{-22.5em}\hspace{14.0em}
            (b)
        \end{minipage}
        \vspace{-2em}
    \end{minipage}
    
        \hspace{3em}
        \includegraphics[width=0.40\linewidth]{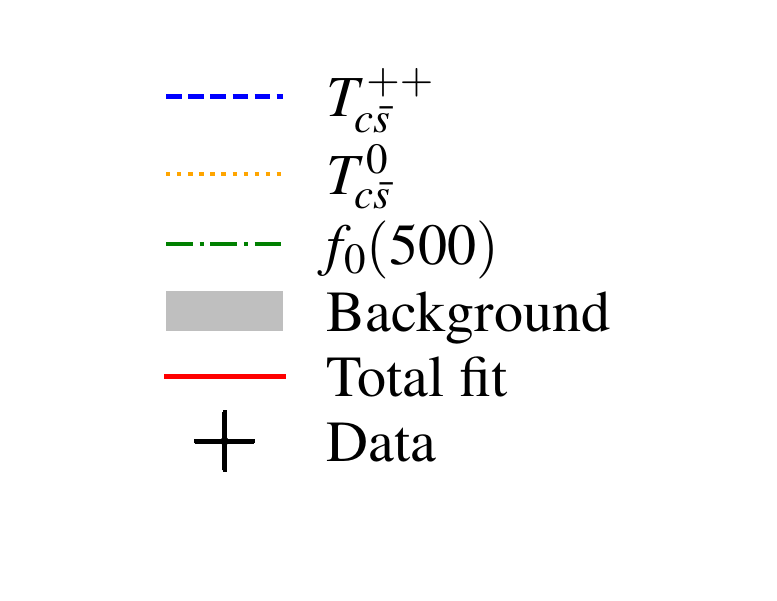}
        \hspace{-1.2em}
    \begin{minipage}{0.45\linewidth}
        \vspace{-9em}
        \centering
        \includegraphics[width=\linewidth]{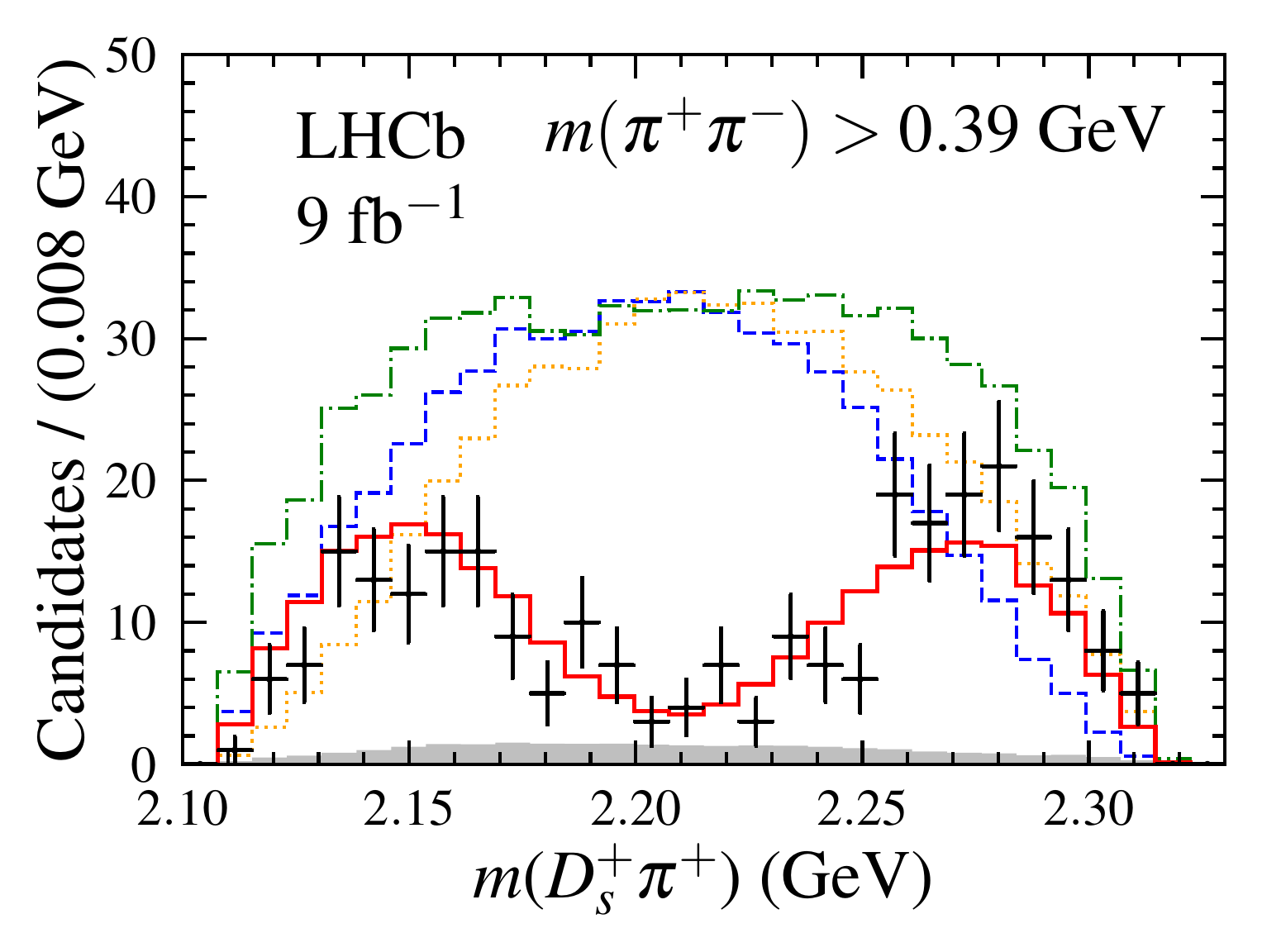}
        \begin{minipage}{\linewidth}
            \vspace{-22.5em}\hspace{14.0em}
            (c)
        \end{minipage}
    \end{minipage}
    \vspace{-2em}
    \caption{Comparison between data (black dots with error bars) and results of the fit with the $f_0(500)+\text{RBW}~T_{c\bar{s}}(0^{+})$ model. The distributions are for the three channels combined in (a)~$m(\pi^{+}\pi^{-})$, (b)~$m(D_{s}^{+}\pi^{+})$, and (c)~$m(D_{s}^{+}\pi^{+})$ requiring $m(\pi^{+}\pi^{-})>0.39\,\mathrm{GeV}$. Individual components, corresponding to the background contribution estimated from $m(D_{s}^{+}\pi^{+}\pi^{-})$ sideband regions (gray-filled) and the different resonant contributions (coloured dashed lines), are also shown as indicated in the legend. }
    \label{fig:model_sgm_tcsbar_2}
\end{figure}

\begin{figure}[tb]
    \centering
    \begin{minipage}{0.45\linewidth}
        \centering
        \includegraphics[width=\linewidth]{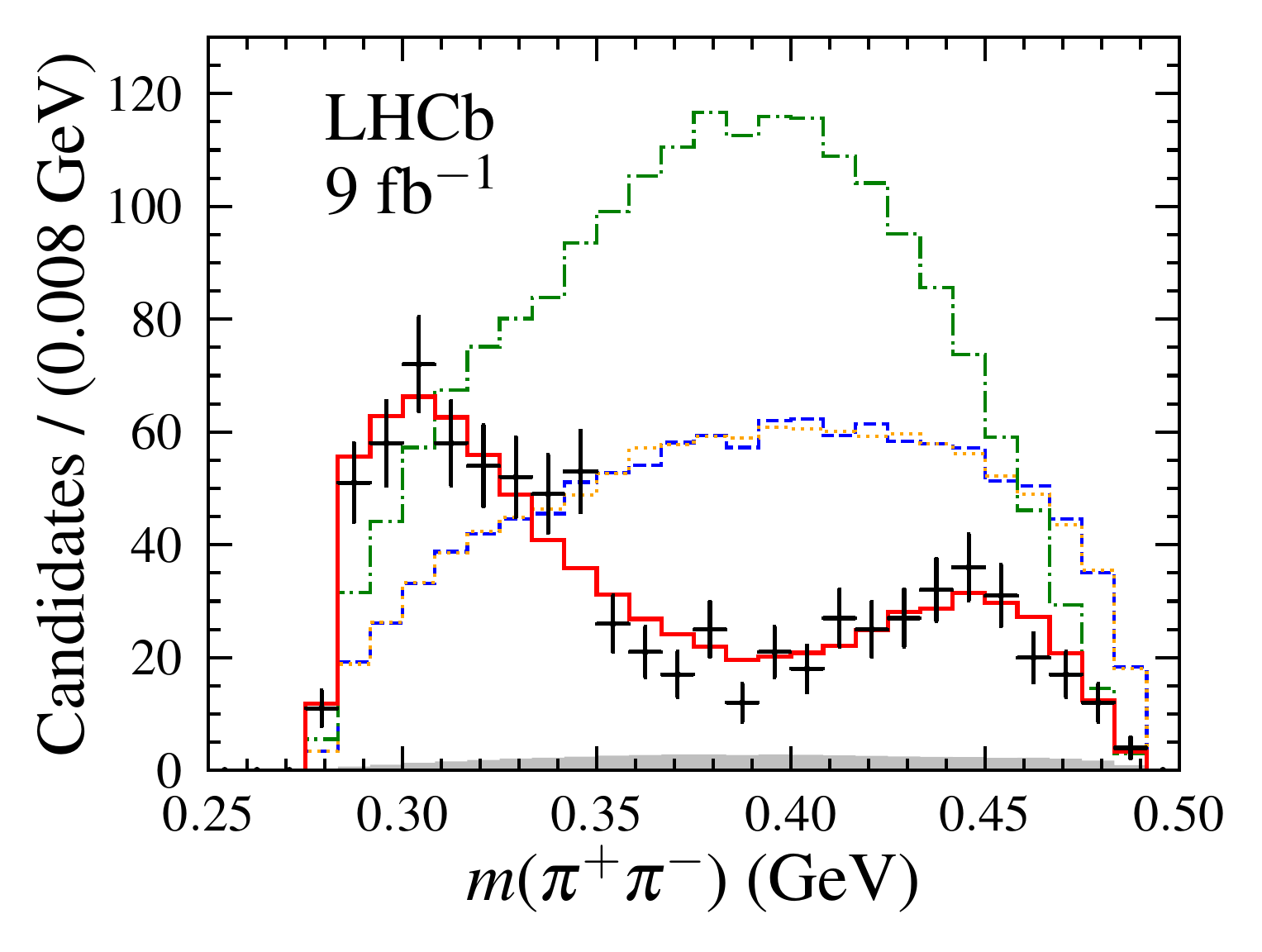}
        \begin{minipage}{\linewidth}
            \vspace{-22.5em}\hspace{14.0em}
            (a)
        \end{minipage}
        \vspace{-2em}
    \end{minipage}
    \begin{minipage}{0.45\linewidth}
        \centering
        \includegraphics[width=\linewidth]{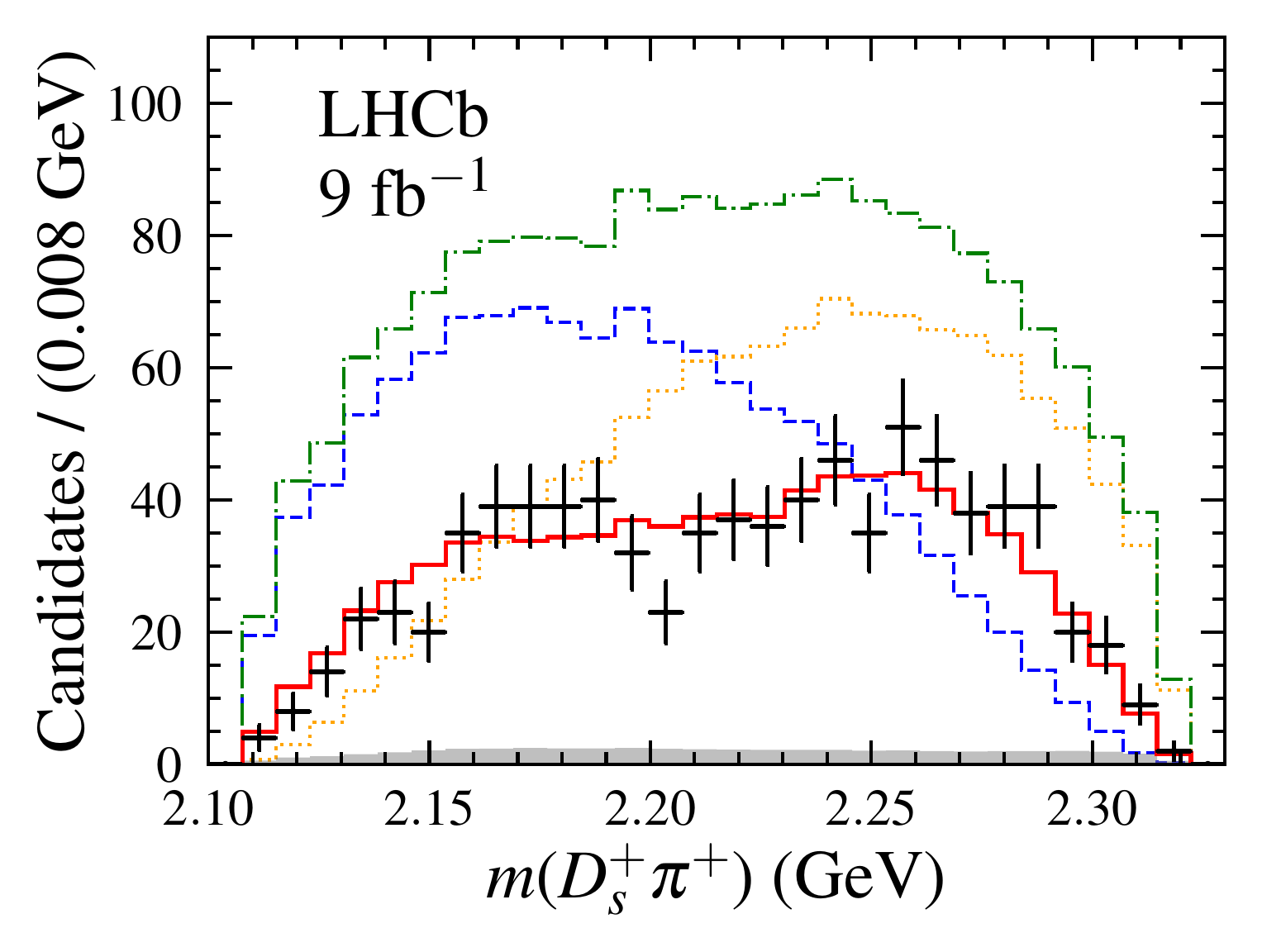}
        \begin{minipage}{\linewidth}
            \vspace{-22.5em}\hspace{14.0em}
            (b)
        \end{minipage}
        \vspace{-2em}
    \end{minipage}
    
        \hspace{3em}
        \includegraphics[width=0.40\linewidth]{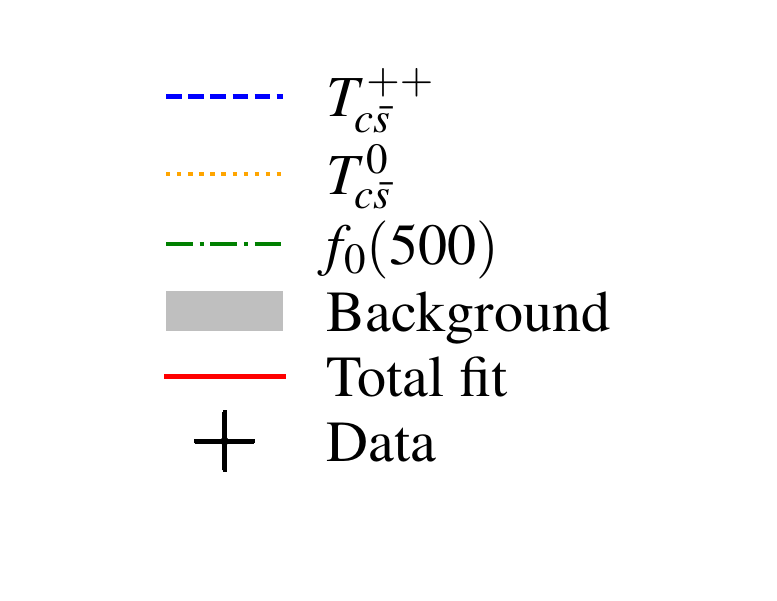}
        \hspace{-1.2em}
    \begin{minipage}{0.45\linewidth}
        \vspace{-9em}
        \centering
        \includegraphics[width=\linewidth]{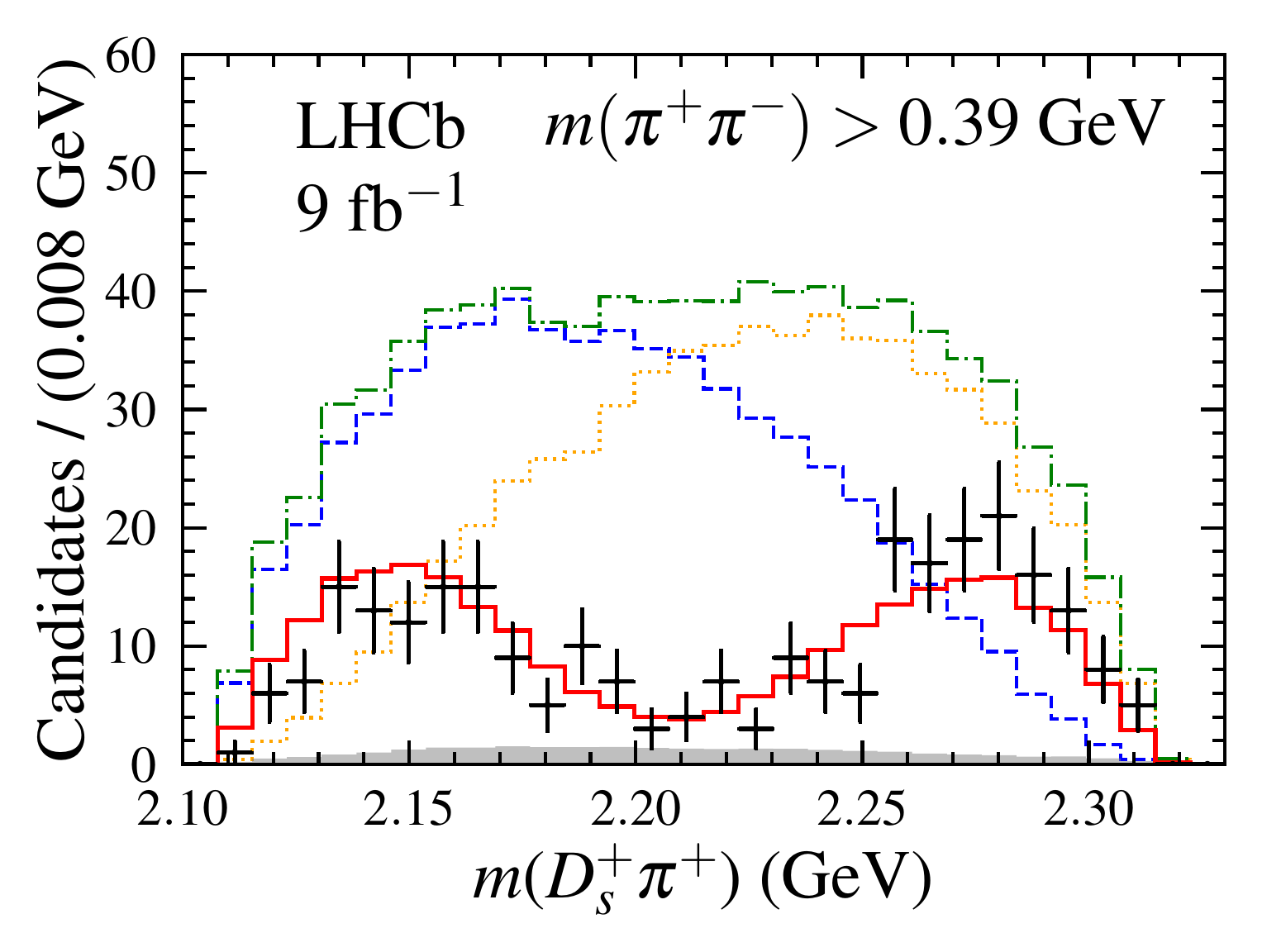}
        \begin{minipage}{\linewidth}
            \vspace{-22.5em}\hspace{14.0em}
            (c)
        \end{minipage}
    \end{minipage}
    \vspace{-2em}
    \caption{Comparison between data (black dots with error bars) and results of the fit with the $f_0(500)+K\text{-matrix}~T_{c\bar{s}}(0^+)$ model (red solid line). The distributions are for the three channels combined in (a)~$m(\pi^{+}\pi^{-})$, (b)~$m(D_{s}^{+}\pi^{+})$, and (c)~$m(D_{s}^{+}\pi^{+})$ requiring $m(\pi^{+}\pi^{-})>0.39\,\mathrm{GeV}$. Individual components, corresponding to the background contribution estimated from $m(D_{s}^{+}\pi^{+}\pi^{-})$ sideband regions (gray-filled) and the different resonant contributions (coloured dashed lines), are also shown as indicated in the legend. }
    \label{fig:model_sgm_Kmatrix}
\end{figure}

As seen in the lower section of Table~\ref{tab:fit_res_sum}, models with only $f_0(500)$ and $T_{c\bar{s}}$ states with spin-parity $J^P = 0^+$ give approximately as good descriptions of the data as the best (but, as discussed above, arguably implausible) models without $T_{c\bar{s}}$ states.
Two possible $T_{c\bar{s}}$ lineshapes, RBW and $K$-matrix, are considered and give similar fit quality. 
The projections of the fit results are given in Figs.~\ref{fig:model_sgm_tcsbar_2} and \ref{fig:model_sgm_Kmatrix}. 
The corresponding plots in the $m^2(\pip\pim)-m^2(\Dsp\pip)$ and $\phi_1-\cos{(\theta_1)}$ planes with the RBW and $K$-matrix models are shown in Fig.~\ref{fig:model_sgm_tcsbar_2_chisq} and~\ref{fig:model_sgm_Kmatrix_chisq}, respectively, in the supplemental material. A second solution with similar $\Delta$NLL is also obtained, but is quite unstable and therefore is not discussed further.

\begin{figure}[bp]
    \centering
    \begin{subfigure}[t]{0.55\linewidth}
        \includegraphics[width=1.0\textwidth]{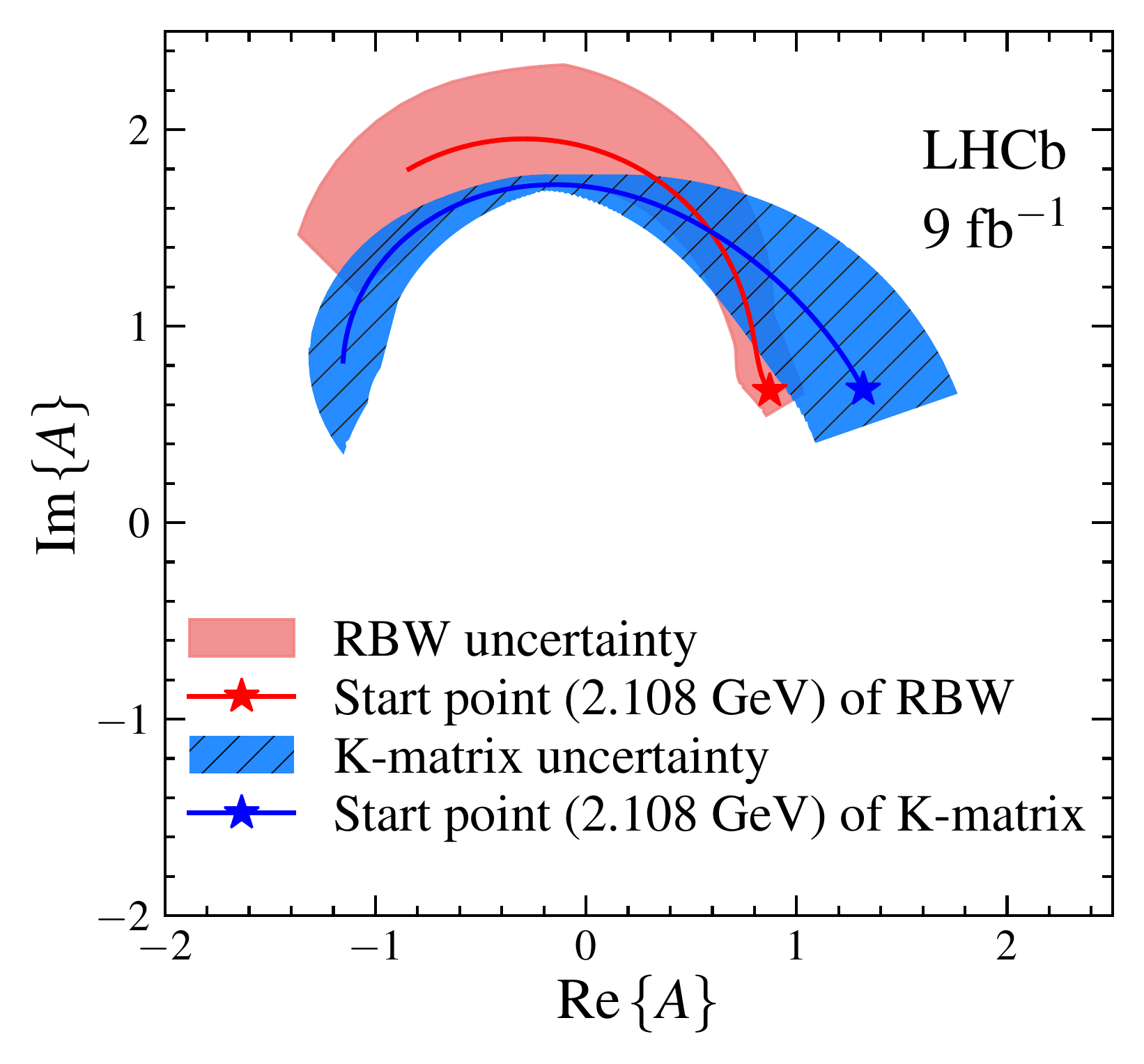}
    \end{subfigure}
    \caption{Argand diagrams for the variation with mass of the $T_{c\bar{s}}$ states amplitude with both RBW and $K$-matrix models. The red and blue stars denote the lower $m(D_{s}\pi)$ kinematic limit (denoted ``start point''). The statistical uncertainties for the RBW and $K$-matrix models are marked as red solid and blue slashed bands, respectively. }
    \label{fig:argand}
\end{figure}

As seen from the large fit fractions in Table~\ref{tab:fit_res_sum_2}, fits with these models have similarly large destructive interference effects as in models without the $T_{c\bar{s}}$ states. 
It may also be noted that the fitted $f_0(500)$ mass and width values are now in better agreement with previous measurements~\cite{PDG2024}. 
The mass of the $T_{c\bar{s}}$ states is comparable between the RBW and $K$-matrix models but a large variation in the width is found. 
For the $K$-matrix model, the $\gamma_2$ parameter in Eq.~\eqref{eq:kscatk} is fixed to 0 as it is expected that the coupling to the $D_s\pi$ channel is weak. 
Values of $\beta = 153 \pm 12$ and $\gamma = -259 \pm 21$ are obtained, from which the scattering length is calculated to be $-0.86(\pm 0.07) + 0.44(\pm 0.07)i \fm$, incompatible with the value predicted in Ref.~\cite{Liu:2012zya}. The mass and width of the $K$-matrix model given in Table~\ref{tab:fit_res_sum_2} are calculated in the second Riemann sheet and are also different from those predicted in Ref.~\cite{Guo:2009ct}. 
When allowed to vary freely in the fit, $\gamma_2 = 47 \pm 41$ is obtained, consistent with the expectation of zero,  while $\beta = 133 \pm 16$ and $\gamma = -244 \pm 17$ are consistent with the values obtained when $\gamma_2$ is fixed to 0. The Argand diagrams~\cite{PDG2024} for the RBW and $K$-matrix descriptions of the $T_{c\bar{s}}$ lineshape are shown in Fig.~\ref{fig:argand}, and are seen to be consistent with each other. 

In order to test the assumption of isospin symmetry, the coupling constants, masses, and widths of the two $T_{c\bar{s}}$ states are allowed to differ.
This is done in both the RBW and $K$-matrix $T_{c\bar{s}}$ models, with results consistent with isospin symmetry in both cases.
For example, with the $K$-matrix $T_{c\bar{s}}$ lineshape, the mass and width for $T_{c\bar{s}}^{++}$ ($T_{c\bar{s}}^{0}$) are measured to be $2325\pm11\mev$ ($2325\pm10\mev$) and $81\pm14\mev$ ($118\pm20\mev$), respectively. 

Models with additional $\rho(770)^0$, $f_0(980)$ and $f_2(1270)$ components are also tested. 
None of these extra contributions are found to be significant, as seen in Table~\ref{tab:fit_res_sum}.
An upper limit on the fit fraction of the isospin-breaking $D_{s1}(2460)^+ \to \Dsp \rho(770)^0$ decay is set at 2.8\% at the 90\% confidence level. This is less restrictive than the upper limit of 1.7\% at the 90\% confidence level obtained if the model does not include any $T_{c\bar{s}}$ component. 
Bearing in mind the large contributions from the $f_0(980)$ and $f_2(1270)$ components in the $f_0(500)+f_0(980)+f_2(1270)$ model, it is interesting to note that the corresponding contributions are small and not significant in the $f_0(500)+f_0(980)+T_{c\bar{s}}$ and $f_0(500)+f_0(1270)+T_{c\bar{s}}$ models.
An attempt is made to fit with the $\pi\pi$ resonances described by the chiral dynamics model together with $T_{c\bar{s}}$ states, but the fit results have unphysically large interference and are not further discussed. 

To estimate the significance of the two $T_{c\bar{s}}$ contributions, samples of pseudoexperiments are generated based on the results of the fit with a model containing $f_0(500)$ and $f_0(980)$ resonances only.  
These pseudoexperiments are each fitted both with and without $T_{c\bar{s}}$ states. 
The distribution of the $2\Delta$NLL values between the two fit results is fitted with a $\chi^{2}$ distribution, and the number of degrees of freedom ($N_{\rm{dof}}$) is determined to be $6.77\pm0.25$. 
Given that the $2\Delta$NLL value from data is 490.4, the significance is estimated to be much larger than 10 standard deviations ($\sigma$). 
The 2$\Delta$NLL distribution is shown in Fig.~\ref{fig:delta_nll_signif} in the supplemental material. 
This significance value implicitly rejects the $f_0(500)+f_0(980)+f_2(1270)$ model.
If the null hypothesis is based on that model, the $T_{c\bar{s}}$ components are not significant.

The spin-parity of the $T_{c\bar{s}}$ states used in the fits is changed to $1^-$ instead of $0^+$. 
This reduces significantly the interference effects, but results in a $\Delta$NLL value about 60 units larger than that obtained with $J^P = 0^+$ $T_{c\bar{s}}$ states.
Pseudoexperiments are generated according to the fit results under this alternative spin-parity assumption to evaluate the significance of this outcome. 
The pseudoexperiments are fitted with both spin-0 and spin-1 models, and the distribution of 2$\Delta$NLL values is obtained. Comparing to the 2$\Delta$NLL value observed in data, the spin-parity $0^+$ is favoured with $10\,\sigma$ significance. The 2$\Delta$NLL distribution is shown in Fig.~\ref{fig:delta_nll_spin} in the supplemental material.

\section{Systematic uncertainties}
Systematic uncertainties are evaluated on the masses, widths and fit fractions of each of the components included in the $f_0(500)+f_0(980)+f_2(1270)$, $f_0(500)+\text{RBW}~T_{c\bar{s}}(0^+)$ and $f_0(500)+K\text{-matrix}~T_{c\bar{s}}(0^+)$ models.
The sources of systematic uncertainty are divided into five categories: the signal fraction, the background model, the efficiency map, the fixed parameters in the amplitude fit, and the choices for the lineshape models. Among them, the dominant systematic uncertainties are from the fixed parameters in the amplitude fit and the choices for lineshape models. 
The total systematic uncertainties presented in Table~\ref{tab:fit_res_sum_2} are determined by combining all contributions in quadrature, and do not include the uncertainty from the choices for the $T_{c\bar{s}}$ lineshape models, which is treated later. 

Most of the systematic uncertainties are estimated by performing several times the fit to data, each time varying the input parameters within their respective uncertainties, such as altering distributions or fixed parameters. 
The root mean squares of the distributions of the fit results are taken as the corresponding measures of systematic uncertainty. 
A further source of uncertainty related to the signal fraction is estimated by changing the signal shape in the $m(\Dsp\pip\pim)$ fit to a Gaussian function and calculating the resulting signal fraction.
The difference between the results in the amplitude fits using the two signal fraction estimations is assigned as an additional uncertainty. 
The background model uncertainty is estimated by changing the background description using a different nonparameterised method to model the two-body invariant masses and helicity angles considering correlations between them. 
The variation in the fit results 
is considered as the systematic uncertainty. 
The efficiency map category accounts for uncertainties related to the size of the simulation sample used to describe the efficiency variation over the phase space, as well as uncertainties due to simulation corrections.
The fixed parameters in the amplitude models include the Blatt--Weisskopf radius parameter and the $f_0(980)$ and $f_2(1270)$ masses and widths.
The former is varied from its default value of $3.0\gev^{-1}$ to $1.5\gev^{-1}$ and $4.5\gev^{-1}$. 
The latter are varied within the uncertainties of previous measurements~\cite{PDG2024,LHCb-PAPER-2013-069}.
Additionally, the effect of allowing the $\gamma_2$ parameter of the $K$-matrix $T_{c\bar{s}}$ model to vary in the fit is assigned as a systematic uncertainty. 

Possible biases in the fit procedure are studied with pseudoexperiments generated from the fit results, and then fitted with the same model. 
The pull distribution for each fit parameter is modelled using a Gaussian function for symmetric distributions or a double-sided Crystal Ball function~\cite{Skwarnicki:1986xj} for asymmetric ones. Almost all pull distributions show deviations from normal distributions that are smaller than $3\,\sigma$. Nonetheless, adjustments are applied to the central values and uncertainties to correct for any potential biases and under- or over-coverage.

For the final results on the mass and width of the $T_{c\bar{s}}$ states, an additional systematic uncertainty is assigned to account for the description of the $T_{c\bar{s}}$ lineshape.
The results with the $f_0(500)+K\text{-matrix}~T_{c\bar{s}}(0^+)$ model are taken as the central values, and an additional asymmetric systematic uncertainty calculated as the difference in the results between the $f_0(500)+\text{RBW}~T_{c\bar{s}}(0^+)$ and $f_0(500)+K\text{-matrix}~T_{c\bar{s}}(0^+)$ models is assigned.
This is the dominant uncertainty on the $T_{c\bar{s}}$ width.
The results for the $T_{c\bar{s}}$ mass and width are $2327\pm13\pm13\mev$ and $96\pm16\,^{+170}_{-23}\mev$, respectively.

\section{Summary}
An amplitude analysis to study the resonant structure of $\decay{D_{s1}(2460)^{+}}{\Dsp\pip\pim}$ decays is performed for the first time.
The analysis is based on exclusively reconstructed $\decay{\Bz}{\Dm D_{s1}(2460)^{+}}$, $\decay{\Bp}{\Dzb D_{s1}(2460)^{+}}$ and $\decay{\Bz}{\Dstarm D_{s1}(2460)^{+}}$ decays obtained from a $pp$ collision sample recorded at centre-of-mass energies of $\sqrt{s}=7,8$ and $13\tev$, corresponding to $9\invfb$ of integrated luminosity. 

A clear double-peak structure is observed in the $m(\pip\pim)$ spectrum of $\decay{D_{s1}(2460)^{+}}{\Dsp\pip\pim}$ decays. 
The data can be described well with a model including only $\pi\pi$ resonances and without $\Ds\pi$ exotic states, but only with implausibly large $f_0(980)$ and $f_2(1270)$ contributions. In addition, this model has large interference between the $f_0(500)$ and $f_0(980)$ states, which is not seen in other processes involving these two contributions in the $\pi\pi$ S-wave.

An alternative model with a new exotic $T_{c\bar{s}}^{++}$ state and its isospin partner $T_{c\bar{s}}^{0}$ is introduced. 
The $T_{c\bar{s}}$ mass and width are determined to be $2327\pm13\pm13\mev$ and $96\pm16\,^{+170}_{-23}\mev$, where the first uncertainties are statistical and the second are systematic. 
The significance of the new states exceeds $10\,\sigma$, evaluated relative to a model containing $f_0(500)$ and $f_0(980)$ contributions only, and the $T_{c\bar{s}}$ spin-parity is found to be $J^P = 0^+$ with a significance of $10\,\sigma$. The $T_{c\bar{s}}$ states can be interpreted as two members of the isotriplet predicted in Ref.~\cite{Maiani:2024quj}, with the masses consistent with their prediction. 
These results complement those obtained on other $T_{cs}$ and $T_{c\bar{s}}$ hadrons~\cite{LHCb-PAPER-2020-024,LHCb-PAPER-2020-025,LHCb-PAPER-2022-026,LHCb-PAPER-2022-027}, and are an important step to probe the nature of the $D_{s1}(2460)^+$ and $D_{s0}^{*}(2317)^+$ resonances.

% Comment this in for paper drafts; do not include this in analysis note, conference and figure reports
\section*{Acknowledgements}
%
% These Acknowledgements valid from 3-May-2019
%
\noindent 
We acknowledge important input from Alex Bondar, which helped to shape the analysis reported here. We express our gratitude to our colleagues in the CERN
accelerator departments for the excellent performance of the LHC. We
thank the technical and administrative staff at the LHCb
institutes.
We acknowledge support from CERN and from the national agencies:
CAPES, CNPq, FAPERJ and FINEP (Brazil);
MOST and NSFC (China);
CNRS/IN2P3 (France);
BMBF, DFG and MPG (Germany);
INFN (Italy);
NWO (Netherlands);
MNiSW and NCN (Poland);
MCID/IFA (Romania);
%MSHE (Russia);
MICIU and AEI (Spain);
SNSF and SER (Switzerland);
NASU (Ukraine);
STFC (United Kingdom);
DOE NP and NSF (USA).
We acknowledge the computing resources that are provided by CERN, IN2P3
(France), KIT and DESY (Germany), INFN (Italy), SURF (Netherlands),
PIC (Spain), GridPP (United Kingdom),
%RRCKI and Yandex LLC (Russia),
CSCS (Switzerland), IFIN-HH (Romania), CBPF (Brazil),
and Polish WLCG (Poland).
We are indebted to the communities behind the multiple open-source
software packages on which we depend.
Individual groups or members have received support from
ARC and ARDC (Australia);
Key Research Program of Frontier Sciences of CAS, CAS PIFI, CAS CCEPP,
Fundamental Research Funds for the Central Universities,
and Sci. \& Tech. Program of Guangzhou (China);
Minciencias (Colombia);
EPLANET, Marie Sk\l{}odowska-Curie Actions, ERC and NextGenerationEU (European Union);
A*MIDEX, ANR, IPhU and Labex P2IO, and R\'{e}gion Auvergne-Rh\^{o}ne-Alpes (France);
%RFBR, RSF and Yandex LLC (Russia);
AvH Foundation (Germany);
ICSC (Italy);
%GVA, XuntaGal, GENCAT, Inditex, InTalent and Prog.~Atracci\'on Talento, CM (Spain);
Severo Ochoa and Mar\'ia de Maeztu Units of Excellence, GVA, XuntaGal, GENCAT, InTalent-Inditex and Prog. ~Atracci\'on Talento CM (Spain);
SRC (Sweden);
the Leverhulme Trust, the Royal Society
 and UKRI (United Kingdom).

\clearpage
\section*{Appendix: Supplemental material}
\appendix
\label{sec:Supplemental}
\section{\texorpdfstring{Definition of angles}{}}

\begin{figure}[!b]
    \centering
    \includegraphics[width=0.55\textwidth]{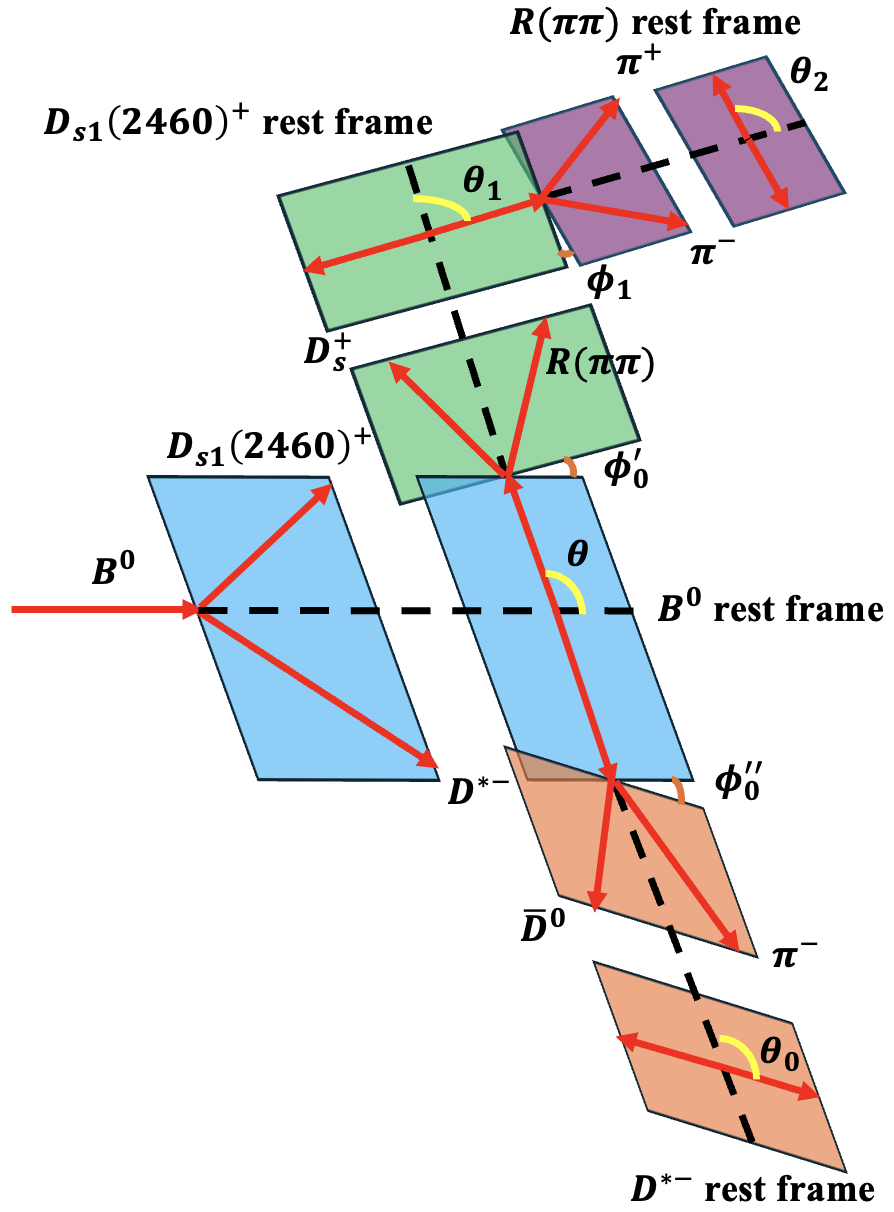}
    \caption{Definitions of the helicity angles for \mbox{$B^0\to D^{*-}D_{s1}(2460)^{+}$} decays, with the intermediate resonance $R$ decaying into $\pi\pi$.}
    \label{fig:helangle}
\end{figure}

Using the decay $\Bz\to\Dstarm D_{s1}(2460)^+$ as an example and naming the intermediate dipion resonance $R(\pi\pi)$, there are four decays accounted for in the amplitude: \mbox{$\Bz\to\Dstarm D_{s1}(2460)^{+}$}, $\Dstarm\to\Dzb\pim$, $D_{s1}(2460)^{+}\to\Dsp R(\pi\pi)$, and $R(\pi\pi)\to\pip\pim$. 
Figure~\ref{fig:helangle} shows the definitions of the helicity angles and the angles between two decay planes used to describe the amplitude. 
The helicity angle of the $B\to\Dstar D_{s1}(2460)^+$ decay is denoted as $\theta$. 
The angle between the $D_{s1}(2460)^+\to\Ds{}R(\pi\pi)$ and $\Dstar\to D\pi$ decay planes is denoted as $\phi_0=\phi_0^{'}+\phi_0^{''}$. 
The helicity angles of the $\Dstar\to D\pi$ and $D_{s1}(2460)^+\to\Ds{}R(\pi\pi)$ decays are denoted as $\theta_0$ and $\theta_1$, respectively. 
The angle between the $D_{s1}(2460)^+\to\Ds{}R(\pi\pi)$ and $R(\pi\pi)\to\pip\pim$ decay planes is denoted as $\phi_1$, and the helicity angle of the $R(\pi\pi)\to\pip\pim$ decay is denoted as $\theta_2$. 

\section{\texorpdfstring{Data distributions}{}}
\label{sup:1d_data}
\begin{figure}[tb]
    \centering
    \begin{minipage}{0.48\linewidth}
        \centering
        \includegraphics[width=\linewidth]{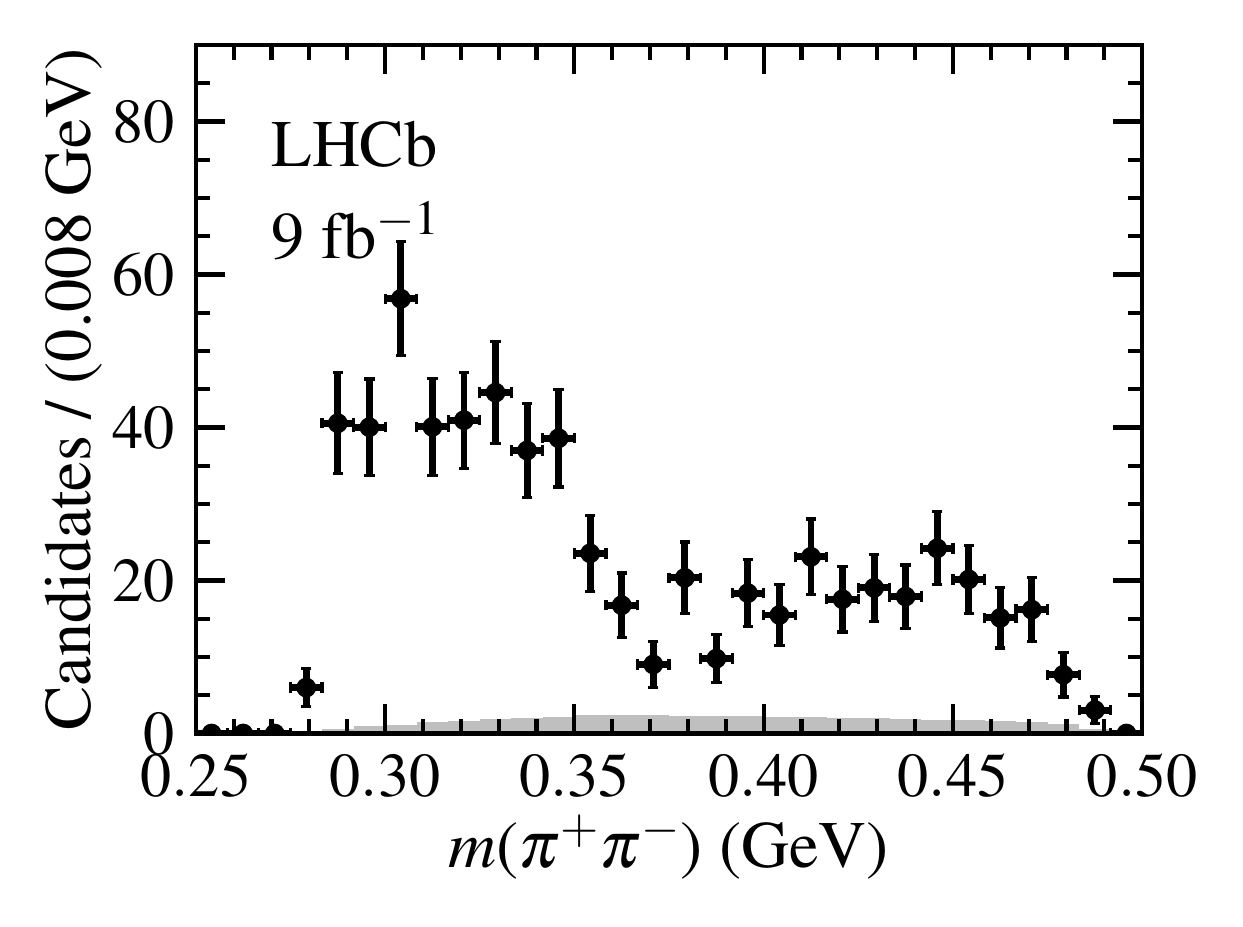}
        \begin{minipage}{\linewidth}
            \vspace{-26.0em}\hspace{14em}
            (a)
        \end{minipage}
    \end{minipage}
    \begin{minipage}{0.48\linewidth}
        \centering
        \includegraphics[width=\linewidth]{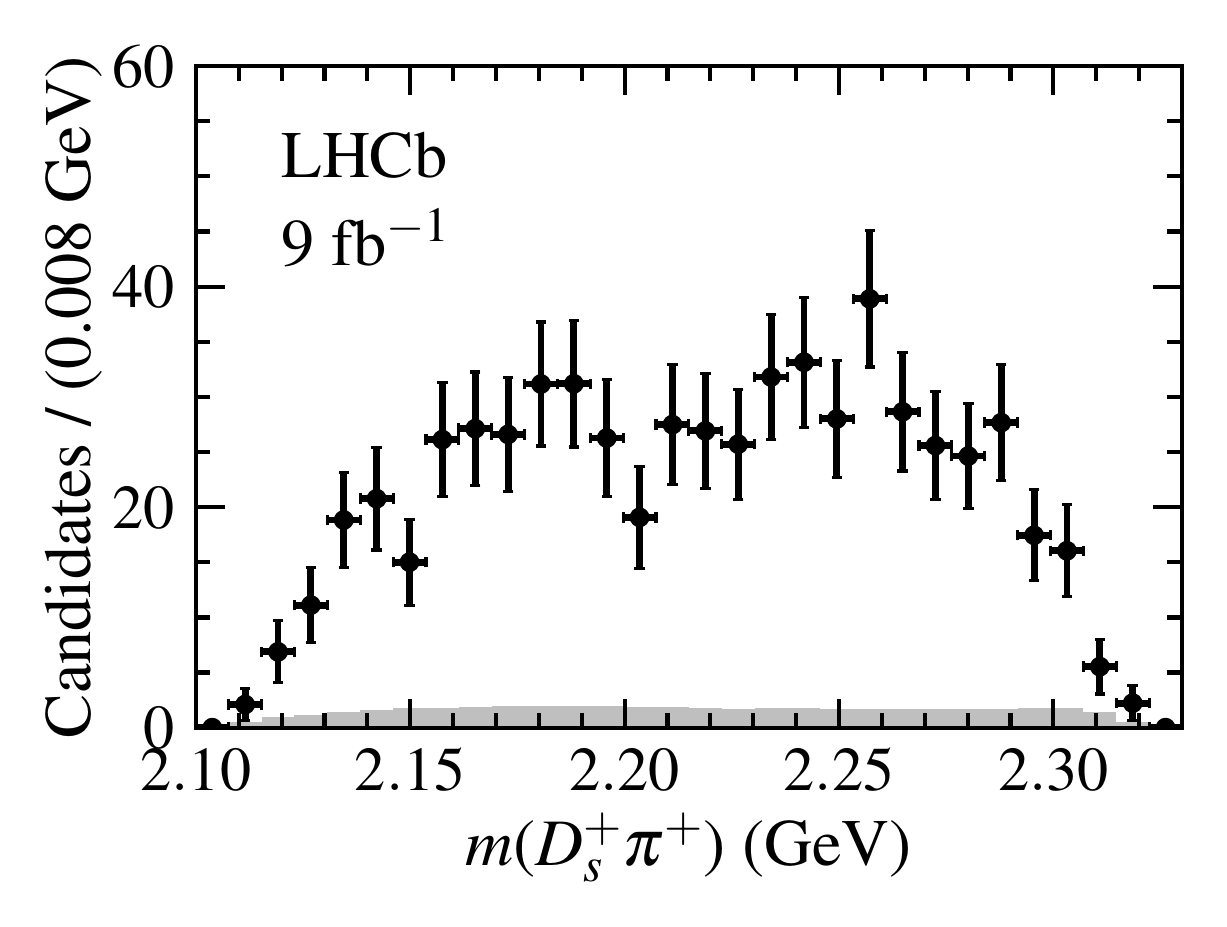}
        \begin{minipage}{\linewidth}
            \vspace{-26.0em}\hspace{14em}
            (b)
        \end{minipage}
    \end{minipage}
    \begin{minipage}{0.48\linewidth}
        \centering
        \includegraphics[width=\linewidth]{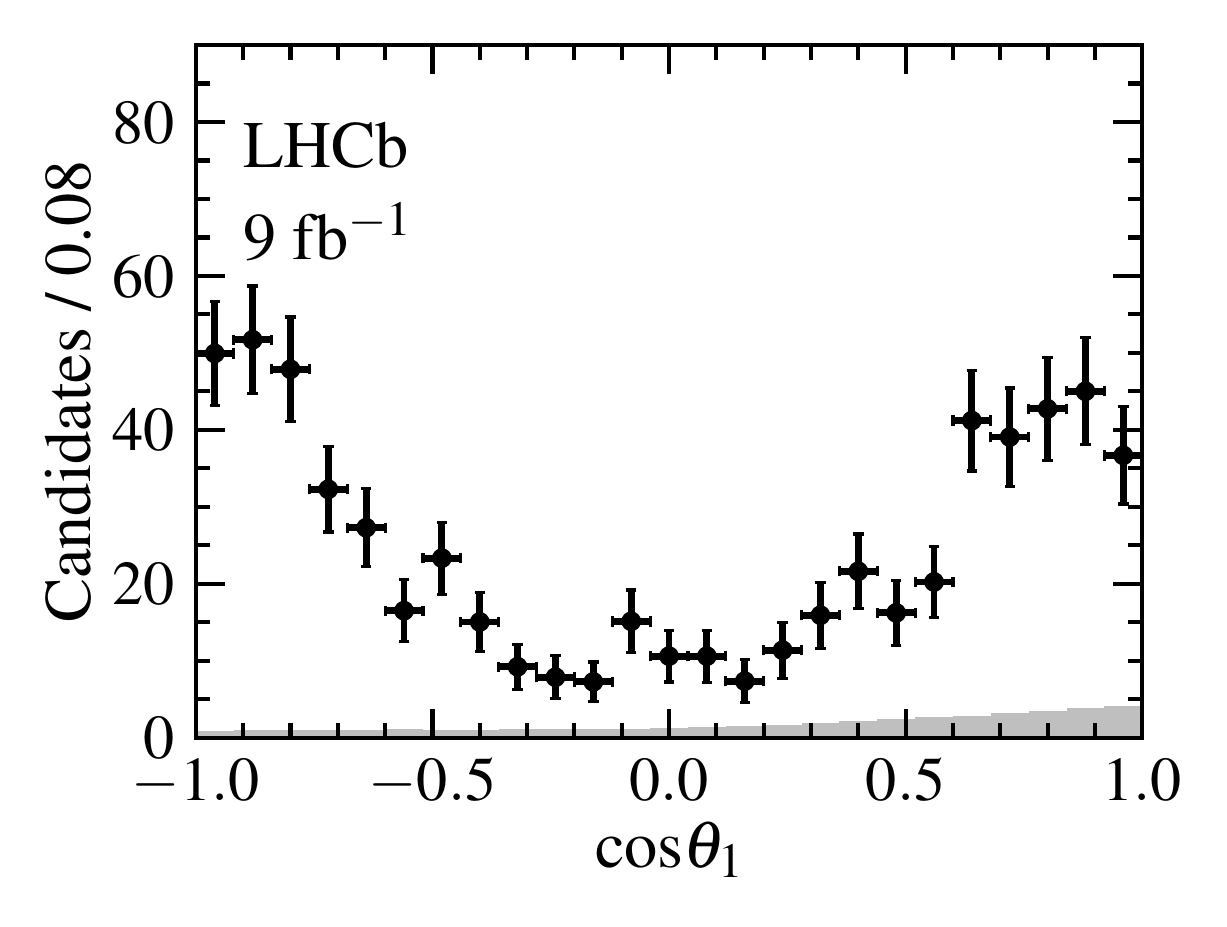}
        \begin{minipage}{\linewidth}
            \vspace{-26.0em}\hspace{14em}
            (c)
        \end{minipage}
    \end{minipage}
    \begin{minipage}{0.48\linewidth}
        \centering
        \includegraphics[width=\linewidth]{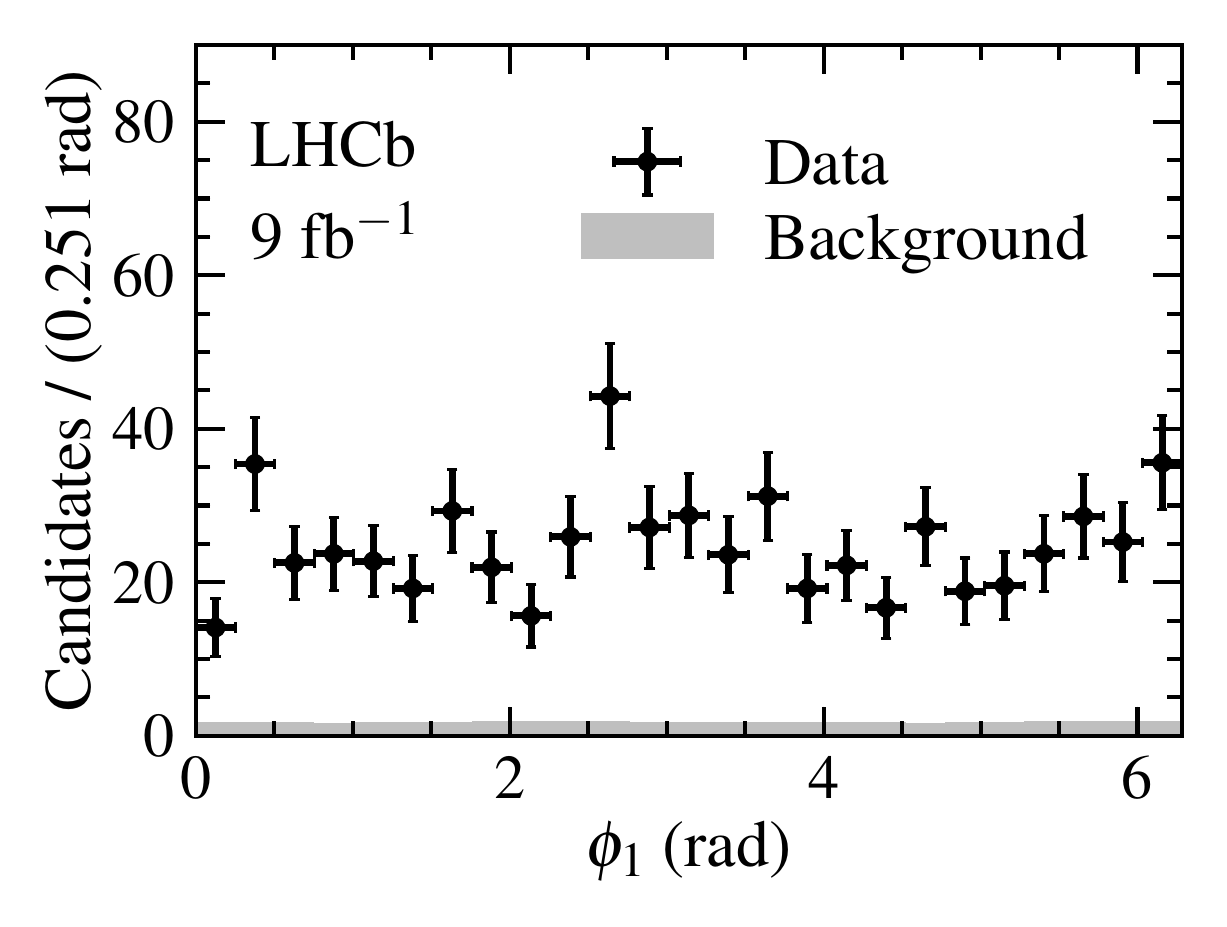}
        \begin{minipage}{\linewidth}
            \vspace{-26.0em}\hspace{14em}
            (d)
        \end{minipage}
    \end{minipage}
    \begin{minipage}{0.48\linewidth}
        \centering
        \includegraphics[width=\linewidth]{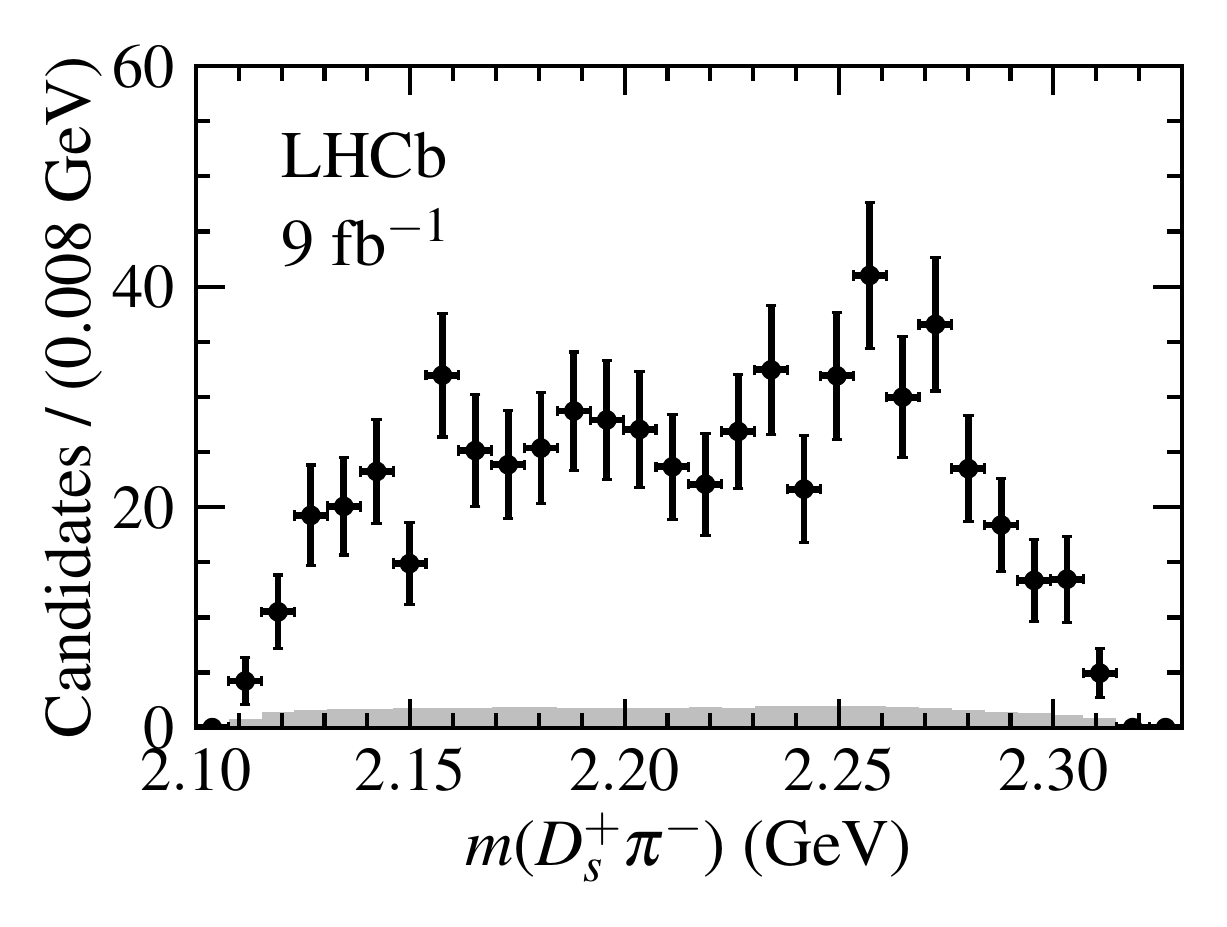}
        \begin{minipage}{\linewidth}
            \vspace{-26.0em}\hspace{14em}
            (e)
        \end{minipage}
    \end{minipage}
\caption{
Efficiency-corrected distributions of the \mbox{$D_{s1}(2460)^+\to D_{s}^{+}\pi^{+}\pi^{-}$} phase-space variables including (a)~$m(\pi^{+}\pi^{-})$, (b)~$m(D_{s}^{+}\pi^{+})$, (c)~$\cos\theta_1$, (d)~$\phi_1$ and (e)~$m(D_{s}^{+}\pi^{-})$ combining the \mbox{$B^{0}\to D^{-} D_{s1}(2460)^+$} and \mbox{$B^{+}\to {{\ensuremath{\offsetoverline{\PD}}}\xspace}{}^{0} D_{s1}(2460)^+$} channels, where black dots with error bars denote data points and gray histograms denote background.}
    \label{fig:noeffD}
\end{figure}

\begin{figure}[!htbp]
    \centering
    \begin{minipage}{0.32\linewidth}
        \centering
        \includegraphics[width=\linewidth]{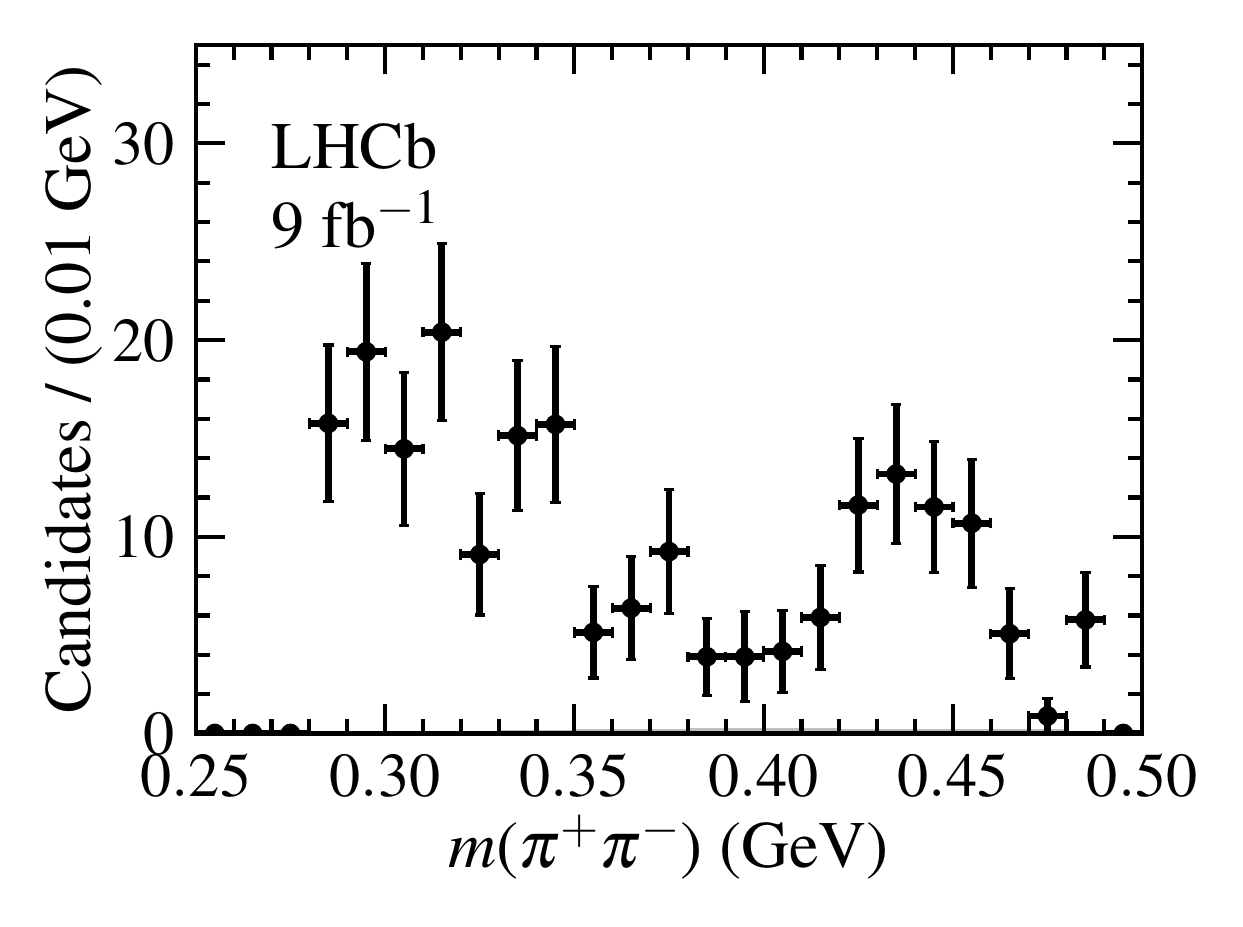}
        \begin{minipage}{5\linewidth}
            \vspace{-17.5em}\hspace{9em}
            (a)
        \end{minipage}
    \end{minipage}
    \begin{minipage}{0.32\linewidth}
        \centering
        \includegraphics[width=\linewidth]{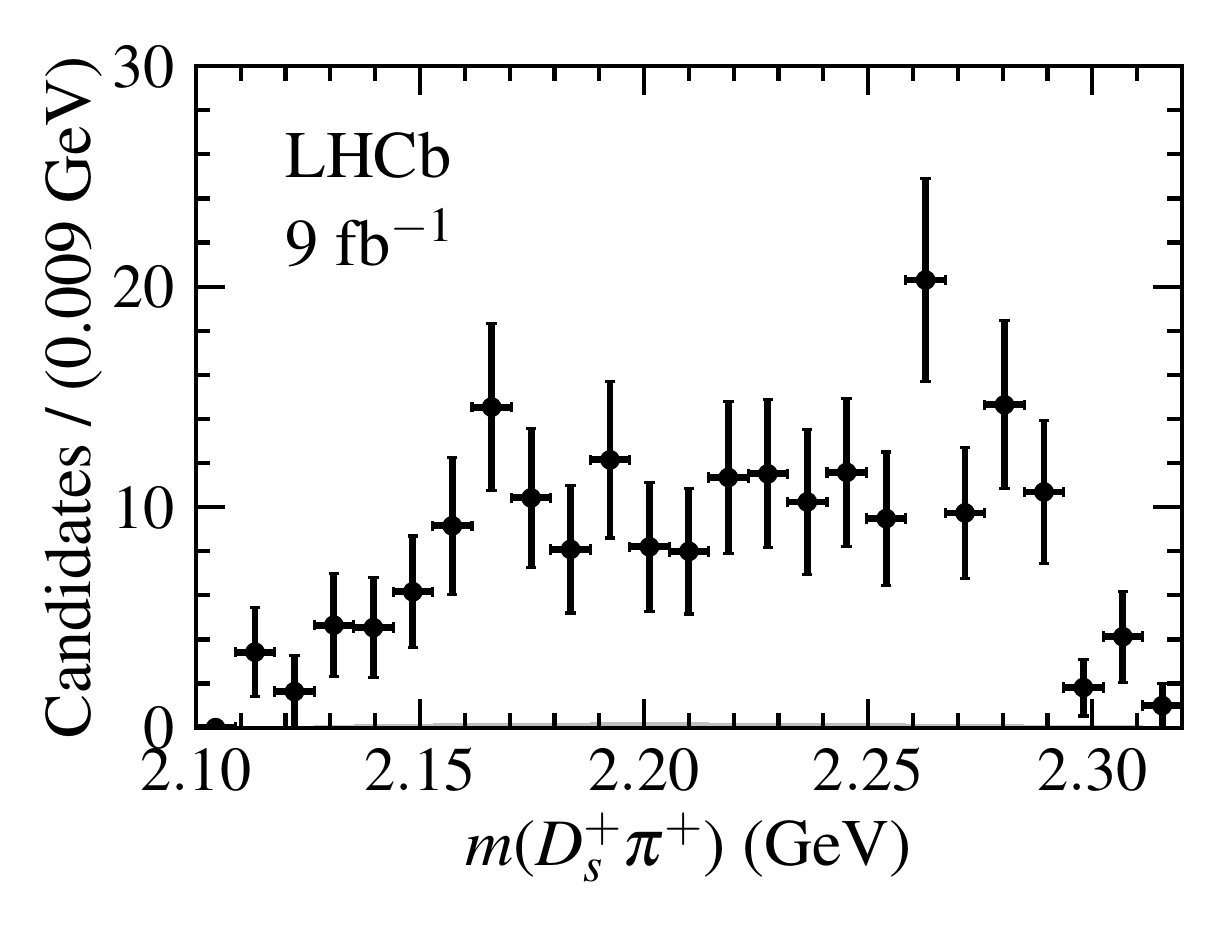}
        \begin{minipage}{5\linewidth}
            \vspace{-17.5em}\hspace{9em}
            (b)
        \end{minipage}
    \end{minipage}
    \begin{minipage}{0.32\linewidth}
        \centering
        \includegraphics[width=\linewidth]{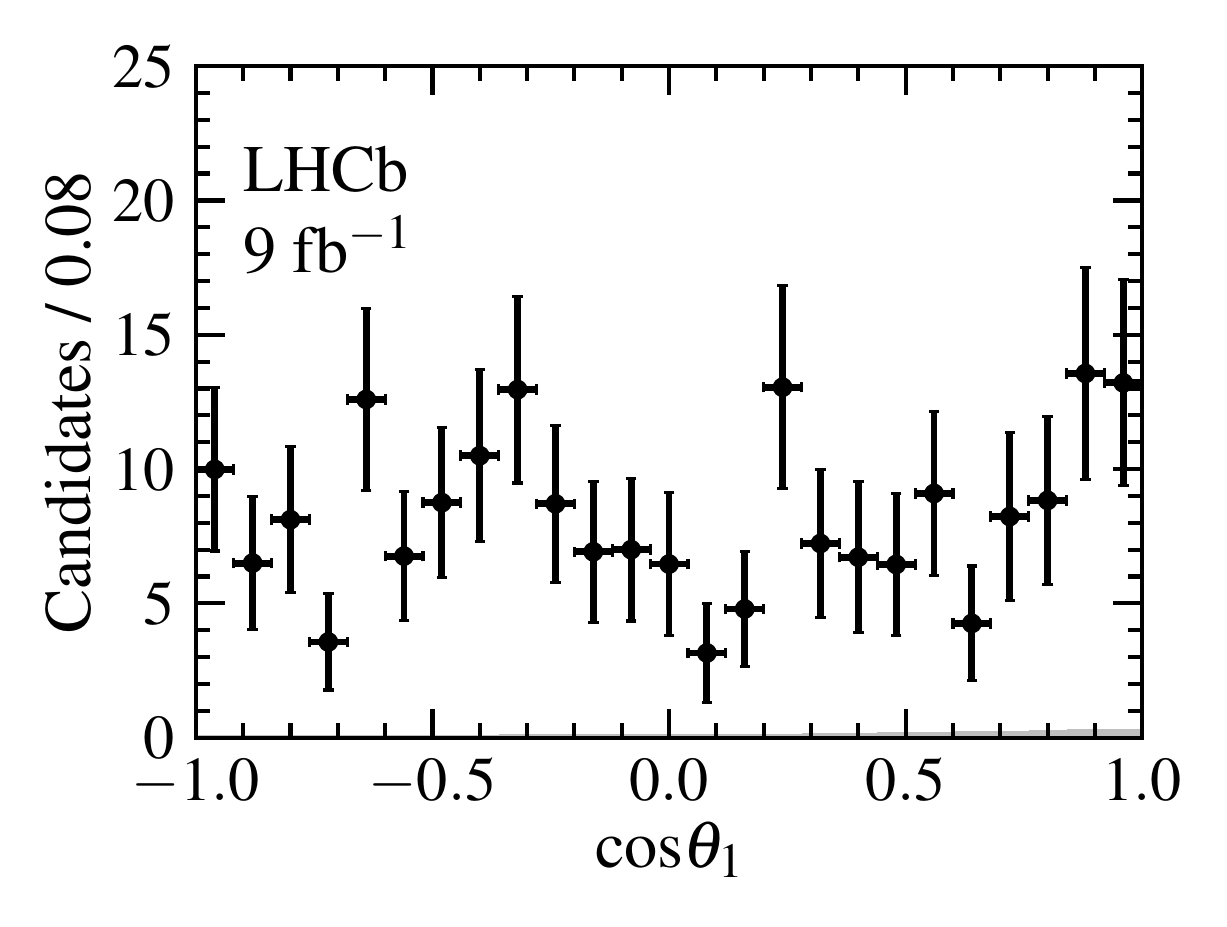}
        \begin{minipage}{5\linewidth}
            \vspace{-17.5em}\hspace{9em}
            (c)
        \end{minipage}
    \end{minipage}
    \begin{minipage}{0.32\linewidth}
        \centering
        \includegraphics[width=\linewidth]{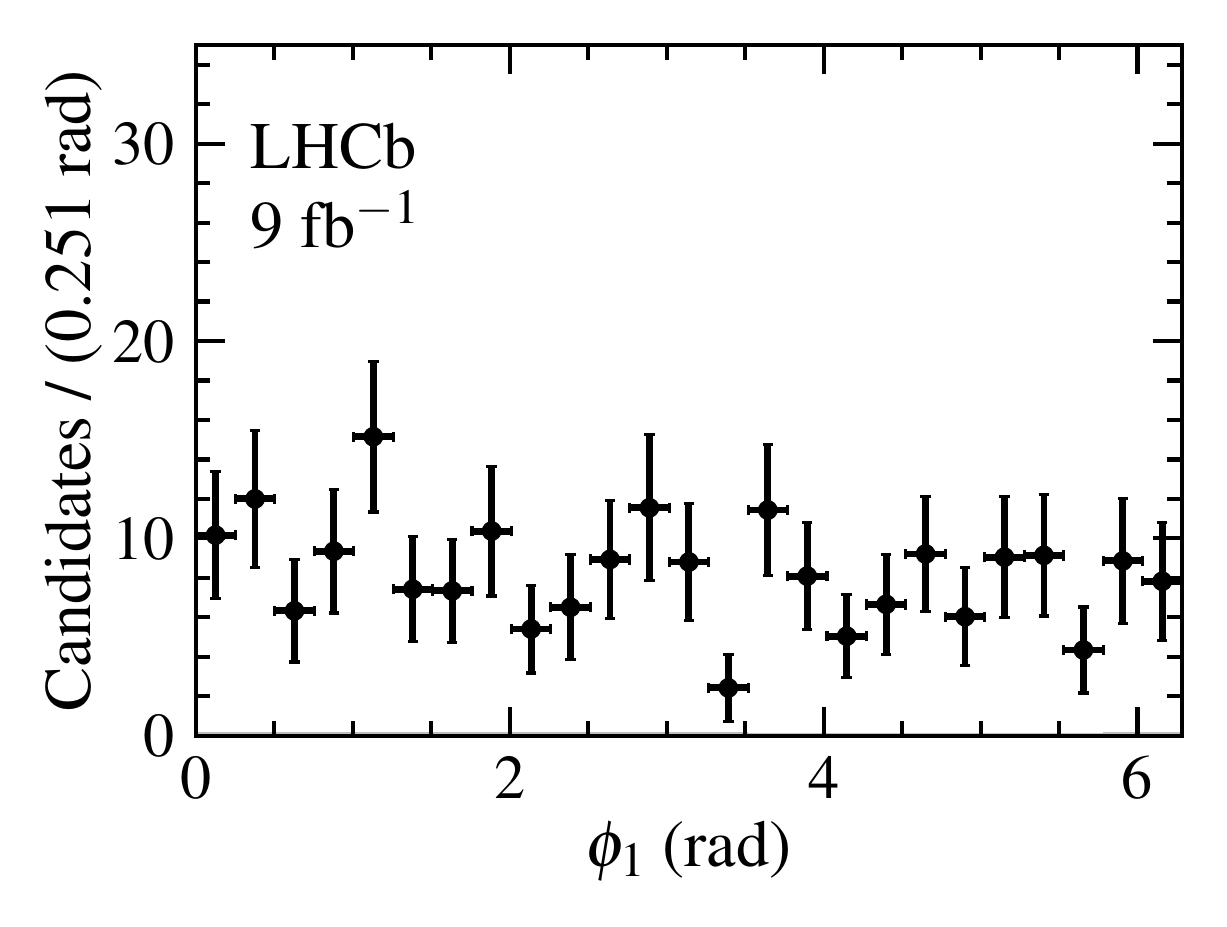}
        \begin{minipage}{5\linewidth}
            \vspace{-17.5em}\hspace{9em}
            (d)
        \end{minipage}
    \end{minipage}
    \begin{minipage}{0.32\linewidth}
        \centering
        \includegraphics[width=\linewidth]{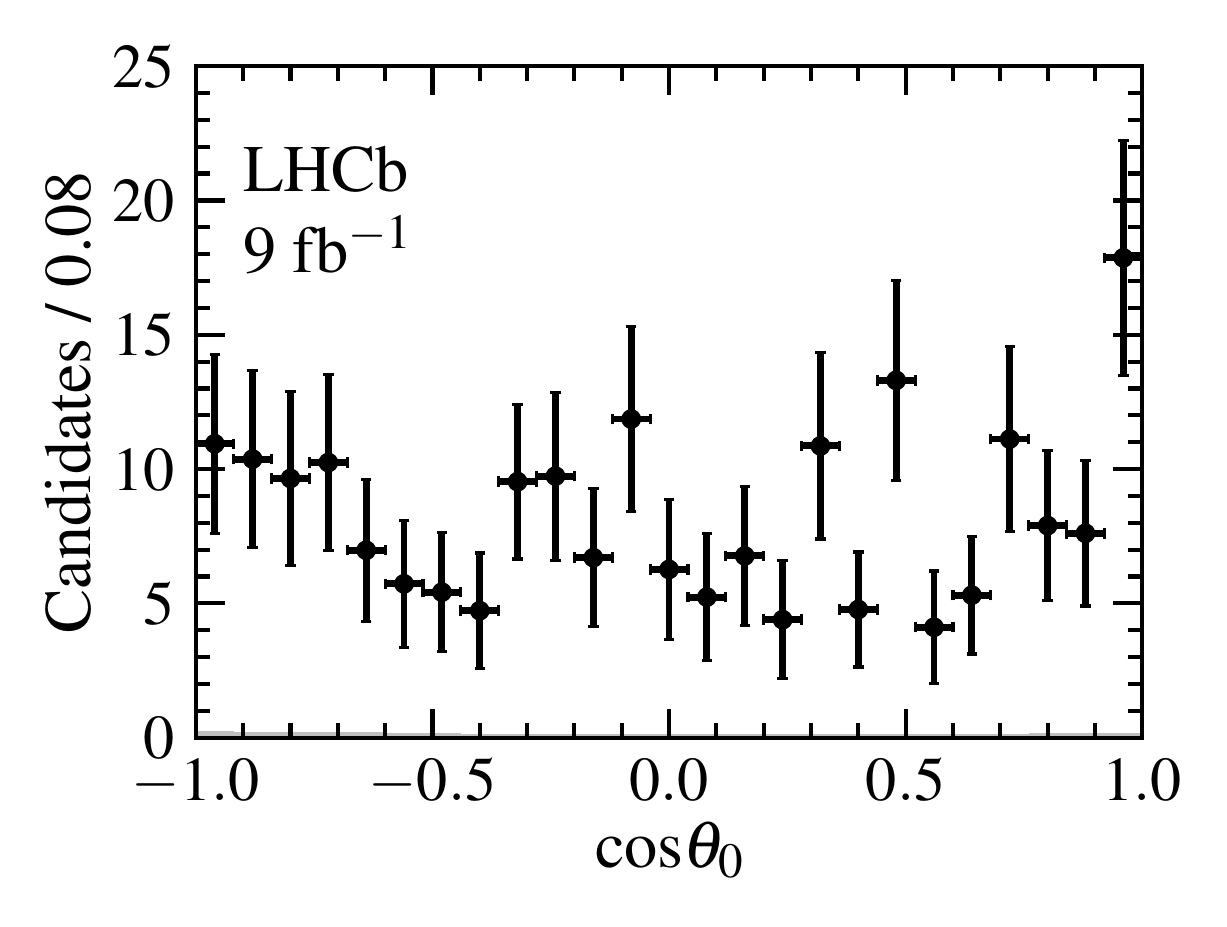}
        \begin{minipage}{5\linewidth}
            \vspace{-17.5em}\hspace{9em}
            (e)
        \end{minipage}
    \end{minipage}
    \begin{minipage}{0.32\linewidth}
        \centering
        \includegraphics[width=\linewidth]{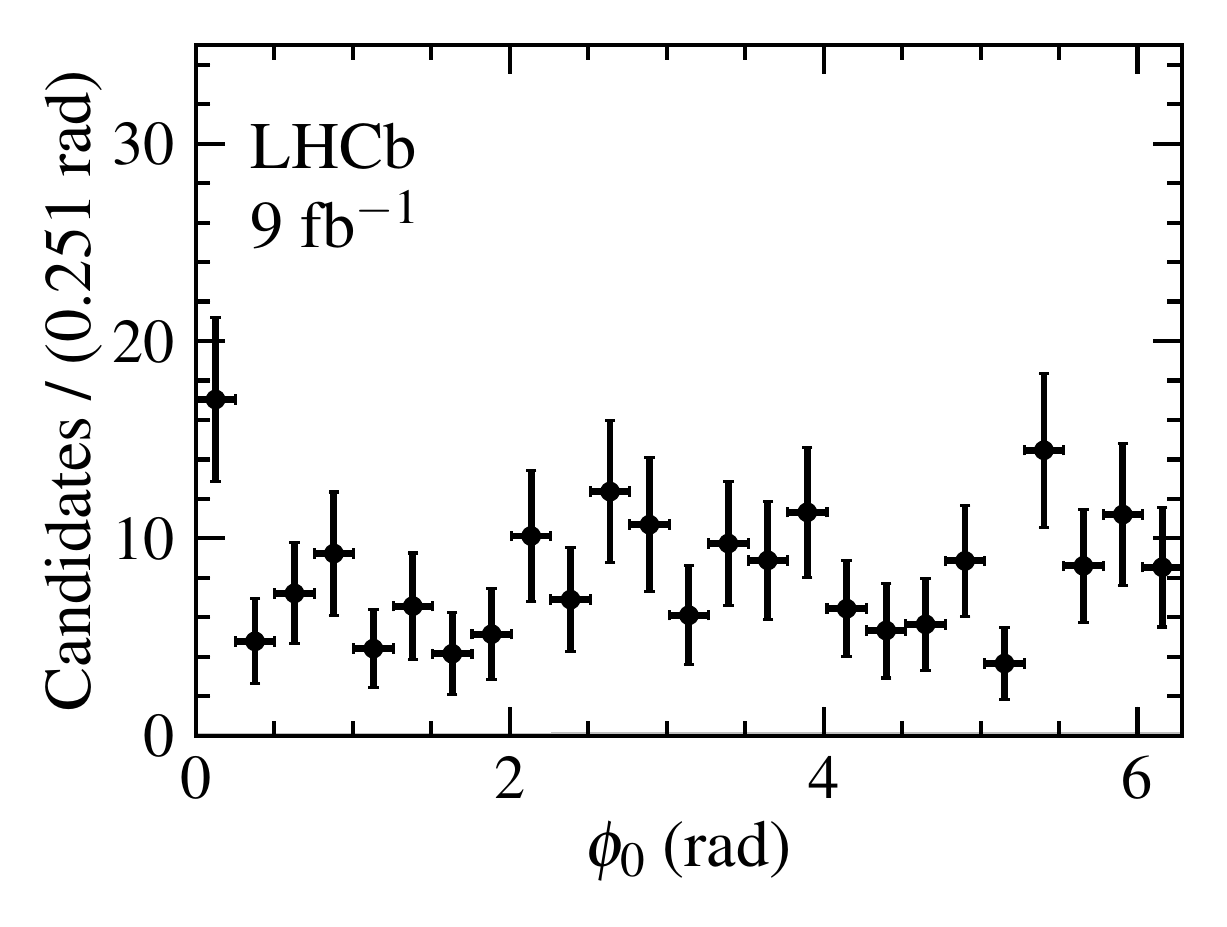}
        \begin{minipage}{5\linewidth}
            \vspace{-17.5em}\hspace{9em}
            (f)
        \end{minipage}
    \end{minipage}
    \begin{minipage}{0.32\linewidth}
        \centering
        \includegraphics[width=\linewidth]{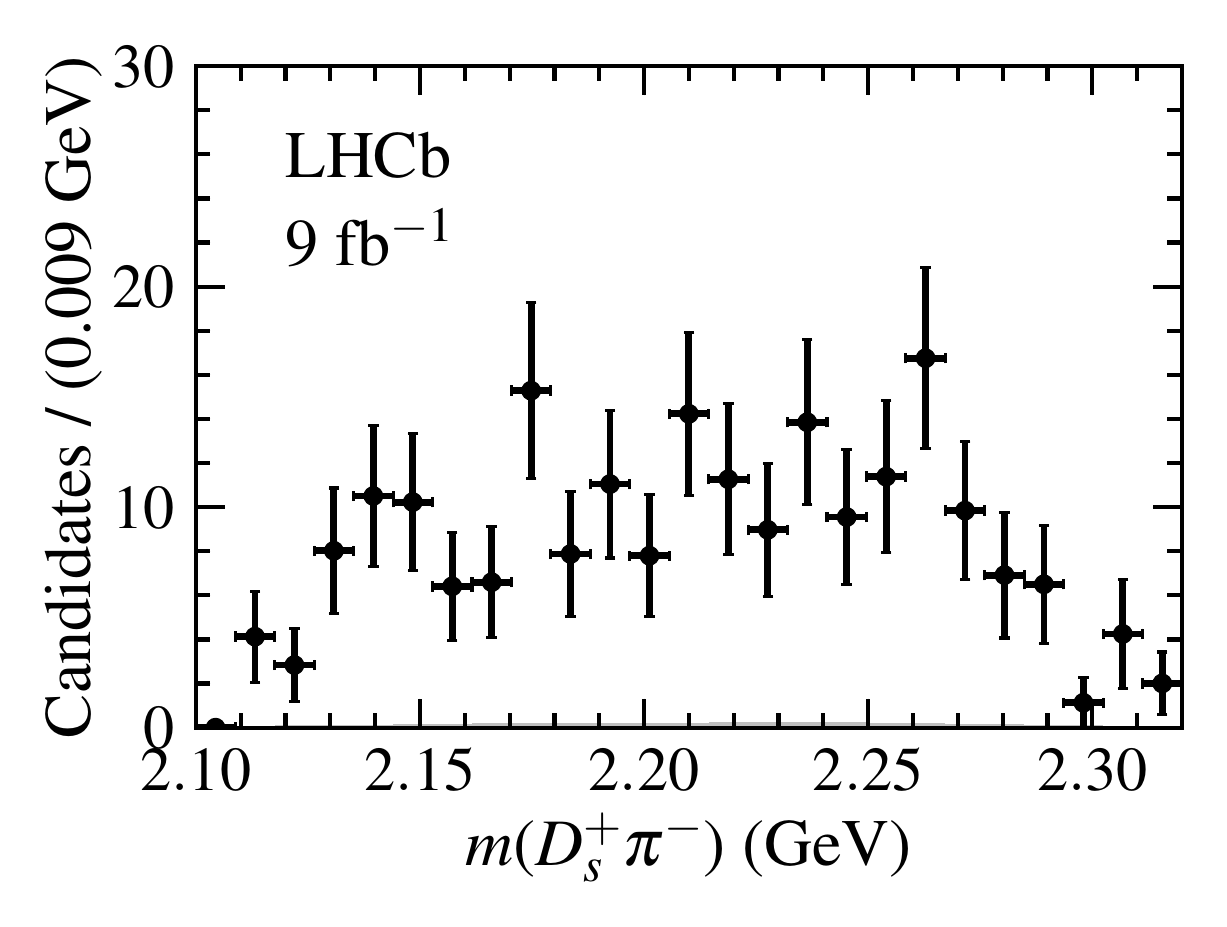}
        \begin{minipage}{5\linewidth}
            \vspace{-17.5em}\hspace{9em}
            (g)
        \end{minipage}
    \end{minipage}
    \caption{Efficiency-corrected distributions of the \mbox{$D_{s1}(2460)^+\to D_{s}^{+}\pi^{+}\pi^{-}$} phase-space variables including (a)~$m(\pi^{+}\pi^{-})$, (b)~$m(D_{s}^{+}\pi^{+})$, (c)~$\cos\theta_1$, (d)~$\phi_1$, (e)~$\cos\theta_0$, (f)~$\phi_0$ and (g)~$m(D_{s}^{+}\pi^{-})$ for the \mbox{$B^0\to D^{*-}D_{s1}(2460)^{+}$} channel, where black dots with error bars denote data points and the background is not shown due to its low contribution.}
    \label{fig:noeffDst}
\end{figure}

\begin{figure}[tb]
    \centering
    \begin{minipage}{0.32\linewidth}
        \centering
        \includegraphics[width=\linewidth]{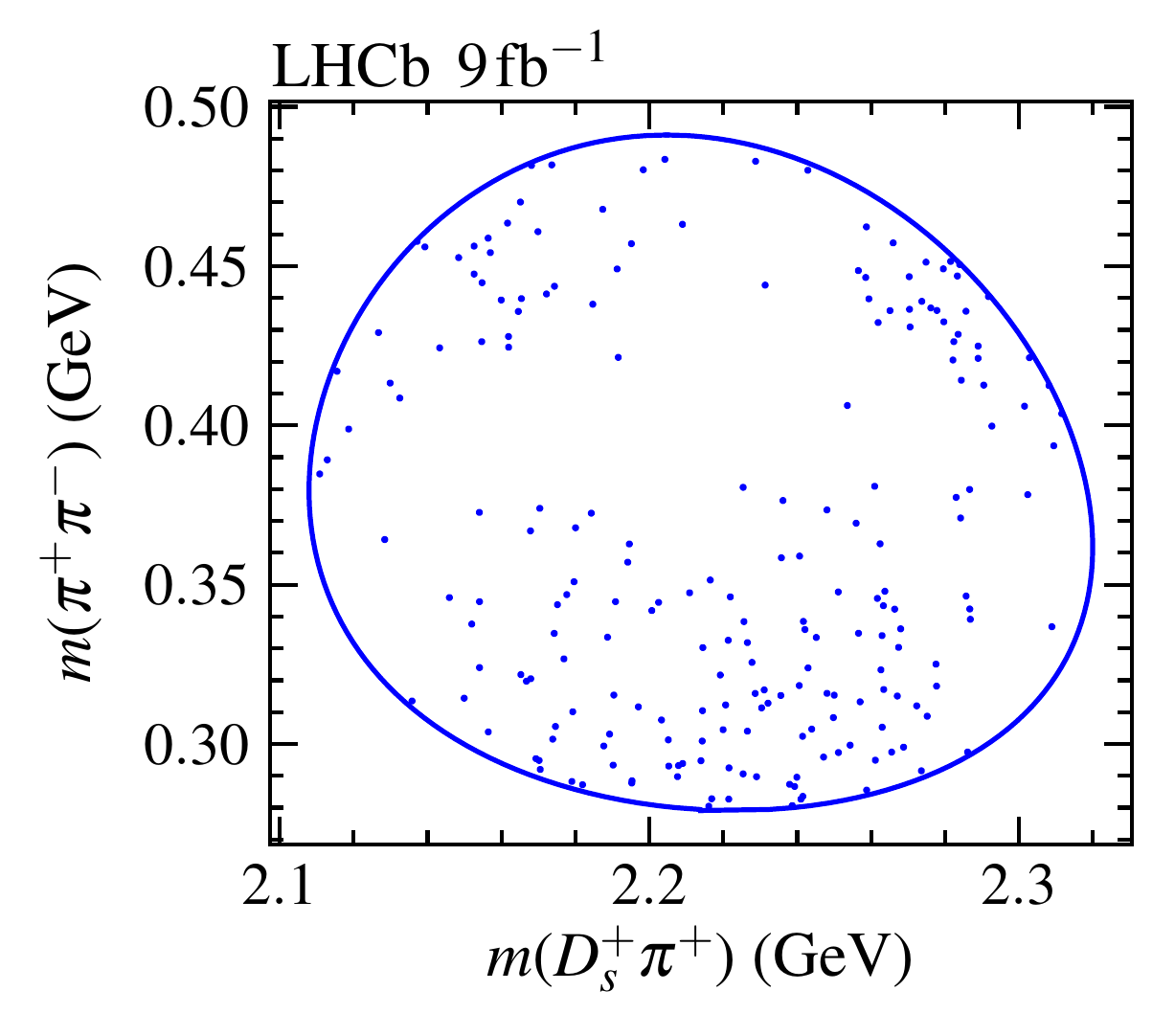}
        \begin{minipage}{5\linewidth}
            \vspace{-22.8em}\hspace{10.5em}
            (a)
        \end{minipage}
    \end{minipage}
    \begin{minipage}{0.32\linewidth}
        \centering
        \includegraphics[width=\linewidth]{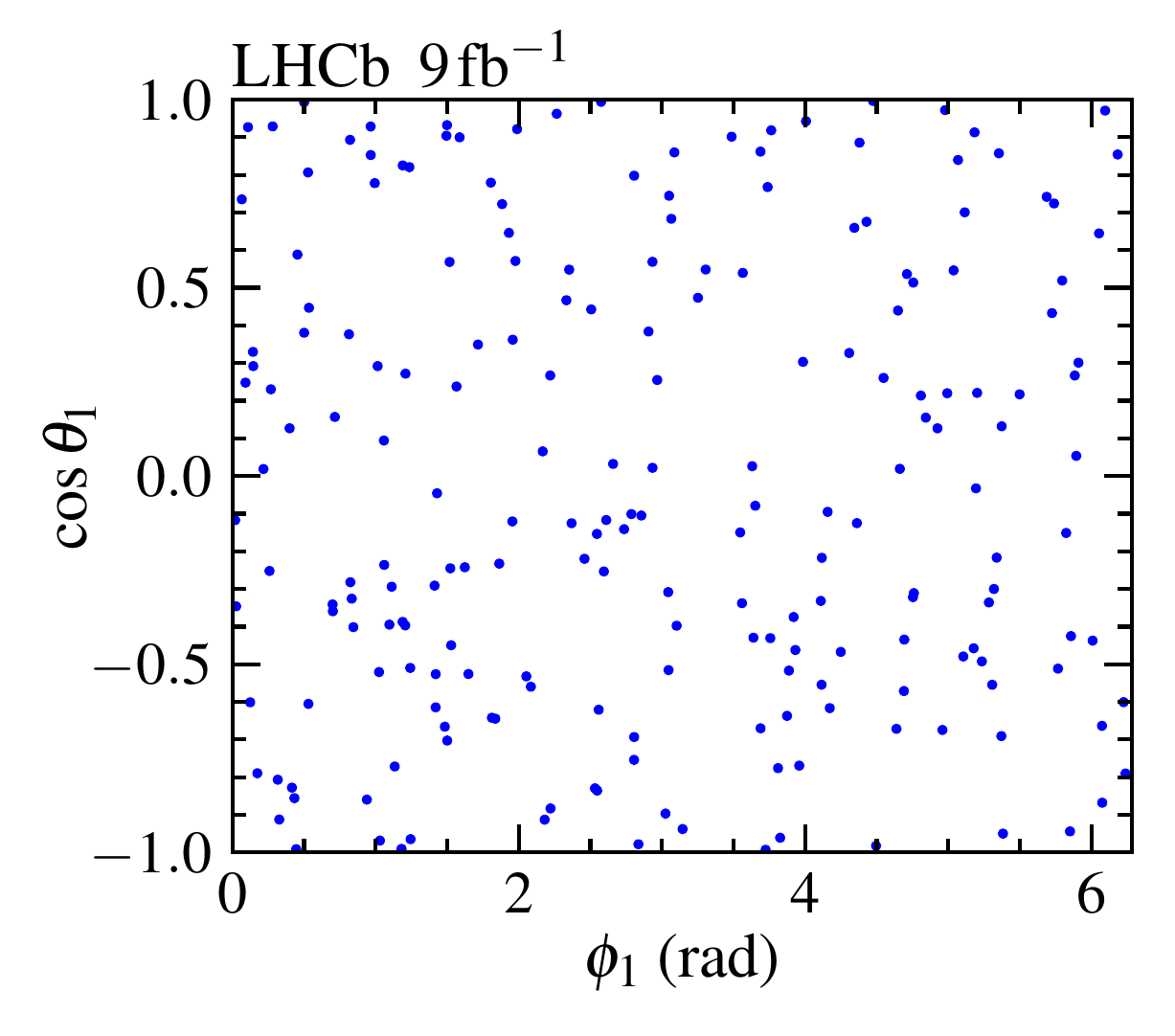}
        \begin{minipage}{5\linewidth}
            \vspace{-22.8em}\hspace{10.5em}
            (b)
        \end{minipage}
    \end{minipage}
    \begin{minipage}{0.32\linewidth}
        \centering
        \includegraphics[width=\linewidth]{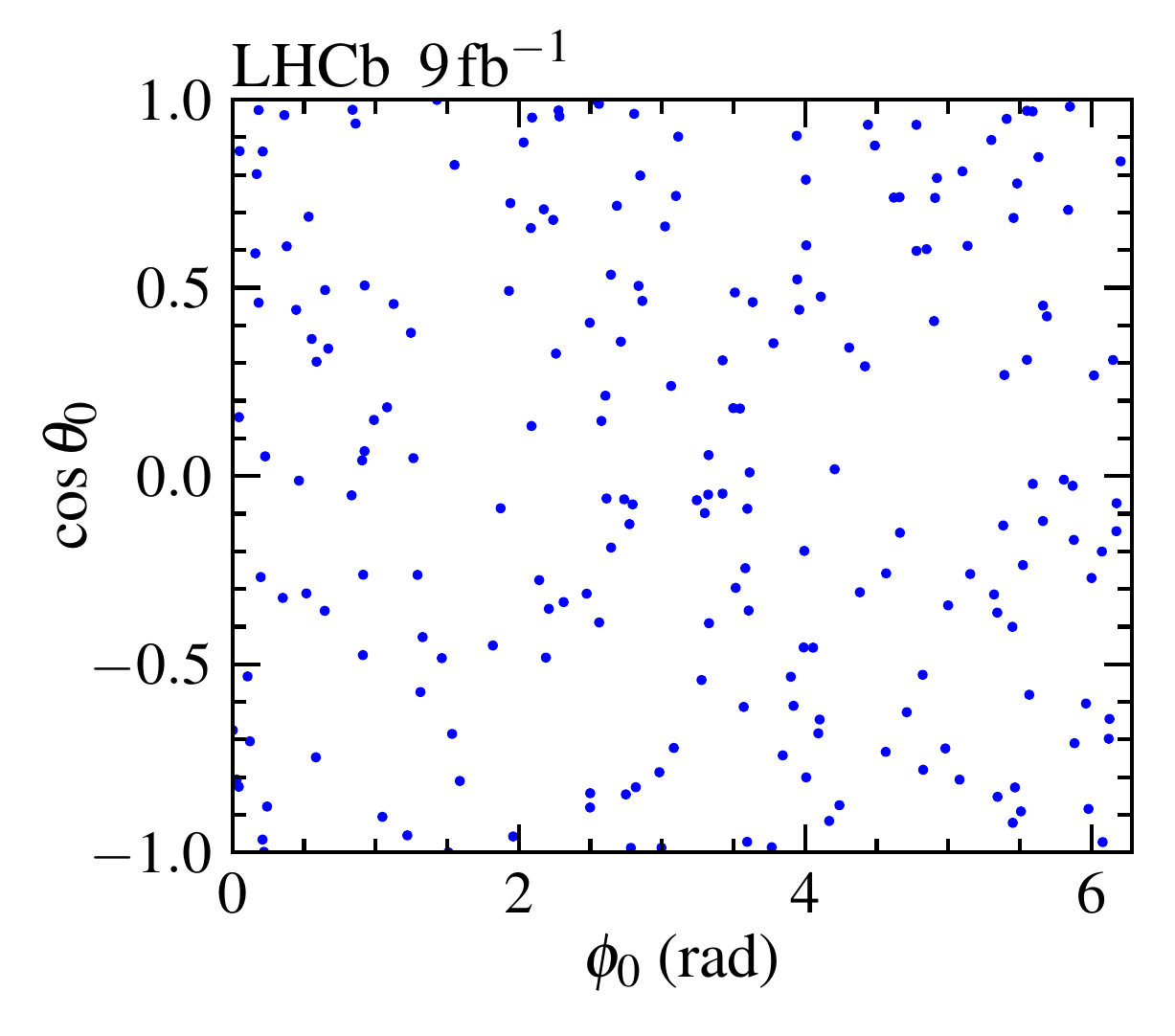}
        \begin{minipage}{5\linewidth}
            \vspace{-22.8em}\hspace{10.5em}
            (c)
        \end{minipage}
    \end{minipage}
\caption{Distributions of selected candidates in the (a)~$m(D_{s}^{+}\pi^{+})-m(\pi^{+}\pi^{-})$ plane, (b)~$\phi_1-\cos\theta_1$ plane and (c)~$\phi_0-\cos\theta_0$ plane for the \mbox{$B^0\to D^{*-}D_{s1}(2460)^{+}$} channel.}
    \label{fig:mpipi_mdspi_dstm}
\end{figure}

The efficiency-corrected data distributions combining $\decay{\Bz}{\Dm D_{s1}(2460)^{+}}$ and $\decay{\Bp}{\Dzb D_{s1}(2460)^{+}}$ channels are shown in Fig.~\ref{fig:noeffD}. 
The corresponding distributions for the $\decay{\Bz}{\Dstarm D_{s1}(2460)^{+}}$ channel are shown in Fig.~\ref{fig:noeffDst}, while two-dimensional distributions for this channel are shown in Fig.~\ref{fig:mpipi_mdspi_dstm}. 

\section{\texorpdfstring{Pull distributions}{}}
\begin{figure}[tb]
    \centering
    \begin{minipage}{0.45\linewidth}
        \centering
        \includegraphics[width=\linewidth]{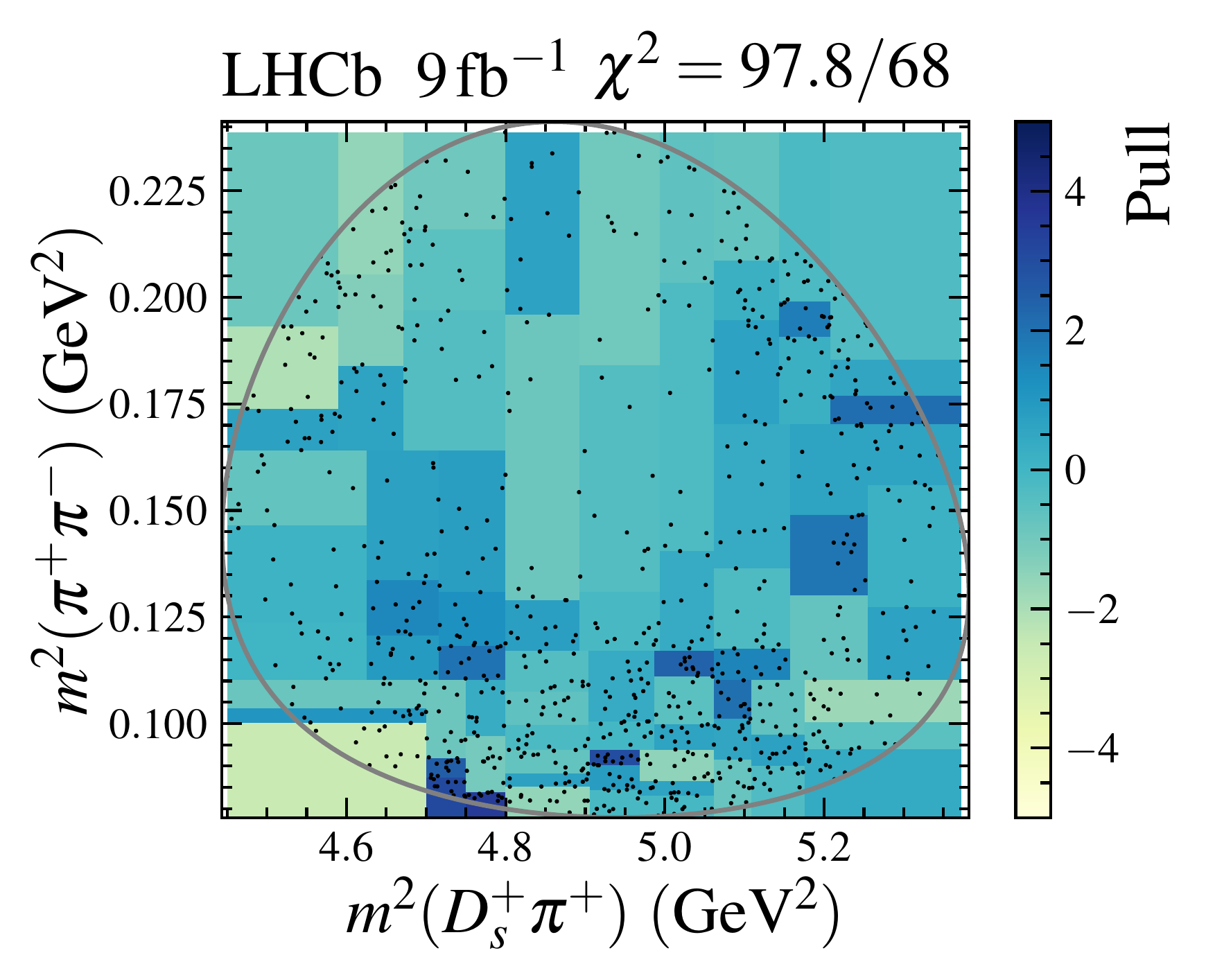}
        \begin{minipage}{\linewidth}
            \vspace{-28.0em}\hspace{13.5em}
            (a)
        \end{minipage}
    \end{minipage}
    \begin{minipage}{0.45\linewidth}
        \centering
        \includegraphics[width=\linewidth]{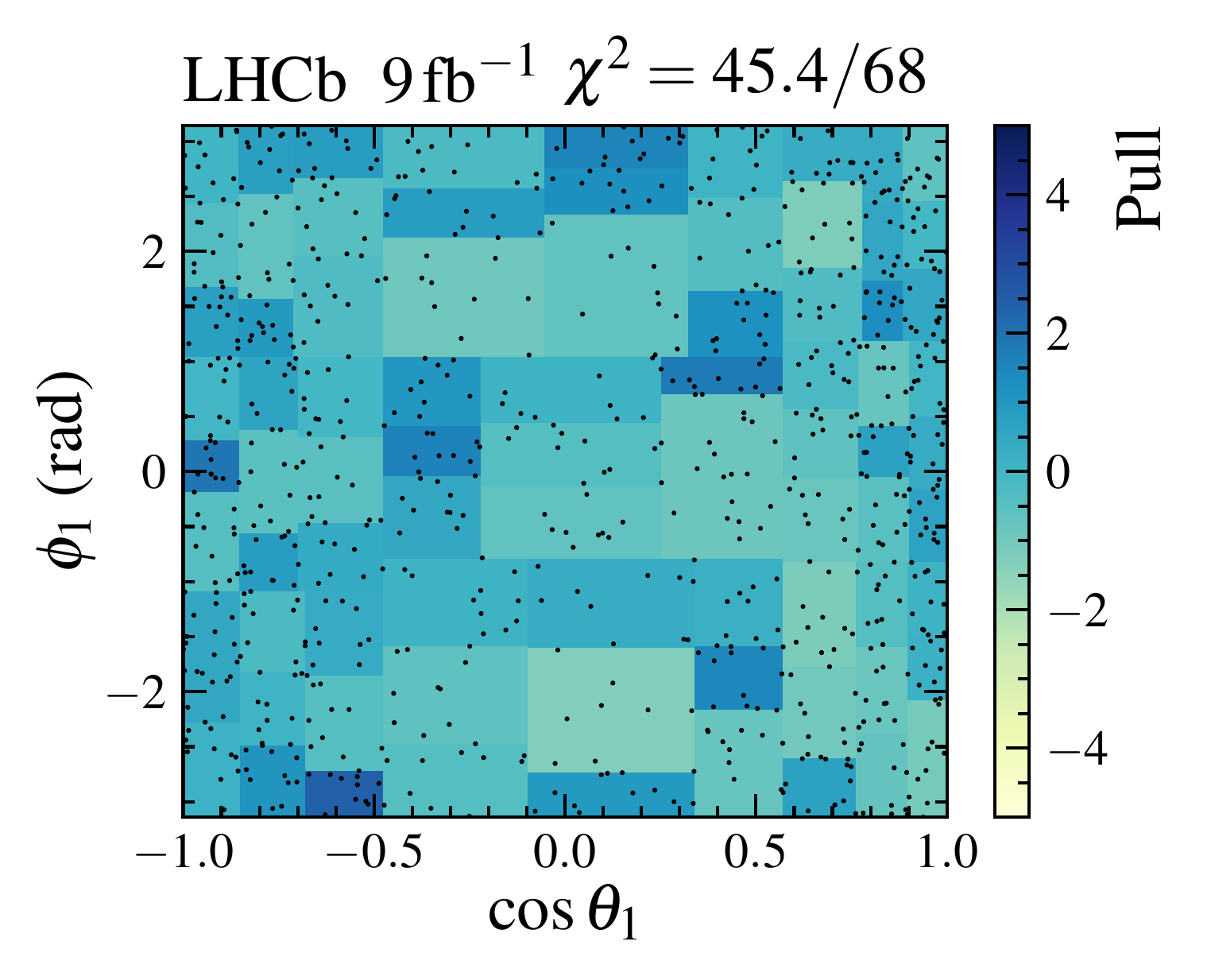}
        \begin{minipage}{\linewidth}
            \vspace{-28.0em}\hspace{13.2em}
            (b)
        \end{minipage}
    \end{minipage}
    \caption{Data distributions in the (a)~$m^{2}(\pi^{+}\pi^{-})-m^{2}(D_{s}^{+}\pi^{+})$ and (b)~$\phi_1-\cos{(\theta_1)}$ planes combining the three signal channels, superimposed on the normalised residuals between adaptive binned data and model (pull) for the $f_0(500)+f_0(980)+f_2(1270)$ model. The gray solid line in (a) denotes the boundary of the \mbox{$D_{s1}(2460)^+\to D_{s}^{+}\pi^{+}\pi^{-}$} Dalitz plot. 
    }
    \label{fig:model_sgm_f0980_f21270_chisq}
\end{figure}

\begin{figure}[tb]
    \centering
    \begin{minipage}{0.45\linewidth}
        \centering
        \includegraphics[width=\linewidth]{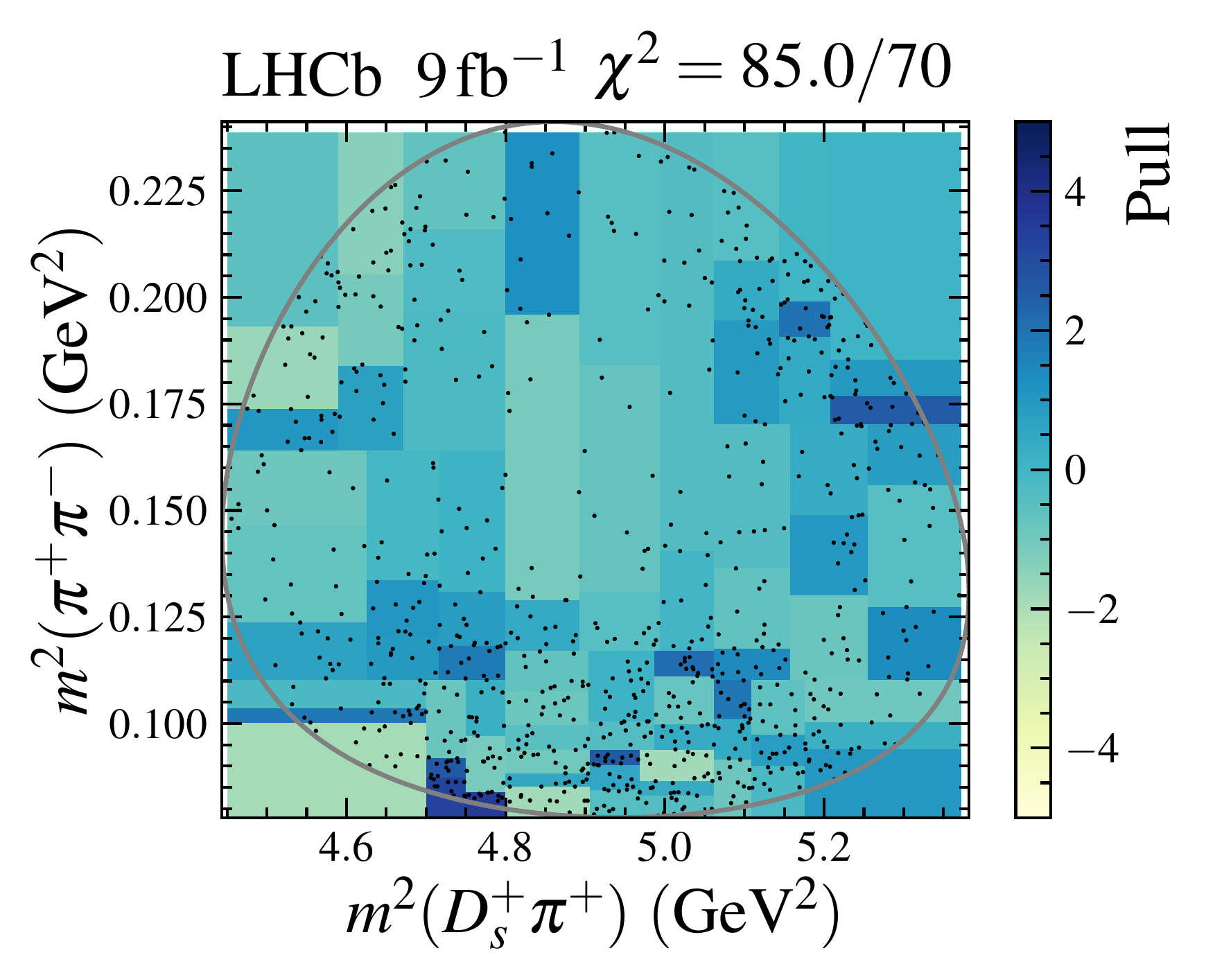}
        \begin{minipage}{\linewidth}
            \vspace{-28.0em}\hspace{13.5em}
            (a)
        \end{minipage}
    \end{minipage}
    \begin{minipage}{0.45\linewidth}
        \centering
        \includegraphics[width=\linewidth]{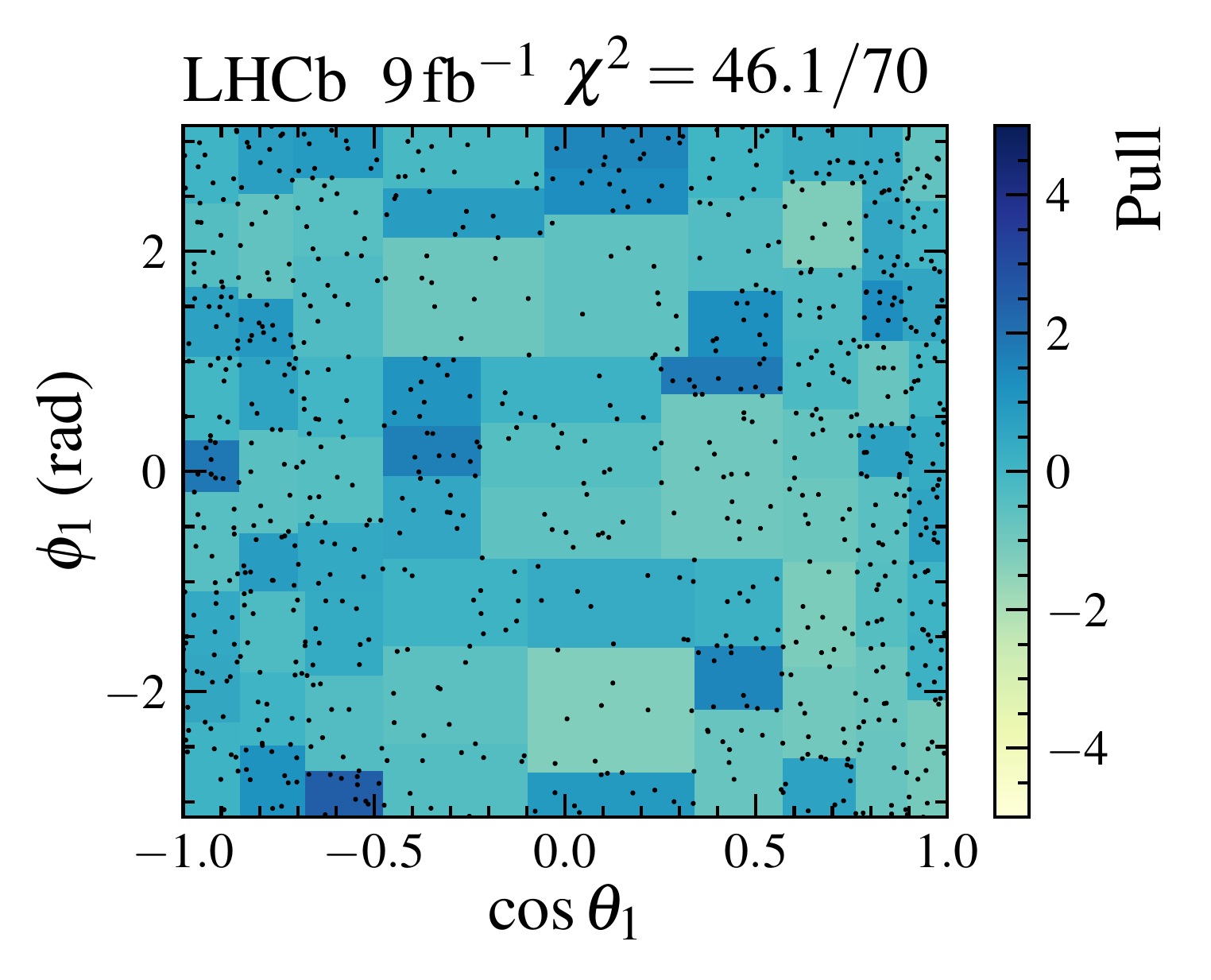}
        \begin{minipage}{\linewidth}
            \vspace{-28.0em}\hspace{13.2em}
            (b)
        \end{minipage}
    \end{minipage}
    \caption{Data distributions in the (a)~$m^{2}(\pi^{+}\pi^{-})-m^{2}(D_{s}^{+}\pi^{+})$ and (b)~$\phi_1-\cos{(\theta_1)}$ planes combining the three signal channels, superimposed on the normalised residuals between adaptive binned data and model (pull) for the $f_0(500)+\text{RBW}~T_{c\bar{s}}^{++}$ model. The gray solid line in (a) denotes the boundary of the \mbox{$D_{s1}(2460)^+\to D_{s}^{+}\pi^{+}\pi^{-}$} Dalitz plot. 
    }
    \label{fig:model_sgm_tcsbar_2_chisq}
\end{figure}

\begin{figure}[tb]
    \centering
    \begin{minipage}{0.45\linewidth}
        \centering
        \includegraphics[width=\linewidth]{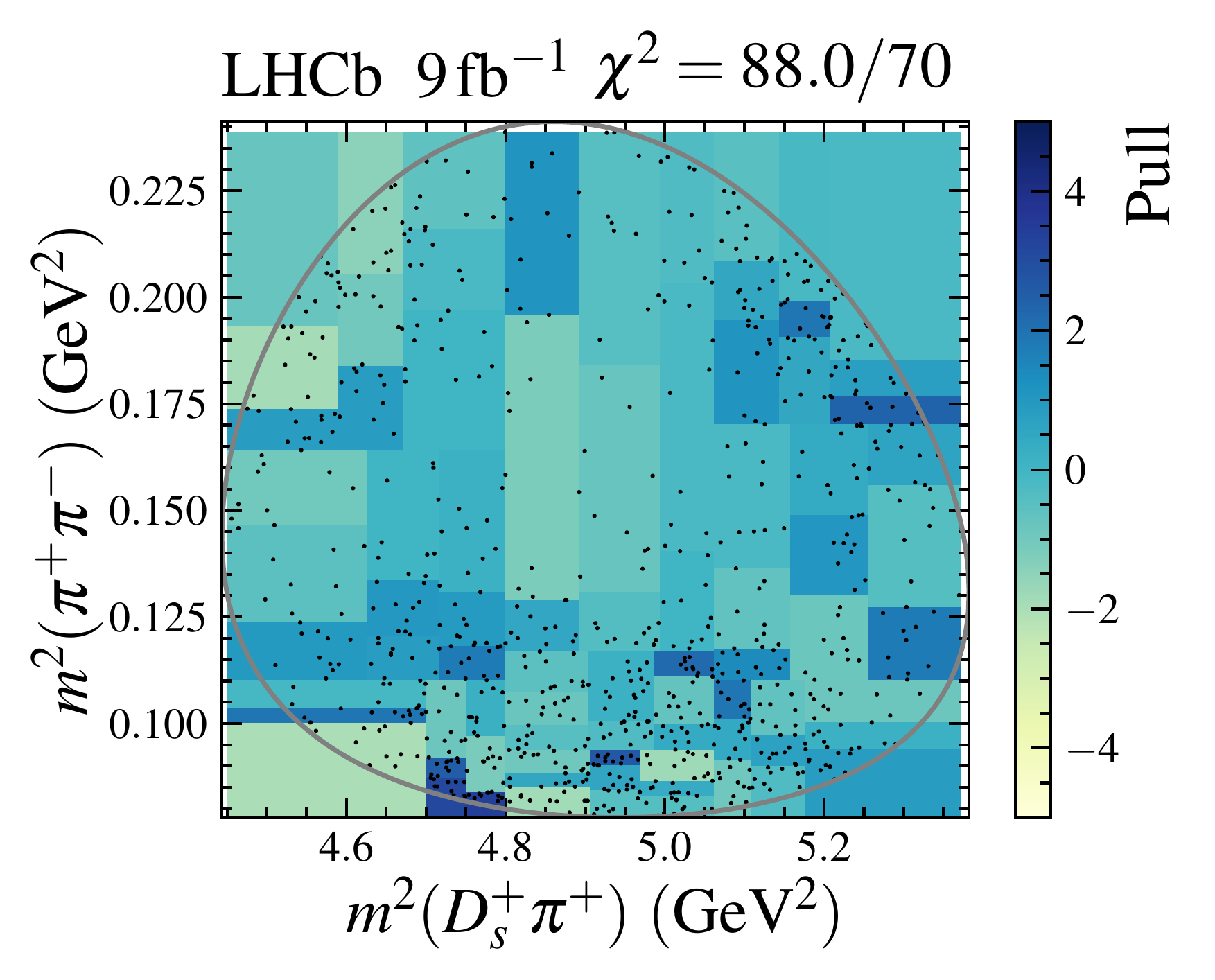}
        \begin{minipage}{\linewidth}
            \vspace{-28.0em}\hspace{13.5em}
            (a)
        \end{minipage}
    \end{minipage}
    \begin{minipage}{0.45\linewidth}
        \centering
        \includegraphics[width=\linewidth]{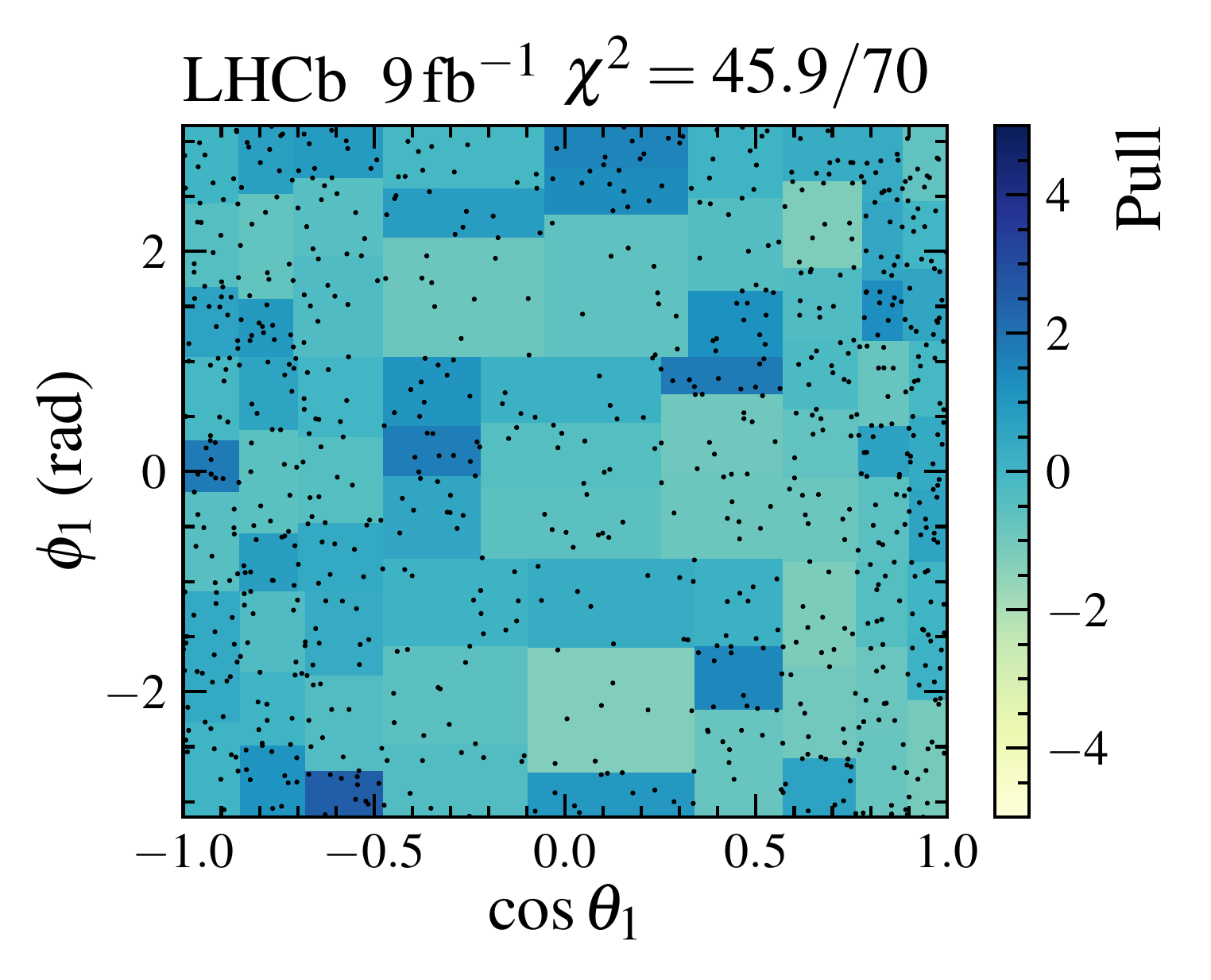}
        \begin{minipage}{\linewidth}
            \vspace{-28.0em}\hspace{13.2em}
            (b)
        \end{minipage}
    \end{minipage}
    \caption{Data distributions in the (a)~$m^{2}(\pi^{+}\pi^{-})-m^{2}(D_{s}^{+}\pi^{+})$ and (b)~$\phi_1-\cos{(\theta_1)}$ planes combining the three signal channels, superimposed on the normalised residuals between adaptive binned data and model (pull) for the $f_0(500)+\text{K-matrix}~T_{c\bar{s}}^{++}$ model. The gray solid line in (a) denotes the boundary of the \mbox{$D_{s1}(2460)^+\to D_{s}^{+}\pi^{+}\pi^{-}$} Dalitz plot. }
    \label{fig:model_sgm_Kmatrix_chisq}
\end{figure}

The distributions in the $m^{2}(\pip\pim)-m^{2}(D_{s}^+\pip)$ and $\phi_1-\cos{(\theta_1)}$ planes, superimposed on the normalised residual between adaptive binned data and model (pull) for the $f_0(500)+f_0(980)+f_2(1270)$ model, the $f_0(500)+\text{RBW}~T_{c\bar{s}}^{++}$ model, and the $f_0(500)+\text{K-matrix}~T_{c\bar{s}}^{++}$ model, are shown in Figs.~\ref{fig:model_sgm_f0980_f21270_chisq},~\ref{fig:model_sgm_tcsbar_2_chisq}, and~\ref{fig:model_sgm_Kmatrix_chisq}, respectively. The adaptive binning is chosen such that each bin contains enough data points, so the $\chi^2$ value can be correctly evaluated. 

\section{\texorpdfstring{Significance test}{}}
\begin{figure}[tb]
    \centering
    \includegraphics[width=0.55\linewidth]{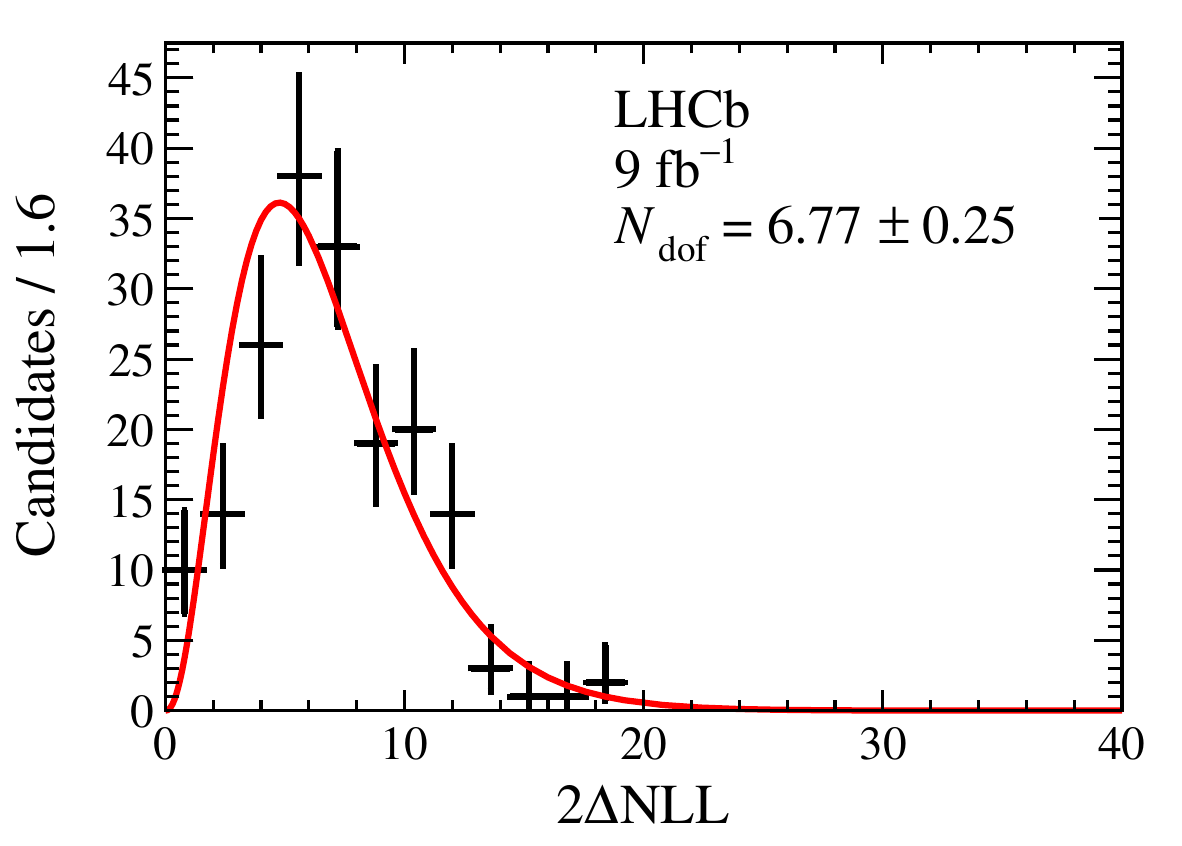}
    \caption{Distribution of 2$\Delta$NLL values used to estimate the significance of two $T_{c\bar{s}}$ contributions, where $\Delta$NLL is the change in negative log likelihood between the fit results with model $f_0(500)+f_0(980)$ and model $f_0(500)+f_0(980)+\text{K-matrix}~T_{c\bar{s}}$. 
    The 2$\Delta$NLL distribution is fitted with a $\chi^2$ distribution shown as red solid line. }
    \label{fig:delta_nll_signif}
\end{figure}

The 2$\Delta$NLL distribution obtained from pseudoexperiments is shown in Fig.~\ref{fig:delta_nll_signif}, and is fitted with a $\chi^2$ distribution. Given that the 2$\Delta$NLL value from data is 490.4, the significance of two $T_{c\bar{s}}$ contributions is estimated to be much larger than 10$\,\sigma$. 

\section{\texorpdfstring{Spin-parity test}{}}
\begin{figure}[tb]
    \centering
    \includegraphics[width=0.55\linewidth]{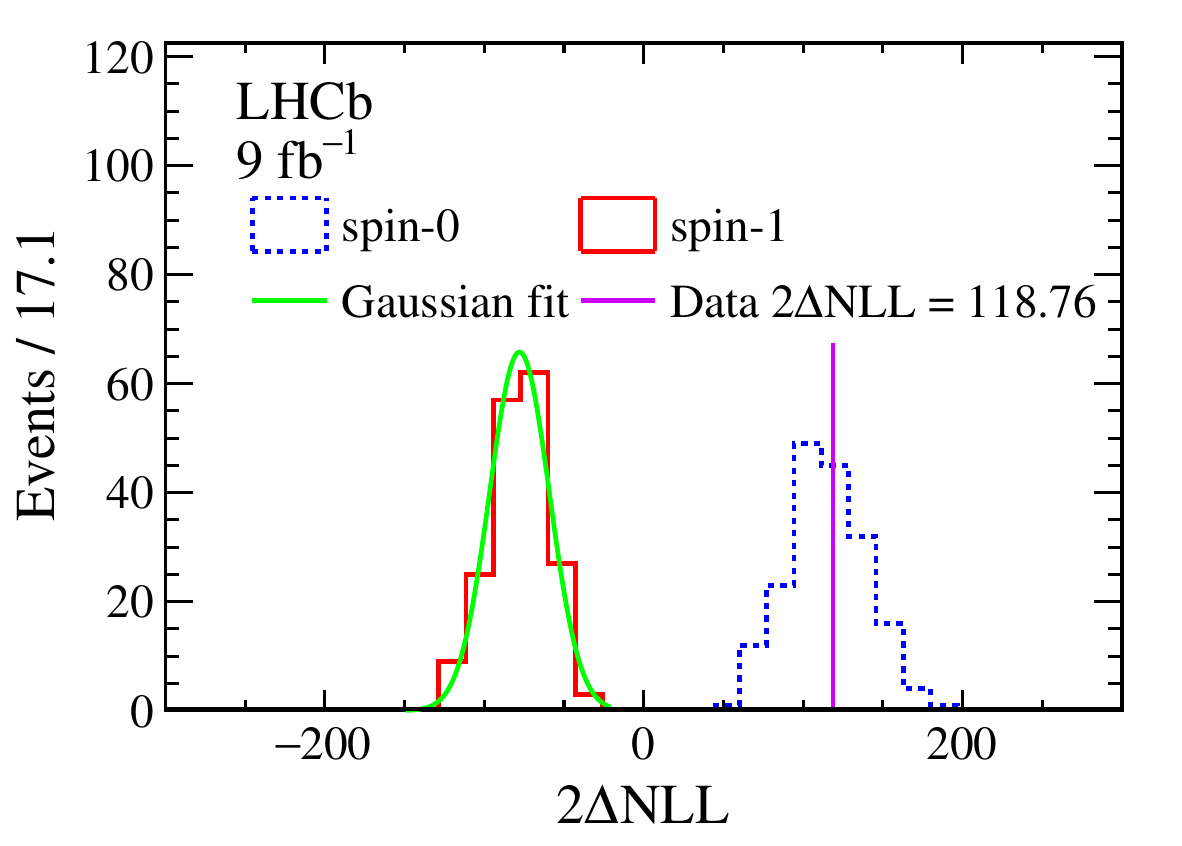}
    \caption{Distributions of 2$\Delta$NLL values used to estimate the significance of the $T_{c\bar{s}}$ spin-parity hypothesis, where $\Delta$NLL is the change in negative log likelihood between the fit results with spin-1 and spin-0 hypotheses.}
    \label{fig:delta_nll_spin}
\end{figure}

The 2$\Delta$NLL distributions obtained from pseudoexperiments are shown in Fig.~\ref{fig:delta_nll_spin}. 
The blue histogram denotes the distribution obtained from an ensemble of pseudoexperiments generated according to the results of the fit to data with the spin-0 hypothesis, which has a mean consistent with the 2$\Delta$NLL value observed in data (violet line). The red histogram denotes the distribution obtained from a corresponding ensemble with the spin-1 hypothesis, and is fitted with a Gaussian function, the result of which is shown (green line). 
The difference between the 2$\Delta$NLL value observed in data and the mean value of the spin-1 pseudoexperiments corresponds to a significance of $10\,\sigma$, demonstrating that $J^P=0^+$ is favoured with high significance.

\section{\texorpdfstring{Fit plots including interference contributions}{}}

\begin{figure}[tb]
    \centering
    \begin{minipage}{0.45\linewidth}
        \centering
        \includegraphics[width=\linewidth]{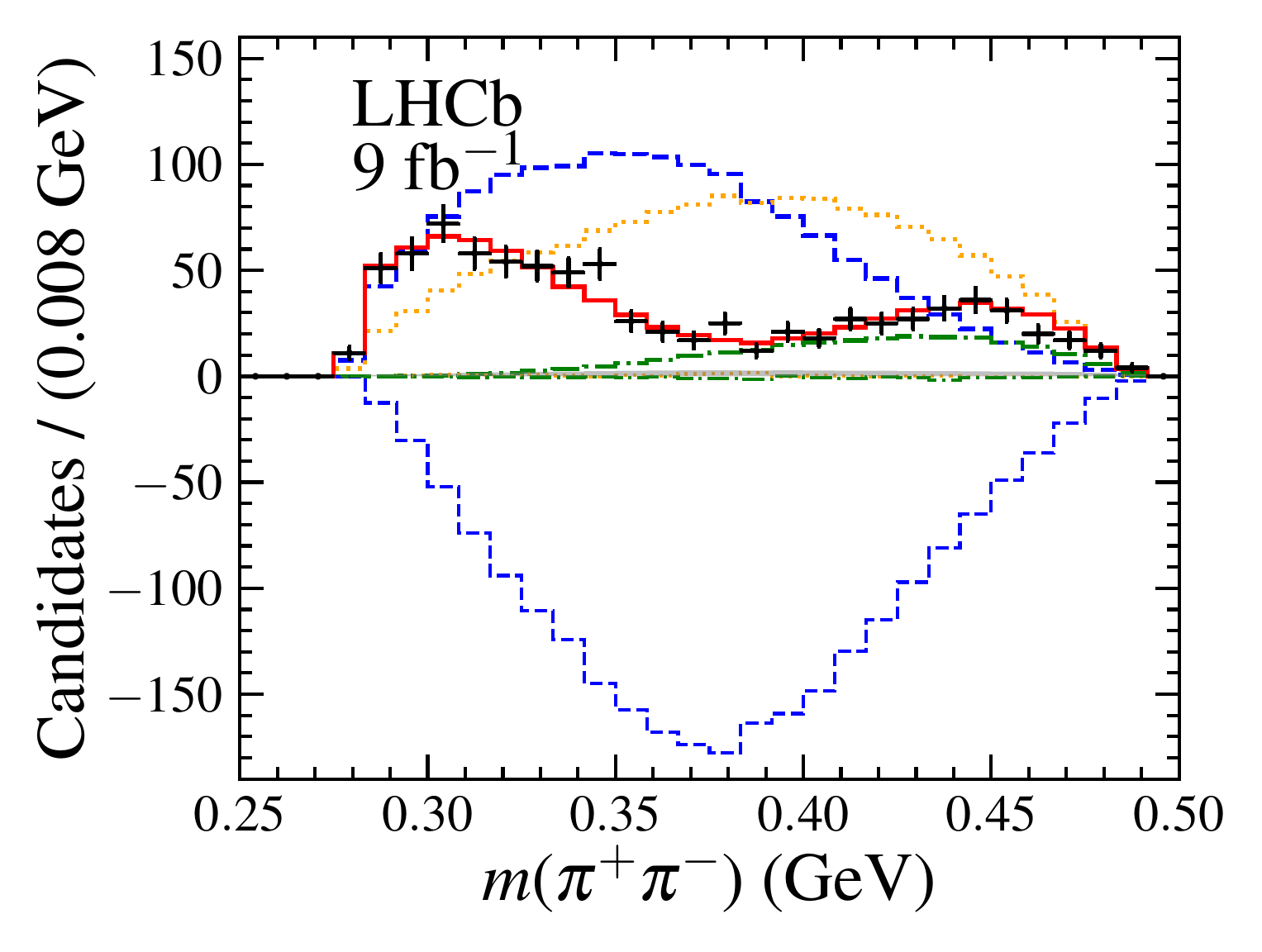}
        \begin{minipage}{\linewidth}
            \vspace{-23.0em}\hspace{14.0em}
            (a)
        \end{minipage}
        \vspace{-3em}
    \end{minipage}
    \begin{minipage}{0.45\linewidth}
        \centering
        \includegraphics[width=\linewidth]{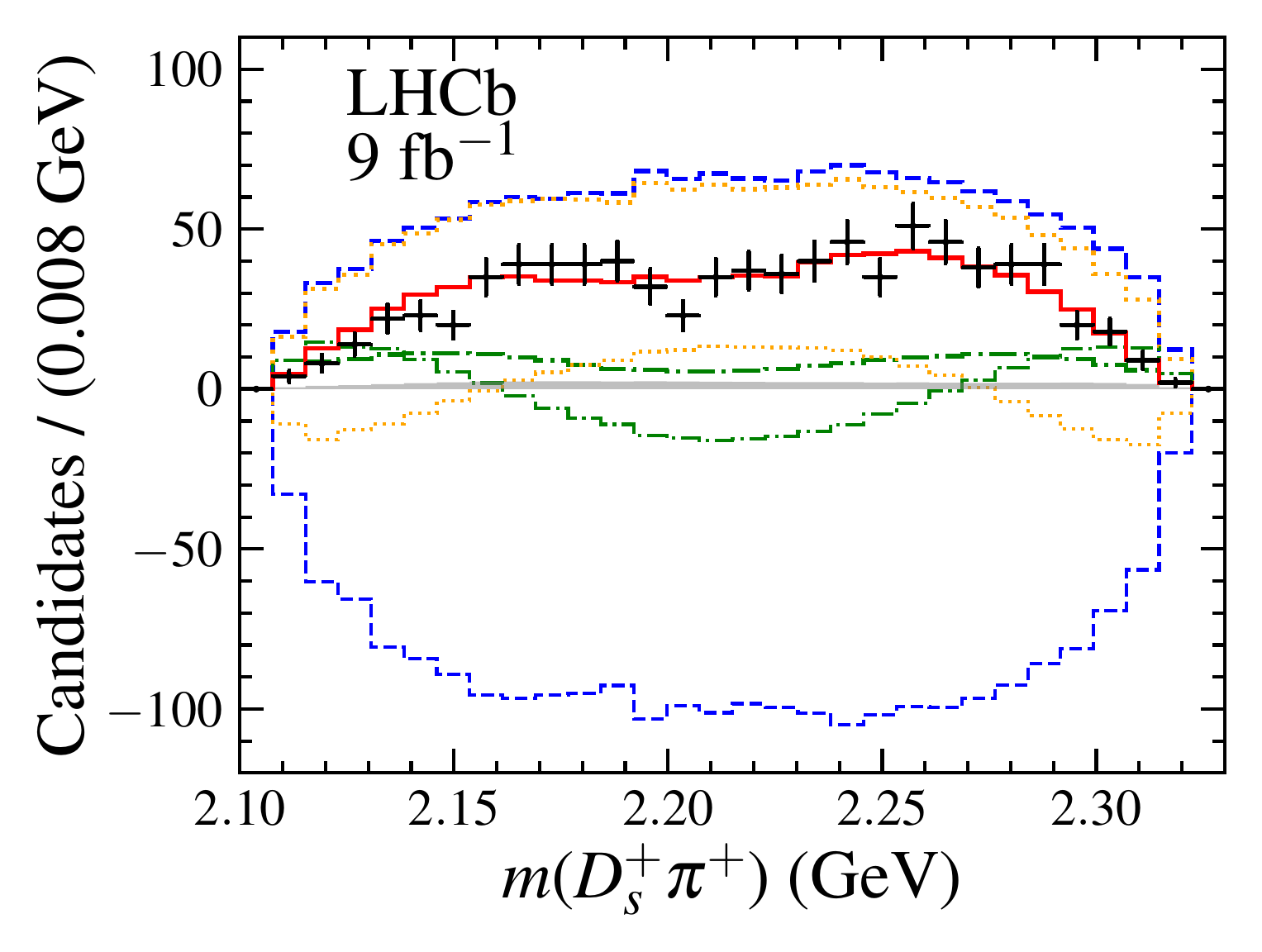}
        \begin{minipage}{\linewidth}
            \vspace{-24.0em}\hspace{14.0em}
            (b)
        \end{minipage}
        \vspace{-3.3em}
    \end{minipage}
    
 %   \begin{minipage}{0.40\linewidth}
        \hspace{3em}
 %       \centering
        \includegraphics[width=0.40\linewidth]{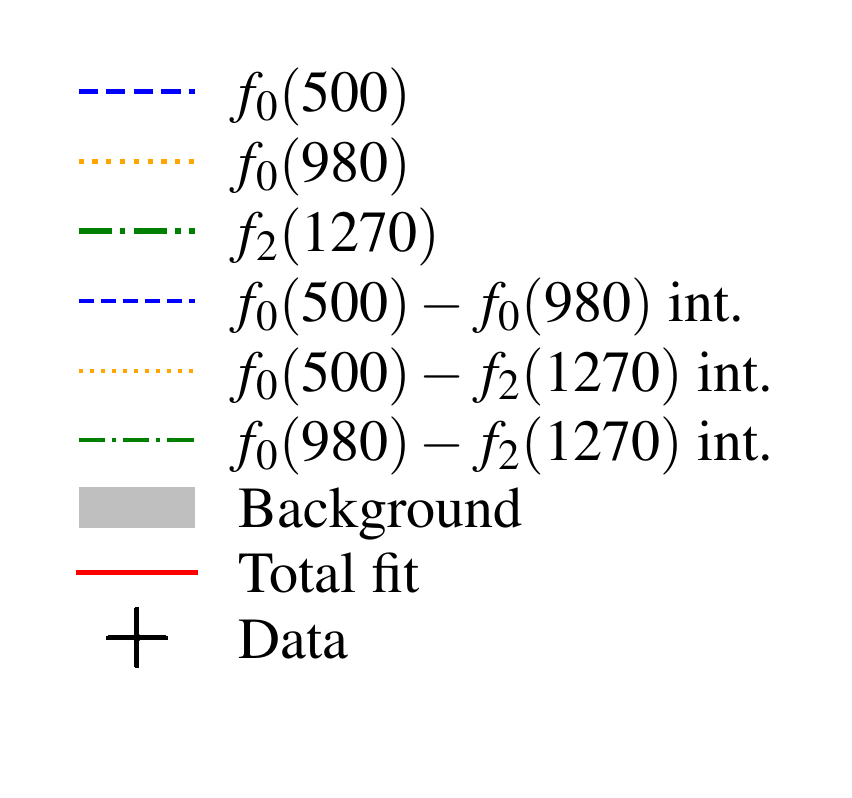}
 %   \end{minipage}
        \hspace{-1.2em}
    \begin{minipage}{0.45\linewidth}
        \vspace{-12em}
        \centering
        \includegraphics[width=\linewidth]{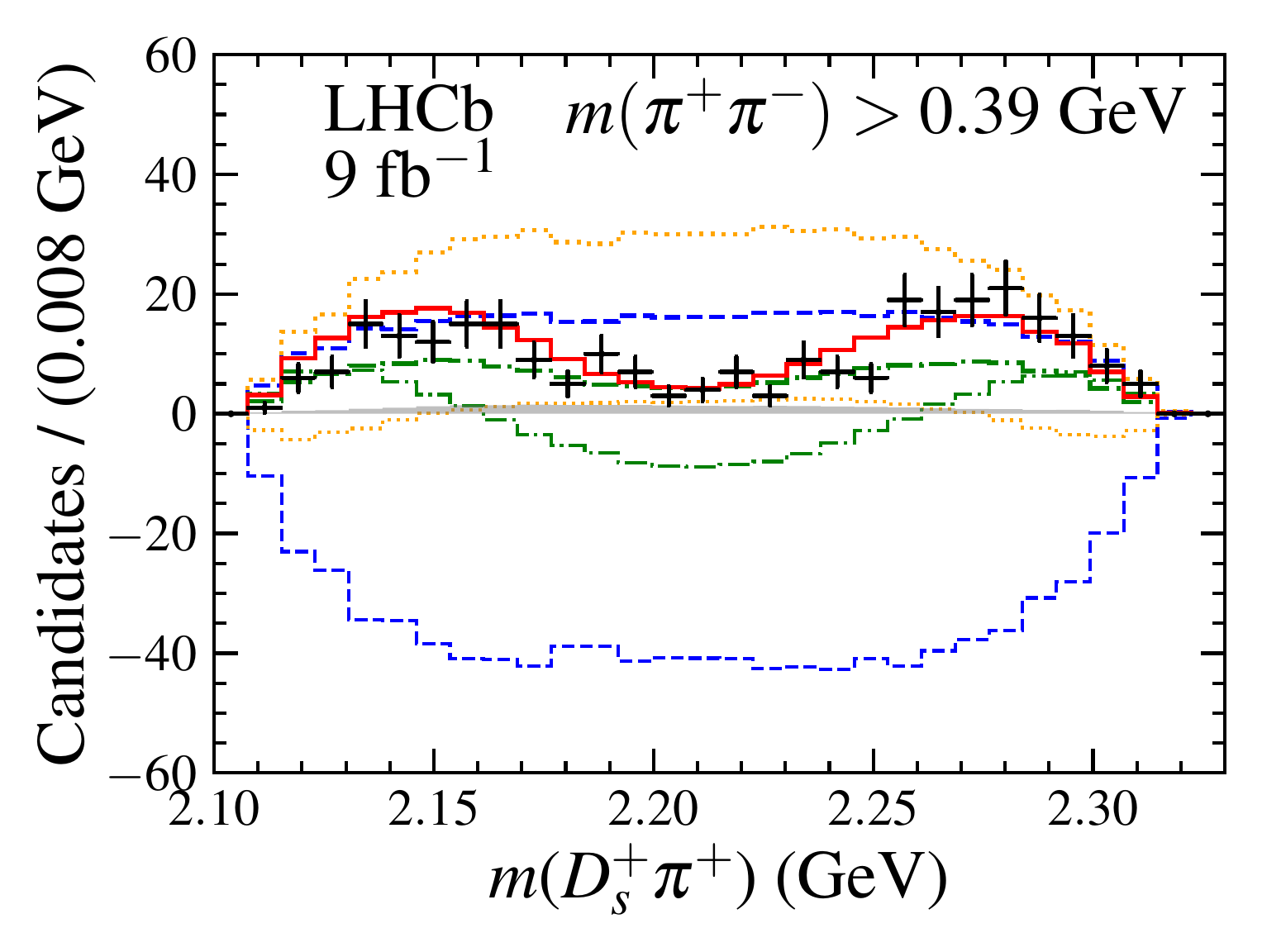}
        \begin{minipage}{\linewidth}
            \vspace{-23.0em}\hspace{14.0em}
            (c)
        \end{minipage}
    \end{minipage}
    \vspace{-2em}
    \caption{Comparison between data (black dots with error bars) and results of the fit with the $f_0(500)+f_0(980)+f_2(1270)$ model (red solid line). The distributions are for the three channels combined in (a)~$m(\pi^{+}\pi^{-})$, (b)~$m(D_{s}^{+}\pi^{+})$, and (c)~$m(D_{s}^{+}\pi^{+})$ requiring $m(\pi^{+}\pi^{-})>0.39\,\mathrm{GeV}$. Individual components, corresponding to the background contribution estimated by $m(\Dsp\pip\pim)$ sideband regions (gray-filled) and different contributions from resonances (coloured dashed lines) and interference between the resonances (coloured dotted lines), are also shown as indicated in the legend. }
    \label{fig:model_sgm_f0980_f21270_itfrc}
\end{figure}

\begin{figure}[tb]
    \centering
    \begin{minipage}{0.45\linewidth}
        \centering
        \includegraphics[width=\linewidth]{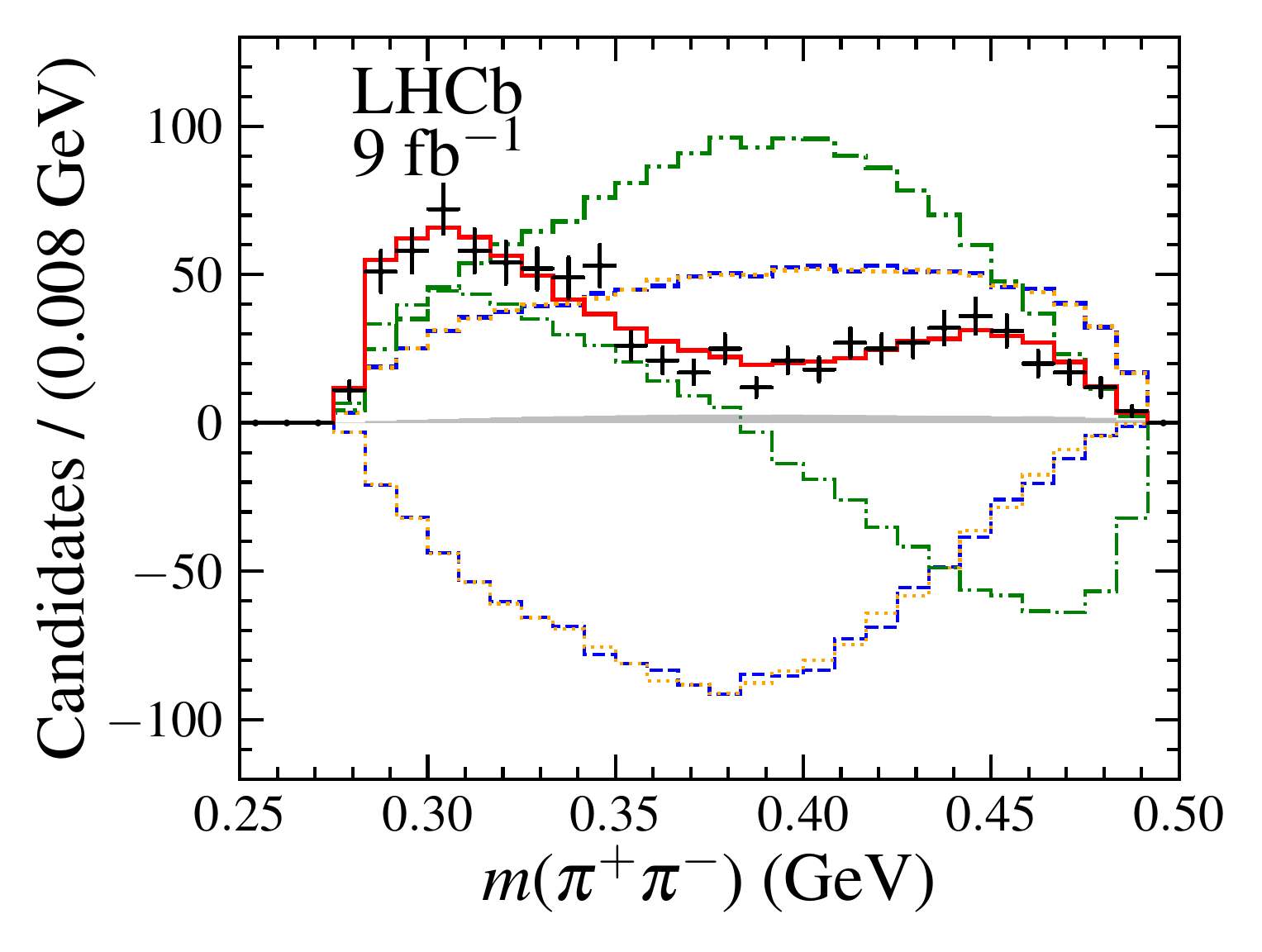}
        \begin{minipage}{\linewidth}
            \vspace{-24.0em}\hspace{14.0em}
            (a)
        \end{minipage}
        \vspace{-3em}
    \end{minipage}
    \begin{minipage}{0.45\linewidth}
        \centering
        \includegraphics[width=\linewidth]{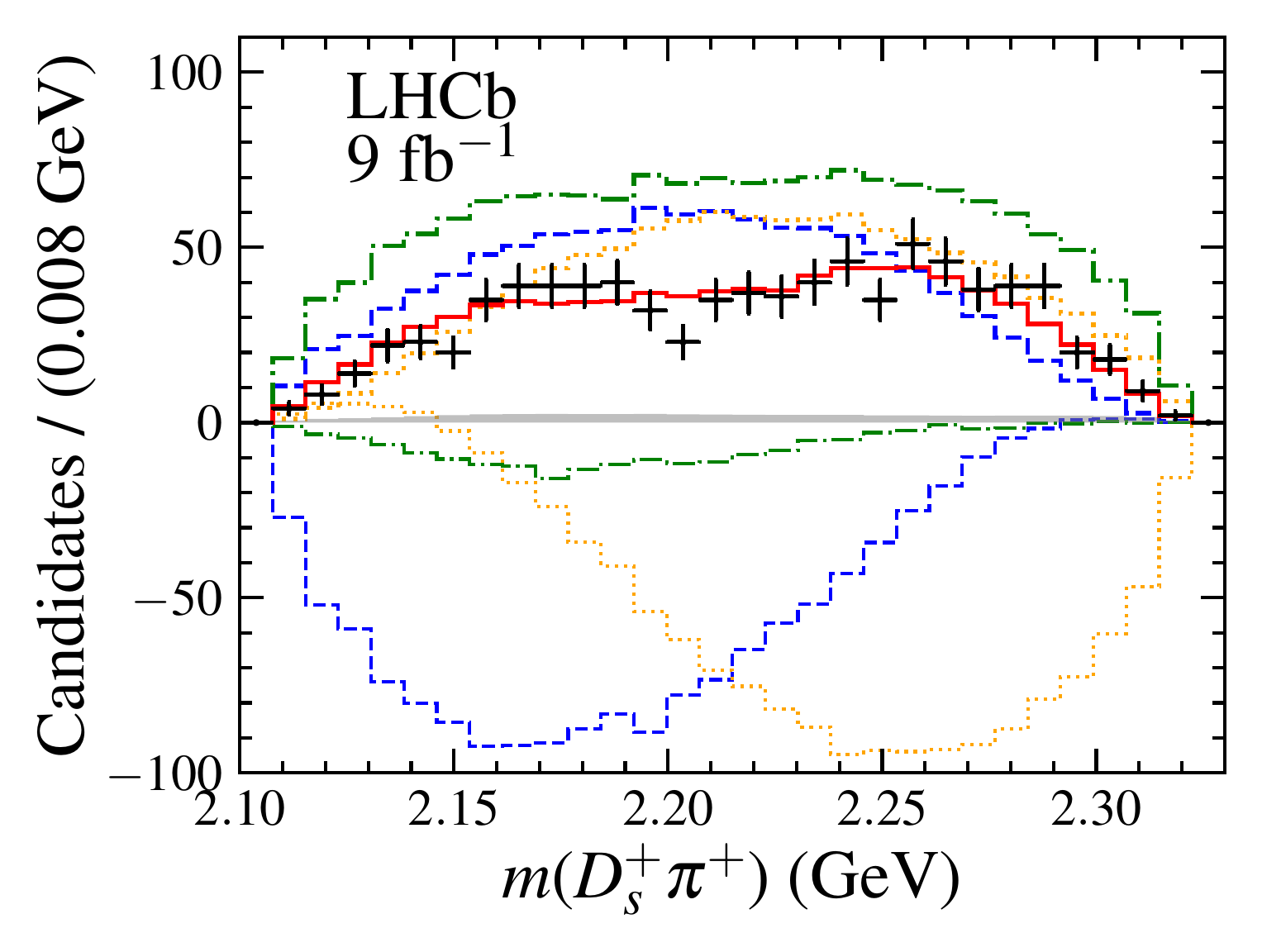}
        \begin{minipage}{\linewidth}
            \vspace{-24.0em}\hspace{14.0em}
            (b)
        \end{minipage}
        \vspace{-3.3em}
    \end{minipage}
    
 %   \begin{minipage}{0.40\linewidth}
        \hspace{3em}
 %       \centering
        \includegraphics[width=0.40\linewidth]{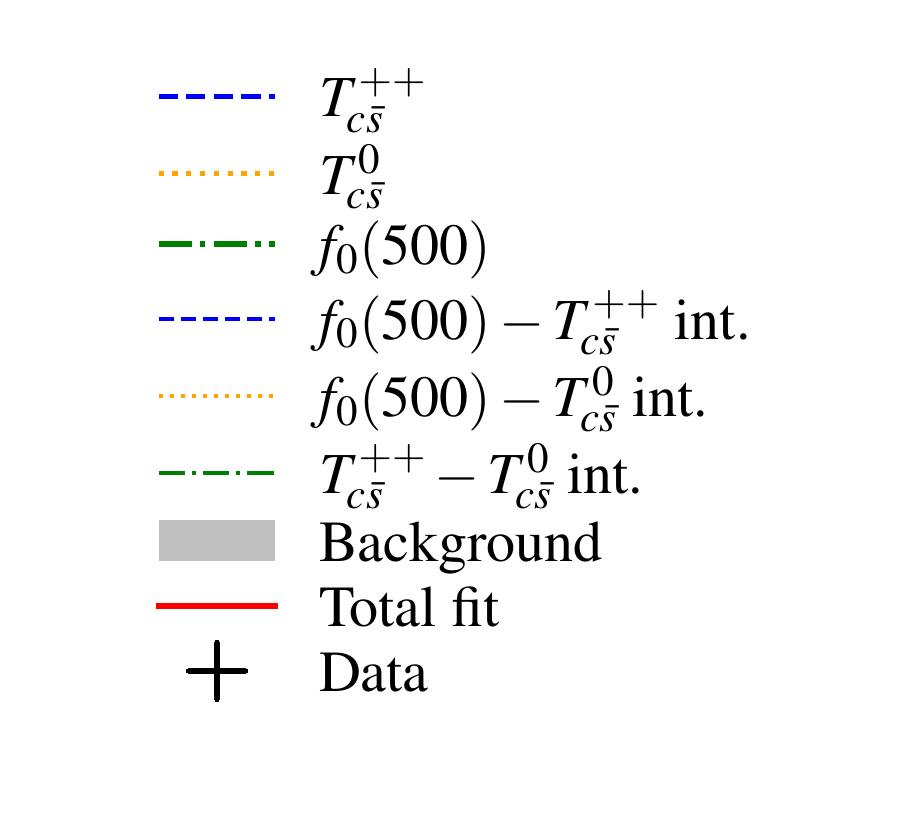}
 %   \end{minipage}
        \hspace{-1.2em}
    \begin{minipage}{0.45\linewidth}
        \vspace{-12em}
        \centering
        \includegraphics[width=\linewidth]{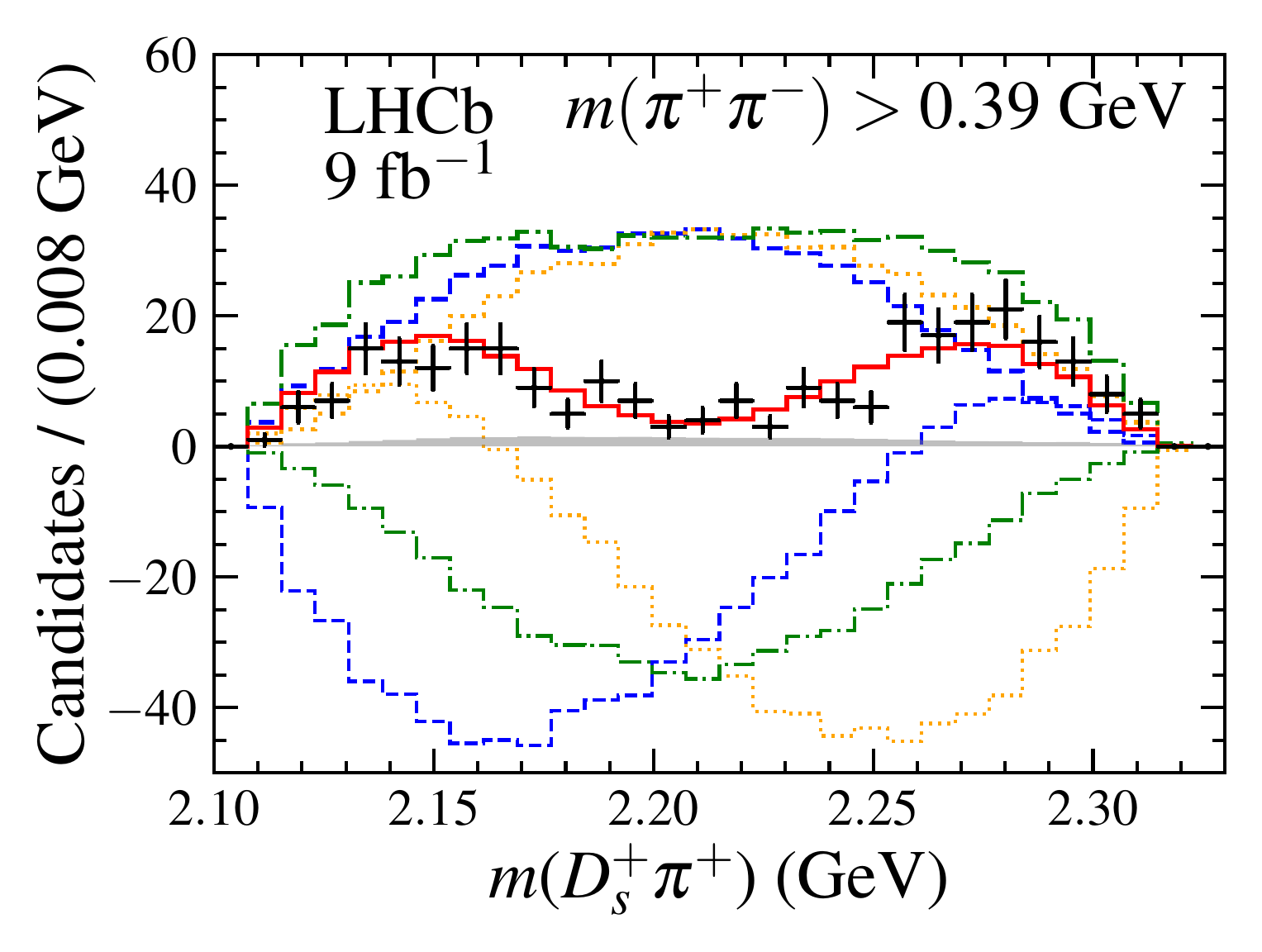}
        \begin{minipage}{\linewidth}
            \vspace{-23.0em}\hspace{14.0em}
            (c)
        \end{minipage}
    \end{minipage}
    \vspace{-2em}
    \caption{Comparison between data (black dots with error bars) and results of the fit with the $f_0(500)+\text{RBW}~T_{c\bar{s}}(0^+)$ model (red solid line). The distributions are for the three channels combined in (a)~$m(\pi^{+}\pi^{-})$, (b)~$m(D_{s}^{+}\pi^{+})$, and (c)~$m(D_{s}^{+}\pi^{+})$ requiring $m(\pi^{+}\pi^{-})>0.39\,\mathrm{GeV}$. Individual components, corresponding to the background contribution estimated by $m(\Dsp\pip\pim)$ sideband regions (gray-filled) and different contributions from resonances (coloured dashed lines) and interference between the resonances (coloured dotted lines), are also shown as indicated in the legend. }
    \label{fig:model_sgm_tcsbar_itfrc}
\end{figure}

\begin{figure}[tb]
    \centering
    \begin{minipage}{0.45\linewidth}
        \centering
        \includegraphics[width=\linewidth]{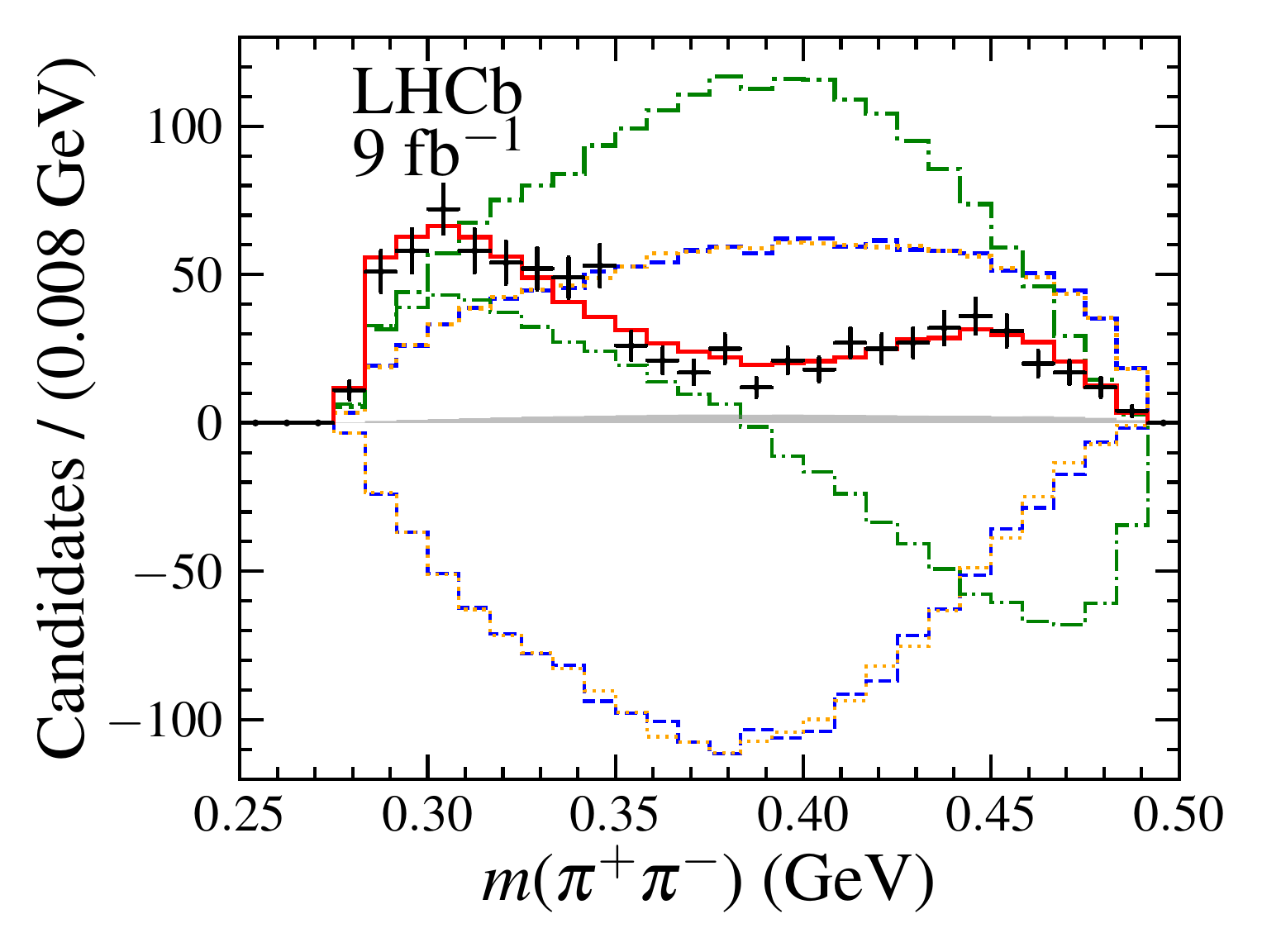}
        \begin{minipage}{\linewidth}
            \vspace{-24.0em}\hspace{14.0em}
            (a)
        \end{minipage}
        \vspace{-3em}
    \end{minipage}
    \begin{minipage}{0.45\linewidth}
        \centering
        \includegraphics[width=\linewidth]{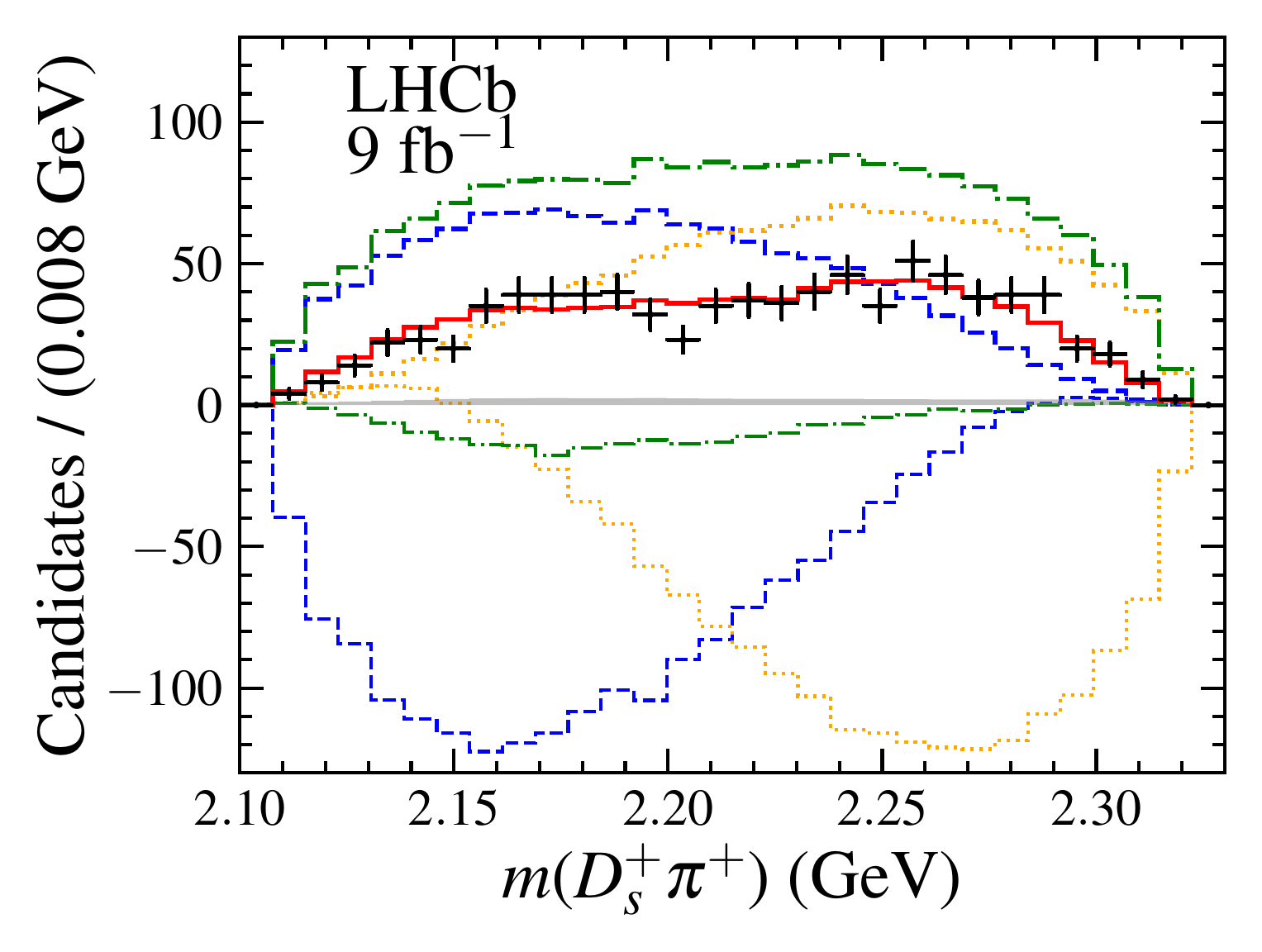}
        \begin{minipage}{\linewidth}
            \vspace{-24.0em}\hspace{14.0em}
            (b)
        \end{minipage}
        \vspace{-3.3em}
    \end{minipage}
    
 %   \begin{minipage}{0.40\linewidth}
        \hspace{3em}
 %       \centering
        \includegraphics[width=0.40\linewidth]{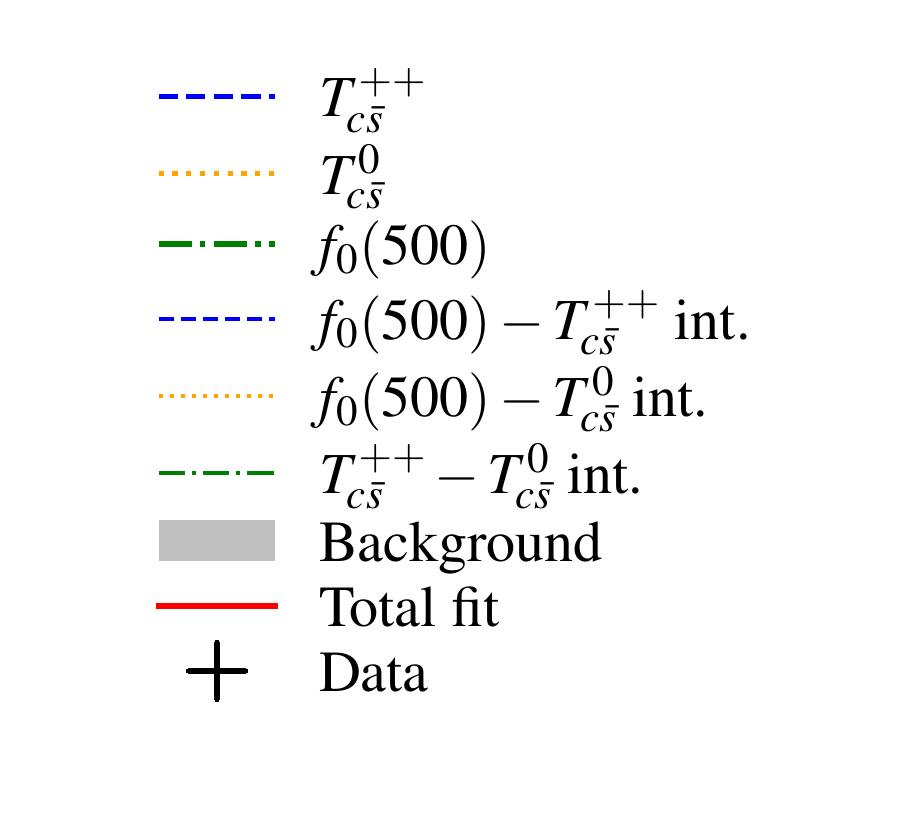}
 %   \end{minipage}
        \hspace{-1.2em}
    \begin{minipage}{0.45\linewidth}
        \vspace{-12em}
        \centering
        \includegraphics[width=\linewidth]{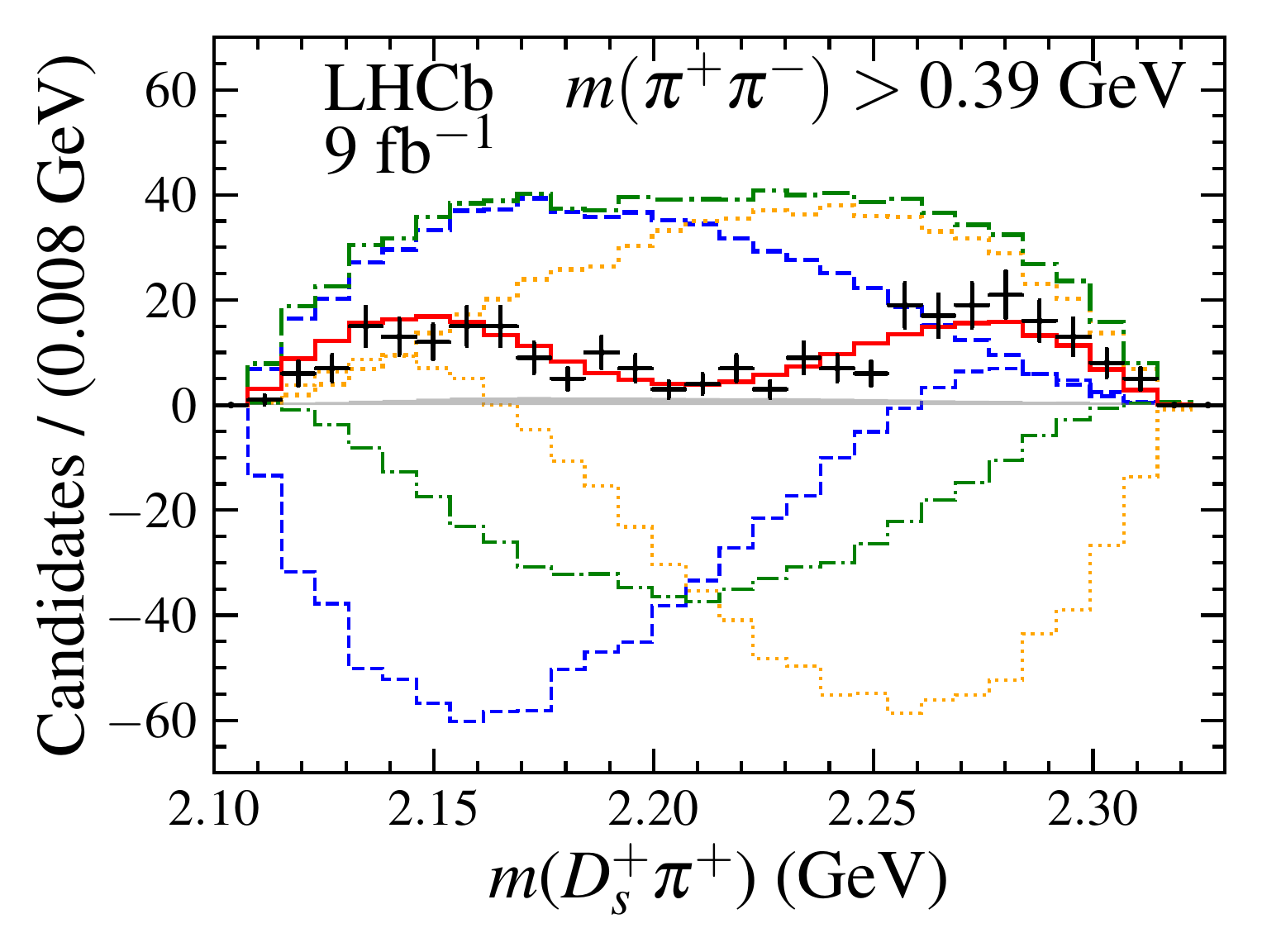}
        \begin{minipage}{\linewidth}
            \vspace{-23.0em}\hspace{14.0em}
            (c)
        \end{minipage}
    \end{minipage}
    \vspace{-2em}
    \caption{Comparison between data (black dots with error bars) and results of the fit with the $f_0(500)+\text{K-matrix}~T_{c\bar{s}}(0^+)$ model (red solid line). The distributions are for the three channels combined in (a)~$m(\pi^{+}\pi^{-})$, (b)~$m(D_{s}^{+}\pi^{+})$, and (c)~$m(D_{s}^{+}\pi^{+})$ requiring $m(\pi^{+}\pi^{-})>0.39\,\mathrm{GeV}$. Individual components, corresponding to the background contribution estimated by $m(\Dsp\pip\pim)$ sideband regions (gray-filled) and different contributions from resonances (coloured dashed lines) and interference between the resonances (coloured dotted lines), are also shown as indicated in the legend. }
    \label{fig:model_sgm_Kmatrix_itfrc}
\end{figure}

The comparisons between data and fit results with different models including the interference contributions are shown in Figs.~\ref{fig:model_sgm_f0980_f21270_itfrc}--\ref{fig:model_sgm_Kmatrix_itfrc}, where the interference contributions are mostly negative. 

\clearpage

% This should be taken out in the final paper
%\input{supplementary-app}

\addcontentsline{toc}{section}{References}
%\setboolean{inbibliography}{true}
\bibliographystyle{LHCb}
\bibliography{main,standard,LHCb-PAPER,LHCb-CONF,LHCb-DP,LHCb-TDR}

\ifx\mcitethebibliography\mciteundefinedmacro
\PackageError{LHCb.bst}{mciteplus.sty has not been loaded}
{This bibstyle requires the use of the mciteplus package.}\fi
\providecommand{\href}[2]{#2}
\begin{mcitethebibliography}{10}
\mciteSetBstSublistMode{n}
\mciteSetBstMaxWidthForm{subitem}{\alph{mcitesubitemcount})}
\mciteSetBstSublistLabelBeginEnd{\mcitemaxwidthsubitemform\space}
{\relax}{\relax}

\bibitem{BaBar:2003oey}
BaBar Collaboration, B.~Aubert {\em et~al.},
  \ifthenelse{\boolean{articletitles}}{\emph{{Observation of a narrow meson
  decaying to $D_s^+ \pi^0$ at a mass of $2.32 \gevcc$}},
  }{}\href{https://doi.org/10.1103/PhysRevLett.90.242001}{Phys.\ Rev.\ Lett.\
  \textbf{90} (2003) 242001}\relax
\mciteBstWouldAddEndPuncttrue
\mciteSetBstMidEndSepPunct{\mcitedefaultmidpunct}
{\mcitedefaultendpunct}{\mcitedefaultseppunct}\relax
\EndOfBibitem
\bibitem{CLEO:2003ggt}
CLEO Collaboration, D.~Besson {\em et~al.},
  \ifthenelse{\boolean{articletitles}}{\emph{{Observation of a narrow resonance
  of mass $2.46\gevcc$ decaying to $D^{*+}_s \pi^0$ and confirmation of the
  $D^*_{sJ}(2317)$ state}},
  }{}\href{https://doi.org/10.1103/PhysRevD.68.032002}{Phys.\ Rev.\
  \textbf{D68} (2003) 032002}, Erratum
  \href{https://doi.org/10.1103/PhysRevD.75.119908}{ibid.\   \textbf{D75}
  (2007) 119908}\relax
\mciteBstWouldAddEndPuncttrue
\mciteSetBstMidEndSepPunct{\mcitedefaultmidpunct}
{\mcitedefaultendpunct}{\mcitedefaultseppunct}\relax
\EndOfBibitem
\bibitem{Chen:2004dy}
Y.-Q. Chen and X.-Q. Li, \ifthenelse{\boolean{articletitles}}{\emph{{A
  comprehensive four-quark interpretation of $D_s(2317)$, $D_s(2457)$ and
  $D_s(2632)$}}, }{}\href{https://doi.org/10.1103/PhysRevLett.93.232001}{Phys.\
  Rev.\ Lett.\  \textbf{93} (2004) 232001}\relax
\mciteBstWouldAddEndPuncttrue
\mciteSetBstMidEndSepPunct{\mcitedefaultmidpunct}
{\mcitedefaultendpunct}{\mcitedefaultseppunct}\relax
\EndOfBibitem
\bibitem{Guo:2006rp}
F.-K. Guo, P.-N. Shen, and H.-C. Chiang,
  \ifthenelse{\boolean{articletitles}}{\emph{{Dynamically generated $1^+$ heavy
  mesons}}, }{}\href{https://doi.org/10.1016/j.physletb.2007.01.050}{Phys.\
  Lett.\  \textbf{B647} (2007) 133}\relax
\mciteBstWouldAddEndPuncttrue
\mciteSetBstMidEndSepPunct{\mcitedefaultmidpunct}
{\mcitedefaultendpunct}{\mcitedefaultseppunct}\relax
\EndOfBibitem
\bibitem{Lutz:2007sk}
M.~F.~M. Lutz and M.~Soyeur,
  \ifthenelse{\boolean{articletitles}}{\emph{{Radiative and isospin-violating
  decays of $D_s$ mesons in the hadrogenesis conjecture}},
  }{}\href{https://doi.org/10.1016/j.nuclphysa.2008.09.003}{Nucl.\ Phys.\
  \textbf{A813} (2008) 14}\relax
\mciteBstWouldAddEndPuncttrue
\mciteSetBstMidEndSepPunct{\mcitedefaultmidpunct}
{\mcitedefaultendpunct}{\mcitedefaultseppunct}\relax
\EndOfBibitem
\bibitem{Rosner:2006vc}
J.~L. Rosner, \ifthenelse{\boolean{articletitles}}{\emph{{Effects of S-wave
  thresholds}}, }{}\href{https://doi.org/10.1103/PhysRevD.74.076006}{Phys.\
  Rev.\  \textbf{D74} (2006) 076006}\relax
\mciteBstWouldAddEndPuncttrue
\mciteSetBstMidEndSepPunct{\mcitedefaultmidpunct}
{\mcitedefaultendpunct}{\mcitedefaultseppunct}\relax
\EndOfBibitem
\bibitem{Feng:2012zze}
G.-Q. Feng, X.-H. Guo, and Z.-H. Zhang,
  \ifthenelse{\boolean{articletitles}}{\emph{{Studying the $D^*K$ molecular
  structure of $D_s(2460)$ in the Bethe-Salpeter approach}},
  }{}\href{https://doi.org/10.1140/epjc/s10052-012-2033-y}{Eur.\ Phys.\ J.\
  \textbf{C72} (2012) 2033}\relax
\mciteBstWouldAddEndPuncttrue
\mciteSetBstMidEndSepPunct{\mcitedefaultmidpunct}
{\mcitedefaultendpunct}{\mcitedefaultseppunct}\relax
\EndOfBibitem
\bibitem{Ortega:2016mms}
P.~G. Ortega, J.~Segovia, D.~R. Entem {\em et~al.},
  \ifthenelse{\boolean{articletitles}}{\emph{{Molecular components in P-wave
  charmed-strange mesons}},
  }{}\href{https://doi.org/10.1103/PhysRevD.94.074037}{Phys.\ Rev.\
  \textbf{D94} (2016) 074037}\relax
\mciteBstWouldAddEndPuncttrue
\mciteSetBstMidEndSepPunct{\mcitedefaultmidpunct}
{\mcitedefaultendpunct}{\mcitedefaultseppunct}\relax
\EndOfBibitem
\bibitem{Zhang:2009pn}
D.~Zhang, Q.-Y. Zhao, and Q.-Y. Zhang,
  \ifthenelse{\boolean{articletitles}}{\emph{{A study of S-wave $DK$
  interactions in the chiral SU(3) quark model}},
  }{}\href{https://doi.org/10.1088/0256-307X/26/9/091201}{Chin.\ Phys.\ Lett.\
  \textbf{26} (2009) 091201}\relax
\mciteBstWouldAddEndPuncttrue
\mciteSetBstMidEndSepPunct{\mcitedefaultmidpunct}
{\mcitedefaultendpunct}{\mcitedefaultseppunct}\relax
\EndOfBibitem
\bibitem{Mehen:2004uj}
T.~Mehen and R.~P. Springer,
  \ifthenelse{\boolean{articletitles}}{\emph{{Heavy-quark symmetry and the
  electromagnetic decays of excited charmed strange mesons}},
  }{}\href{https://doi.org/10.1103/PhysRevD.70.074014}{Phys.\ Rev.\
  \textbf{D70} (2004) 074014}\relax
\mciteBstWouldAddEndPuncttrue
\mciteSetBstMidEndSepPunct{\mcitedefaultmidpunct}
{\mcitedefaultendpunct}{\mcitedefaultseppunct}\relax
\EndOfBibitem
\bibitem{PhysRevD.91.092011}
Belle Collaboration, S.-K. Choi {\em et~al.},
  \ifthenelse{\boolean{articletitles}}{\emph{Measurements of
  ${B}\ensuremath{\rightarrow}\overline{D}{D}_{s0}^{*+}(2317)$ decay rates and
  a search for isospin partners of the ${D}_{s0}^{*+}(2317)$},
  }{}\href{https://doi.org/10.1103/PhysRevD.91.092011}{Phys.\ Rev.\
  \textbf{D91} (2015) 092011}, Erratum
  \href{https://doi.org/10.1103/PhysRevD.92.039905}{ibid.\   \textbf{D92}
  (2015) 039905}\relax
\mciteBstWouldAddEndPuncttrue
\mciteSetBstMidEndSepPunct{\mcitedefaultmidpunct}
{\mcitedefaultendpunct}{\mcitedefaultseppunct}\relax
\EndOfBibitem
\bibitem{Yeo:2024chk}
Hadron Spectrum Collaboration, J.~D.~E. Yeo {\em et~al.},
  \ifthenelse{\boolean{articletitles}}{\emph{{DK/D\ensuremath{\pi} scattering
  and an exotic virtual bound state at the SU(3) flavour symmetric point from
  lattice QCD}}, }{}\href{https://doi.org/10.1007/JHEP07(2024)012}{J.\ High
  Energ.\ Phys.\  \textbf{07} (2024) 012}\relax
\mciteBstWouldAddEndPuncttrue
\mciteSetBstMidEndSepPunct{\mcitedefaultmidpunct}
{\mcitedefaultendpunct}{\mcitedefaultseppunct}\relax
\EndOfBibitem
\bibitem{Godfrey:1985xj}
S.~Godfrey and N.~Isgur, \ifthenelse{\boolean{articletitles}}{\emph{{Mesons in
  a relativized quark model with chromodynamics}},
  }{}\href{https://doi.org/10.1103/PhysRevD.32.189}{Phys.\ Rev.\  \textbf{D32}
  (1985) 189}\relax
\mciteBstWouldAddEndPuncttrue
\mciteSetBstMidEndSepPunct{\mcitedefaultmidpunct}
{\mcitedefaultendpunct}{\mcitedefaultseppunct}\relax
\EndOfBibitem
\bibitem{PhysRevD.43.1679}
S.~Godfrey and R.~Kokoski,
  \ifthenelse{\boolean{articletitles}}{\emph{{Properties of P-wave mesons with
  one heavy quark}}, }{}\href{https://doi.org/10.1103/PhysRevD.43.1679}{Phys.\
  Rev.\  \textbf{D43} (1991) 1679}\relax
\mciteBstWouldAddEndPuncttrue
\mciteSetBstMidEndSepPunct{\mcitedefaultmidpunct}
{\mcitedefaultendpunct}{\mcitedefaultseppunct}\relax
\EndOfBibitem
\bibitem{PhysRevD.64.114004}
M.~Di~Pierro and E.~Eichten, \ifthenelse{\boolean{articletitles}}{\emph{Excited
  heavy-light systems and hadronic transitions},
  }{}\href{https://doi.org/10.1103/PhysRevD.64.114004}{Phys.\ Rev.\
  \textbf{D64} (2001) 114004}\relax
\mciteBstWouldAddEndPuncttrue
\mciteSetBstMidEndSepPunct{\mcitedefaultmidpunct}
{\mcitedefaultendpunct}{\mcitedefaultseppunct}\relax
\EndOfBibitem
\bibitem{PDG2024}
Particle Data Group, S.~Navas {\em et~al.},
  \ifthenelse{\boolean{articletitles}}{\emph{{\href{http://pdg.lbl.gov/}{Review
  of particle physics}}},
  }{}\href{https://doi.org/10.1103/PhysRevD.110.030001}{Phys.\ Rev.\
  \textbf{D110} (2024) 030001}\relax
\mciteBstWouldAddEndPuncttrue
\mciteSetBstMidEndSepPunct{\mcitedefaultmidpunct}
{\mcitedefaultendpunct}{\mcitedefaultseppunct}\relax
\EndOfBibitem
\bibitem{BaBar:2006eep}
BaBar Collaboration, B.~Aubert {\em et~al.},
  \ifthenelse{\boolean{articletitles}}{\emph{{A study of the
  $D^*_{sJ}(2317)^{+}$ and $D_{sJ}(2460)^{+}$ mesons in inclusive $c\bar{c}$
  production near $\sqrt{s} = 10.6 \gev$}},
  }{}\href{https://doi.org/10.1103/PhysRevD.74.032007}{Phys.\ Rev.\
  \textbf{D74} (2006) 032007}\relax
\mciteBstWouldAddEndPuncttrue
\mciteSetBstMidEndSepPunct{\mcitedefaultmidpunct}
{\mcitedefaultendpunct}{\mcitedefaultseppunct}\relax
\EndOfBibitem
\bibitem{Belle:2003kup}
Belle Collaboration, Y.~Mikami {\em et~al.},
  \ifthenelse{\boolean{articletitles}}{\emph{{Measurements of the $D_{sJ}$
  resonance properties}},
  }{}\href{https://doi.org/10.1103/PhysRevLett.92.012002}{Phys.\ Rev.\ Lett.\
  \textbf{92} (2004) 012002}\relax
\mciteBstWouldAddEndPuncttrue
\mciteSetBstMidEndSepPunct{\mcitedefaultmidpunct}
{\mcitedefaultendpunct}{\mcitedefaultseppunct}\relax
\EndOfBibitem
\bibitem{Tang:2023yls}
M.-N. Tang, Y.-H. Lin, F.-K. Guo {\em et~al.},
  \ifthenelse{\boolean{articletitles}}{\emph{{Isospin-conserving hadronic decay
  of the $D_{s1}(2460)$ into $D_{s}\pi^+\pi^-$}},
  }{}\href{https://doi.org/10.1088/1572-9494/accc1f}{Commun.\ Theor.\ Phys.\
  \textbf{75} (2023) 055203}\relax
\mciteBstWouldAddEndPuncttrue
\mciteSetBstMidEndSepPunct{\mcitedefaultmidpunct}
{\mcitedefaultendpunct}{\mcitedefaultseppunct}\relax
\EndOfBibitem
\bibitem{LHCb-PAPER-2020-024}
LHCb Collaboration, R.~Aaij {\em et~al.},
  \ifthenelse{\boolean{articletitles}}{\emph{{Model-independent study of
  structure in \mbox{$\Bp \to \Dp \Dm \Kp$} decays}},
  }{}\href{https://doi.org/10.1103/PhysRevLett.125.242001}{Phys.\ Rev.\ Lett.\
  \textbf{125} (2020) 242001}\relax
\mciteBstWouldAddEndPuncttrue
\mciteSetBstMidEndSepPunct{\mcitedefaultmidpunct}
{\mcitedefaultendpunct}{\mcitedefaultseppunct}\relax
\EndOfBibitem
\bibitem{LHCb-PAPER-2020-025}
LHCb Collaboration, R.~Aaij {\em et~al.},
  \ifthenelse{\boolean{articletitles}}{\emph{{Amplitude analysis of the $\Bp
  \to \Dp \Dm \Kp$ decay}},
  }{}\href{https://doi.org/10.1103/PhysRevD.102.112003}{Phys.\ Rev.\
  \textbf{D102} (2020) 112003}\relax
\mciteBstWouldAddEndPuncttrue
\mciteSetBstMidEndSepPunct{\mcitedefaultmidpunct}
{\mcitedefaultendpunct}{\mcitedefaultseppunct}\relax
\EndOfBibitem
\bibitem{LHCb-PAPER-2022-026}
LHCb Collaboration, R.~Aaij {\em et~al.},
  \ifthenelse{\boolean{articletitles}}{\emph{{First observation of a doubly
  charged tetraquark candidate and its neutral partner}},
  }{}\href{https://doi.org/10.1103/PhysRevLett.131.041902}{Phys.\ Rev.\ Lett.\
  \textbf{131} (2023) 041902}\relax
\mciteBstWouldAddEndPuncttrue
\mciteSetBstMidEndSepPunct{\mcitedefaultmidpunct}
{\mcitedefaultendpunct}{\mcitedefaultseppunct}\relax
\EndOfBibitem
\bibitem{LHCb-PAPER-2022-027}
LHCb Collaboration, R.~Aaij {\em et~al.},
  \ifthenelse{\boolean{articletitles}}{\emph{{Amplitude analysis of $\Bz
  \rightarrow \Dzb \Dsp \pim$ and $\Bp \rightarrow \Dm \Dsp\pip$ decays}},
  }{}\href{https://doi.org/10.1103/PhysRevD.108.012017}{Phys.\ Rev.\
  \textbf{D108} (2023) 012017}\relax
\mciteBstWouldAddEndPuncttrue
\mciteSetBstMidEndSepPunct{\mcitedefaultmidpunct}
{\mcitedefaultendpunct}{\mcitedefaultseppunct}\relax
\EndOfBibitem
\bibitem{Ke:2022ocs}
H.-W. Ke, Y.-F. Shi, X.-H. Liu {\em et~al.},
  \ifthenelse{\boolean{articletitles}}{\emph{{Possible molecular states of
  $\Dbar{}^*K^*$ ($D^*K^*$) and new exotic states $X_0(2900)$, $X_1(2900)$,
  $T_{cs0}^a(2900)^0$ and $T_{cs0}^a(2900)^{++}$}},
  }{}\href{https://doi.org/10.1103/PhysRevD.106.114032}{Phys.\ Rev.\
  \textbf{D106} (2022) 114032}\relax
\mciteBstWouldAddEndPuncttrue
\mciteSetBstMidEndSepPunct{\mcitedefaultmidpunct}
{\mcitedefaultendpunct}{\mcitedefaultseppunct}\relax
\EndOfBibitem
\bibitem{Agaev:2022eyk}
S.~S. Agaev, K.~Azizi, and H.~Sundu,
  \ifthenelse{\boolean{articletitles}}{\emph{{Modeling the resonance
  $T_{cs0}^a(2900)^{++}$ as a hadronic molecule $D^{*+}K^{*+}$}},
  }{}\href{https://doi.org/10.1103/PhysRevD.107.094019}{Phys.\ Rev.\
  \textbf{D107} (2023) 094019}\relax
\mciteBstWouldAddEndPuncttrue
\mciteSetBstMidEndSepPunct{\mcitedefaultmidpunct}
{\mcitedefaultendpunct}{\mcitedefaultseppunct}\relax
\EndOfBibitem
\bibitem{Duan:2023lcj}
M.-Y. Duan, M.-L. Du, Z.-H. Guo {\em et~al.},
  \ifthenelse{\boolean{articletitles}}{\emph{{Coupled-channel
  $D^*K^*$-$D_s^*\rho$ interactions and the origin of $T_{c\bar{s}0}(2900)$}},
  }{}\href{https://doi.org/10.1103/PhysRevD.108.074006}{Phys.\ Rev.\
  \textbf{D108} (2023) 074006}\relax
\mciteBstWouldAddEndPuncttrue
\mciteSetBstMidEndSepPunct{\mcitedefaultmidpunct}
{\mcitedefaultendpunct}{\mcitedefaultseppunct}\relax
\EndOfBibitem
\bibitem{Maiani:2024quj}
L.~Maiani, A.~D. Polosa, and V.~Riquer,
  \ifthenelse{\boolean{articletitles}}{\emph{{Open charm tetraquarks in broken
  SU(3)$_{F}$ symmetry}},
  }{}\href{https://doi.org/10.1103/PhysRevD.110.034014}{Phys.\ Rev.\
  \textbf{D110} (2024) 034014}\relax
\mciteBstWouldAddEndPuncttrue
\mciteSetBstMidEndSepPunct{\mcitedefaultmidpunct}
{\mcitedefaultendpunct}{\mcitedefaultseppunct}\relax
\EndOfBibitem
\bibitem{Terasaki:2003qa}
K.~Terasaki, \ifthenelse{\boolean{articletitles}}{\emph{{BABAR resonance as a
  new window of hadron physics}},
  }{}\href{https://doi.org/10.1103/PhysRevD.68.011501}{Phys.\ Rev.\
  \textbf{D68} (2003) 011501}\relax
\mciteBstWouldAddEndPuncttrue
\mciteSetBstMidEndSepPunct{\mcitedefaultmidpunct}
{\mcitedefaultendpunct}{\mcitedefaultseppunct}\relax
\EndOfBibitem
\bibitem{LHCb-DP-2008-001}
LHCb Collaboration, A.~A. Alves~Jr.\ {\em et~al.},
  \ifthenelse{\boolean{articletitles}}{\emph{{The \lhcb detector at the LHC}},
  }{}\href{https://doi.org/10.1088/1748-0221/3/08/S08005}{Journal of
  Instrumentation \textbf{3} (2008) S08005}\relax
\mciteBstWouldAddEndPuncttrue
\mciteSetBstMidEndSepPunct{\mcitedefaultmidpunct}
{\mcitedefaultendpunct}{\mcitedefaultseppunct}\relax
\EndOfBibitem
\bibitem{LHCb-DP-2014-002}
LHCb Collaboration, R.~Aaij {\em et~al.},
  \ifthenelse{\boolean{articletitles}}{\emph{{LHCb detector performance}},
  }{}\href{https://doi.org/10.1142/S0217751X15300227}{Int.\ J.\ Mod.\ Phys.\
  \textbf{A30} (2015) 1530022}\relax
\mciteBstWouldAddEndPuncttrue
\mciteSetBstMidEndSepPunct{\mcitedefaultmidpunct}
{\mcitedefaultendpunct}{\mcitedefaultseppunct}\relax
\EndOfBibitem
\bibitem{Sjostrand:2007gs}
T.~Sj\"{o}strand, S.~Mrenna, and P.~Skands,
  \ifthenelse{\boolean{articletitles}}{\emph{{A brief introduction to PYTHIA
  8.1}}, }{}\href{https://doi.org/10.1016/j.cpc.2008.01.036}{Comput.\ Phys.\
  Commun.\  \textbf{178} (2008) 852}\relax
\mciteBstWouldAddEndPuncttrue
\mciteSetBstMidEndSepPunct{\mcitedefaultmidpunct}
{\mcitedefaultendpunct}{\mcitedefaultseppunct}\relax
\EndOfBibitem
\bibitem{Sjostrand:2006za}
T.~Sj\"{o}strand, S.~Mrenna, and P.~Skands,
  \ifthenelse{\boolean{articletitles}}{\emph{{PYTHIA 6.4 physics and manual}},
  }{}\href{https://doi.org/10.1088/1126-6708/2006/05/026}{J.\ High Energ.\
  Phys.\  \textbf{05} (2006) 026}\relax
\mciteBstWouldAddEndPuncttrue
\mciteSetBstMidEndSepPunct{\mcitedefaultmidpunct}
{\mcitedefaultendpunct}{\mcitedefaultseppunct}\relax
\EndOfBibitem
\bibitem{LHCb-PROC-2010-056}
I.~Belyaev, T.~Brambach, N.~H. Brook {\em et~al.},
  \ifthenelse{\boolean{articletitles}}{\emph{{Handling of the generation of
  primary events in Gauss, the LHCb simulation framework}},
  }{}\href{https://doi.org/10.1088/1742-6596/331/3/032047}{J.\ Phys.\ Conf.\
  Ser.\  \textbf{331} (2011) 032047}\relax
\mciteBstWouldAddEndPuncttrue
\mciteSetBstMidEndSepPunct{\mcitedefaultmidpunct}
{\mcitedefaultendpunct}{\mcitedefaultseppunct}\relax
\EndOfBibitem
\bibitem{Lange:2001uf}
D.~J. Lange, \ifthenelse{\boolean{articletitles}}{\emph{{The EvtGen particle
  decay simulation package}},
  }{}\href{https://doi.org/10.1016/S0168-9002(01)00089-4}{Nucl.\ Instrum.\
  Meth.\  \textbf{A462} (2001) 152}\relax
\mciteBstWouldAddEndPuncttrue
\mciteSetBstMidEndSepPunct{\mcitedefaultmidpunct}
{\mcitedefaultendpunct}{\mcitedefaultseppunct}\relax
\EndOfBibitem
\bibitem{davidson2015photos}
N.~Davidson, T.~Przedzinski, and Z.~Was,
  \ifthenelse{\boolean{articletitles}}{\emph{{PHOTOS interface in C++:
  Technical and physics documentation}},
  }{}\href{https://doi.org/https://doi.org/10.1016/j.cpc.2015.09.013}{Comp.\
  Phys.\ Comm.\  \textbf{199} (2016) 86}\relax
\mciteBstWouldAddEndPuncttrue
\mciteSetBstMidEndSepPunct{\mcitedefaultmidpunct}
{\mcitedefaultendpunct}{\mcitedefaultseppunct}\relax
\EndOfBibitem
\bibitem{Allison:2006ve}
Geant4 Collaboration, J.~Allison {\em et~al.},
  \ifthenelse{\boolean{articletitles}}{\emph{{Geant4 developments and
  applications}}, }{}\href{https://doi.org/10.1109/TNS.2006.869826}{IEEE
  Trans.\ Nucl.\ Sci.\  \textbf{53} (2006) 270}\relax
\mciteBstWouldAddEndPuncttrue
\mciteSetBstMidEndSepPunct{\mcitedefaultmidpunct}
{\mcitedefaultendpunct}{\mcitedefaultseppunct}\relax
\EndOfBibitem
\bibitem{Agostinelli:2002hh}
Geant4 Collaboration, S.~Agostinelli {\em et~al.},
  \ifthenelse{\boolean{articletitles}}{\emph{{Geant4: A simulation toolkit}},
  }{}\href{https://doi.org/10.1016/S0168-9002(03)01368-8}{Nucl.\ Instrum.\
  Meth.\  \textbf{A506} (2003) 250}\relax
\mciteBstWouldAddEndPuncttrue
\mciteSetBstMidEndSepPunct{\mcitedefaultmidpunct}
{\mcitedefaultendpunct}{\mcitedefaultseppunct}\relax
\EndOfBibitem
\bibitem{LHCb-PROC-2011-006}
M.~Clemencic, G.~Corti, S.~Easo {\em et~al.},
  \ifthenelse{\boolean{articletitles}}{\emph{{The \lhcb simulation application,
  Gauss: Design, evolution and experience}},
  }{}\href{https://doi.org/10.1088/1742-6596/331/3/032023}{J.\ Phys.\ Conf.\
  Ser.\  \textbf{331} (2011) 032023}\relax
\mciteBstWouldAddEndPuncttrue
\mciteSetBstMidEndSepPunct{\mcitedefaultmidpunct}
{\mcitedefaultendpunct}{\mcitedefaultseppunct}\relax
\EndOfBibitem
\bibitem{LHCb-DP-2018-004}
D.~M{\"u}ller, M.~Clemencic, G.~Corti {\em et~al.},
  \ifthenelse{\boolean{articletitles}}{\emph{{ReDecay: A novel approach to
  speed up the simulation at LHCb}},
  }{}\href{https://doi.org/10.1140/epjc/s10052-018-6469-6}{Eur.\ Phys.\ J.\
  \textbf{C78} (2018) 1009}\relax
\mciteBstWouldAddEndPuncttrue
\mciteSetBstMidEndSepPunct{\mcitedefaultmidpunct}
{\mcitedefaultendpunct}{\mcitedefaultseppunct}\relax
\EndOfBibitem
\bibitem{Breiman}
L.~Breiman, J.~H. Friedman, R.~A. Olshen {\em et~al.}, {\em Classification and
  regression trees}, \href{https://doi.org/10.1201/9781315139470}{ Wadsworth
  international group, Belmont, California, USA, 1984}\relax
\mciteBstWouldAddEndPuncttrue
\mciteSetBstMidEndSepPunct{\mcitedefaultmidpunct}
{\mcitedefaultendpunct}{\mcitedefaultseppunct}\relax
\EndOfBibitem
\bibitem{AdaBoost}
Y.~Freund and R.~E. Schapire, \ifthenelse{\boolean{articletitles}}{\emph{A
  decision-theoretic generalization of on-line learning and an application to
  boosting}, }{}\href{https://doi.org/10.1006/jcss.1997.1504}{J.\ Comput.\
  Syst.\ Sci.\  \textbf{55} (1997) 119}\relax
\mciteBstWouldAddEndPuncttrue
\mciteSetBstMidEndSepPunct{\mcitedefaultmidpunct}
{\mcitedefaultendpunct}{\mcitedefaultseppunct}\relax
\EndOfBibitem
\bibitem{Hocker:2007ht}
H.~Voss, A.~Hoecker, J.~Stelzer {\em et~al.},
  \ifthenelse{\boolean{articletitles}}{\emph{{TMVA - Toolkit for Multivariate
  Data Analysis with ROOT}}, }{}\href{https://doi.org/10.22323/1.050.0040}{PoS
  \textbf{ACAT} (2007) 040}\relax
\mciteBstWouldAddEndPuncttrue
\mciteSetBstMidEndSepPunct{\mcitedefaultmidpunct}
{\mcitedefaultendpunct}{\mcitedefaultseppunct}\relax
\EndOfBibitem
\bibitem{TMVA4}
A.~Hoecker, P.~Speckmayer, J.~Stelzer {\em et~al.},
  \ifthenelse{\boolean{articletitles}}{\emph{{TMVA 4 --- Toolkit for
  Multivariate Data Analysis with ROOT. Users Guide.}},
  }{}\href{http://arxiv.org/abs/physics/0703039}{{\normalfont\ttfamily
  arXiv:physics/0703039}}\relax
\mciteBstWouldAddEndPuncttrue
\mciteSetBstMidEndSepPunct{\mcitedefaultmidpunct}
{\mcitedefaultendpunct}{\mcitedefaultseppunct}\relax
\EndOfBibitem
\bibitem{Hulsbergen:2005pu}
W.~D. Hulsbergen, \ifthenelse{\boolean{articletitles}}{\emph{{Decay chain
  fitting with a Kalman filter}},
  }{}\href{https://doi.org/10.1016/j.nima.2005.06.078}{Nucl.\ Instrum.\ Meth.\
  \textbf{A552} (2005) 566}\relax
\mciteBstWouldAddEndPuncttrue
\mciteSetBstMidEndSepPunct{\mcitedefaultmidpunct}
{\mcitedefaultendpunct}{\mcitedefaultseppunct}\relax
\EndOfBibitem
\bibitem{ARGUS:1990hfq}
ARGUS Collaboration, H.~Albrecht {\em et~al.},
  \ifthenelse{\boolean{articletitles}}{\emph{{Search for hadronic $b \to u$
  decays}}, }{}\href{https://doi.org/10.1016/0370-2693(90)91293-K}{Phys.\
  Lett.\  \textbf{B241} (1990) 278}\relax
\mciteBstWouldAddEndPuncttrue
\mciteSetBstMidEndSepPunct{\mcitedefaultmidpunct}
{\mcitedefaultendpunct}{\mcitedefaultseppunct}\relax
\EndOfBibitem
\bibitem{PhysRev.135.B551}
G.~N. Fleming, \ifthenelse{\boolean{articletitles}}{\emph{{Recoupling effects
  in the isobar model. I. General formalism for three-pion scattering}},
  }{}\href{https://doi.org/10.1103/PhysRev.135.B551}{Phys.\ Rev.\  \textbf{135}
  (1964) B551}\relax
\mciteBstWouldAddEndPuncttrue
\mciteSetBstMidEndSepPunct{\mcitedefaultmidpunct}
{\mcitedefaultendpunct}{\mcitedefaultseppunct}\relax
\EndOfBibitem
\bibitem{PhysRev.166.1731}
D.~Morgan, \ifthenelse{\boolean{articletitles}}{\emph{{Phenomenological
  analysis of $I=\frac{1}{2}$ single-pion production processes in the energy
  range 500 to 700 MeV}},
  }{}\href{https://doi.org/10.1103/PhysRev.166.1731}{Phys.\ Rev.\  \textbf{166}
  (1968) 1731}\relax
\mciteBstWouldAddEndPuncttrue
\mciteSetBstMidEndSepPunct{\mcitedefaultmidpunct}
{\mcitedefaultendpunct}{\mcitedefaultseppunct}\relax
\EndOfBibitem
\bibitem{PhysRevD.11.3165}
D.~J. Herndon, P.~S\"oding, and R.~J. Cashmore,
  \ifthenelse{\boolean{articletitles}}{\emph{Generalized isobar model
  formalism}, }{}\href{https://doi.org/10.1103/PhysRevD.11.3165}{Phys.\ Rev.\
  \textbf{D11} (1975) 3165}\relax
\mciteBstWouldAddEndPuncttrue
\mciteSetBstMidEndSepPunct{\mcitedefaultmidpunct}
{\mcitedefaultendpunct}{\mcitedefaultseppunct}\relax
\EndOfBibitem
\bibitem{FLATTE1976224}
S.~M. Flatt\'{e}, \ifthenelse{\boolean{articletitles}}{\emph{{Coupled-channel
  analysis of the $\pi\eta$ and $K\bar{K}$ systems near $K\bar{K}$ threshold}},
  }{}\href{https://doi.org/https://doi.org/10.1016/0370-2693(76)90654-7}{Phys.\
  Lett.\  \textbf{B63} (1976) 224}\relax
\mciteBstWouldAddEndPuncttrue
\mciteSetBstMidEndSepPunct{\mcitedefaultmidpunct}
{\mcitedefaultendpunct}{\mcitedefaultseppunct}\relax
\EndOfBibitem
\bibitem{Bugg:2008ig}
D.~V. Bugg, \ifthenelse{\boolean{articletitles}}{\emph{{Reanalysis of data on
  ${a}_{0}(1450)$ and ${a}_{0}(980)$}},
  }{}\href{https://doi.org/10.1103/PhysRevD.78.074023}{Phys.\ Rev.\
  \textbf{D78} (2008) 074023}\relax
\mciteBstWouldAddEndPuncttrue
\mciteSetBstMidEndSepPunct{\mcitedefaultmidpunct}
{\mcitedefaultendpunct}{\mcitedefaultseppunct}\relax
\EndOfBibitem
\bibitem{LHCb-PAPER-2013-069}
LHCb Collaboration, R.~Aaij {\em et~al.},
  \ifthenelse{\boolean{articletitles}}{\emph{{Measurement of resonant and \CP
  components in \mbox{\decay{\Bsb}{\jpsi\pip\pim}} decays}},
  }{}\href{https://doi.org/10.1103/PhysRevD.89.092006}{Phys.\ Rev.\
  \textbf{D89} (2014) 092006}\relax
\mciteBstWouldAddEndPuncttrue
\mciteSetBstMidEndSepPunct{\mcitedefaultmidpunct}
{\mcitedefaultendpunct}{\mcitedefaultseppunct}\relax
\EndOfBibitem
\bibitem{Aitchison:1972ay}
I.~J.~R. Aitchison, \ifthenelse{\boolean{articletitles}}{\emph{{K-matrix
  formalism for overlapping resonances}},
  }{}\href{https://doi.org/10.1016/0375-9474(72)90305-3}{Nucl.\ Phys.\
  \textbf{A189} (1972) 417}\relax
\mciteBstWouldAddEndPuncttrue
\mciteSetBstMidEndSepPunct{\mcitedefaultmidpunct}
{\mcitedefaultendpunct}{\mcitedefaultseppunct}\relax
\EndOfBibitem
\bibitem{Anisovich:2002ij}
V.~V. Anisovich and A.~V. Sarantsev,
  \ifthenelse{\boolean{articletitles}}{\emph{{K matrix analysis of the
  ($IJ^{PC} = 00^{++}$)-wave in the mass region below 1900 MeV}},
  }{}\href{https://doi.org/10.1140/epja/i2002-10068-x}{Eur.\ Phys.\ J.\
  \textbf{A16} (2003) 229}\relax
\mciteBstWouldAddEndPuncttrue
\mciteSetBstMidEndSepPunct{\mcitedefaultmidpunct}
{\mcitedefaultendpunct}{\mcitedefaultseppunct}\relax
\EndOfBibitem
\bibitem{Fernandez-Ramirez:2019koa}
JPAC Collaboration, C.~Fern\'andez-Ram\'\i{}rez {\em et~al.},
  \ifthenelse{\boolean{articletitles}}{\emph{{Interpretation of the LHCb
  $P_c(4312)^+$ signal}},
  }{}\href{https://doi.org/10.1103/PhysRevLett.123.092001}{Phys.\ Rev.\ Lett.\
  \textbf{123} (2019) 092001}\relax
\mciteBstWouldAddEndPuncttrue
\mciteSetBstMidEndSepPunct{\mcitedefaultmidpunct}
{\mcitedefaultendpunct}{\mcitedefaultseppunct}\relax
\EndOfBibitem
\bibitem{LHCb-DP-2013-002}
LHCb Collaboration, R.~Aaij {\em et~al.},
  \ifthenelse{\boolean{articletitles}}{\emph{{Measurement of the track
  reconstruction efficiency at LHCb}},
  }{}\href{https://doi.org/10.1088/1748-0221/10/02/P02007}{Journal of
  Instrumentation \textbf{10} (2015) P02007}\relax
\mciteBstWouldAddEndPuncttrue
\mciteSetBstMidEndSepPunct{\mcitedefaultmidpunct}
{\mcitedefaultendpunct}{\mcitedefaultseppunct}\relax
\EndOfBibitem
\bibitem{LHCb-DP-2012-004}
R.~Aaij, J.~Albrecht, F.~Alessio {\em et~al.},
  \ifthenelse{\boolean{articletitles}}{\emph{{The \lhcb trigger and its
  performance in 2011}},
  }{}\href{https://doi.org/10.1088/1748-0221/8/04/P04022}{Journal of
  Instrumentation \textbf{8} (2013) P04022}\relax
\mciteBstWouldAddEndPuncttrue
\mciteSetBstMidEndSepPunct{\mcitedefaultmidpunct}
{\mcitedefaultendpunct}{\mcitedefaultseppunct}\relax
\EndOfBibitem
\bibitem{Poluektov:2014rxa}
A.~Poluektov, \ifthenelse{\boolean{articletitles}}{\emph{{Kernel density
  estimation of a multidimensional efficiency profile}},
  }{}\href{https://doi.org/10.1088/1748-0221/10/02/P02011}{Journal of
  Instrumentation \textbf{10} (2015) P02011}\relax
\mciteBstWouldAddEndPuncttrue
\mciteSetBstMidEndSepPunct{\mcitedefaultmidpunct}
{\mcitedefaultendpunct}{\mcitedefaultseppunct}\relax
\EndOfBibitem
\bibitem{LHCb-PAPER-2014-012}
LHCb Collaboration, R.~Aaij {\em et~al.},
  \ifthenelse{\boolean{articletitles}}{\emph{{Measurement of the resonant and
  \CP components in \mbox{\decay{\Bzb}{\jpsi\pip\pim}} decays}},
  }{}\href{https://doi.org/10.1103/PhysRevD.90.012003}{Phys.\ Rev.\
  \textbf{D90} (2014) 012003}\relax
\mciteBstWouldAddEndPuncttrue
\mciteSetBstMidEndSepPunct{\mcitedefaultmidpunct}
{\mcitedefaultendpunct}{\mcitedefaultseppunct}\relax
\EndOfBibitem
\bibitem{LHCb-PAPER-2014-070}
LHCb Collaboration, R.~Aaij {\em et~al.},
  \ifthenelse{\boolean{articletitles}}{\emph{{Dalitz plot analysis of
  \mbox{\decay{\Bz}{\Dzb\pip\pim}} decays}},
  }{}\href{https://doi.org/10.1103/PhysRevD.92.032002}{Phys.\ Rev.\
  \textbf{D92} (2015) 032002}\relax
\mciteBstWouldAddEndPuncttrue
\mciteSetBstMidEndSepPunct{\mcitedefaultmidpunct}
{\mcitedefaultendpunct}{\mcitedefaultseppunct}\relax
\EndOfBibitem
\bibitem{LHCb-PAPER-2019-017}
LHCb Collaboration, R.~Aaij {\em et~al.},
  \ifthenelse{\boolean{articletitles}}{\emph{{Amplitude analysis of the
  \mbox{\decay{\Bp}{\pip\pip\pim}} decay}},
  }{}\href{https://doi.org/10.1103/PhysRevD.101.012006}{Phys.\ Rev.\
  \textbf{D101} (2020) 012006}\relax
\mciteBstWouldAddEndPuncttrue
\mciteSetBstMidEndSepPunct{\mcitedefaultmidpunct}
{\mcitedefaultendpunct}{\mcitedefaultseppunct}\relax
\EndOfBibitem
\bibitem{LHCb-PAPER-2022-016}
LHCb Collaboration, R.~Aaij {\em et~al.},
  \ifthenelse{\boolean{articletitles}}{\emph{{Amplitude analysis of the $\Dp
  \to \pim\pip\pip$ decay and measurement of the $\pim\pip$ S-wave amplitude}},
  }{}\href{https://doi.org/10.1007/JHEP06(2023)044}{J.\ High Energ.\ Phys.\
  \textbf{06} (2023) 044}\relax
\mciteBstWouldAddEndPuncttrue
\mciteSetBstMidEndSepPunct{\mcitedefaultmidpunct}
{\mcitedefaultendpunct}{\mcitedefaultseppunct}\relax
\EndOfBibitem
\bibitem{LHCb-PAPER-2022-030}
LHCb Collaboration, R.~Aaij {\em et~al.},
  \ifthenelse{\boolean{articletitles}}{\emph{{Amplitude analysis of the $\Ds
  \to \pim \pip \pip$ decay}},
  }{}\href{https://doi.org/10.1007/JHEP07(2023)204}{J.\ High Energ.\ Phys.\
  \textbf{07} (2023) 204}\relax
\mciteBstWouldAddEndPuncttrue
\mciteSetBstMidEndSepPunct{\mcitedefaultmidpunct}
{\mcitedefaultendpunct}{\mcitedefaultseppunct}\relax
\EndOfBibitem
\bibitem{ABLIKIM2005243}
BES Collaboration, M.~Ablikim {\em et~al.},
  \ifthenelse{\boolean{articletitles}}{\emph{{Resonances in $\jpsi \to \phi
  \pip\pim$ and $\phi \Kp\Km$}},
  }{}\href{https://doi.org/10.1016/j.physletb.2004.12.041}{Phys.\ Lett.\
  \textbf{B607} (2005) 243}\relax
\mciteBstWouldAddEndPuncttrue
\mciteSetBstMidEndSepPunct{\mcitedefaultmidpunct}
{\mcitedefaultendpunct}{\mcitedefaultseppunct}\relax
\EndOfBibitem
\bibitem{PhysRevLett.118.012001}
BESIII Collaboration, M.~Ablikim {\em et~al.},
  \ifthenelse{\boolean{articletitles}}{\emph{{Amplitude analysis of the decays
  $\etapr \rightarrow \pi^+\pi^-\pi^0$ and $\etapr \rightarrow
  \pi^0\pi^0\pi^0$}},
  }{}\href{https://doi.org/10.1103/PhysRevLett.118.012001}{Phys.\ Rev.\ Lett.\
  \textbf{118} (2017) 012001}\relax
\mciteBstWouldAddEndPuncttrue
\mciteSetBstMidEndSepPunct{\mcitedefaultmidpunct}
{\mcitedefaultendpunct}{\mcitedefaultseppunct}\relax
\EndOfBibitem
\bibitem{Liu:2012zya}
L.~Liu, K.~Orginos, F.-K. Guo {\em et~al.},
  \ifthenelse{\boolean{articletitles}}{\emph{{Interactions of charmed mesons
  with light pseudoscalar mesons from lattice QCD and implications on the
  nature of the $D_{s0}^*(2317)$}},
  }{}\href{https://doi.org/10.1103/PhysRevD.87.014508}{Phys.\ Rev.\
  \textbf{D87} (2013) }\relax
\mciteBstWouldAddEndPuncttrue
\mciteSetBstMidEndSepPunct{\mcitedefaultmidpunct}
{\mcitedefaultendpunct}{\mcitedefaultseppunct}\relax
\EndOfBibitem
\bibitem{Guo:2009ct}
F.-K. Guo, C.~Hanhart, and U.-G. Meissner,
  \ifthenelse{\boolean{articletitles}}{\emph{{Interactions between heavy mesons
  and Goldstone bosons from chiral dynamics}},
  }{}\href{https://doi.org/10.1140/epja/i2009-10762-1}{Eur.\ Phys.\ J.\
  \textbf{A40} (2009) 171}\relax
\mciteBstWouldAddEndPuncttrue
\mciteSetBstMidEndSepPunct{\mcitedefaultmidpunct}
{\mcitedefaultendpunct}{\mcitedefaultseppunct}\relax
\EndOfBibitem
\bibitem{Skwarnicki:1986xj}
T.~Skwarnicki, {\em {A study of the radiative cascade transitions between the
  Upsilon-prime and Upsilon resonances}}, PhD thesis, Institute of Nuclear
  Physics, Krakow, 1986,
  {\href{http://inspirehep.net/record/230779/}{DESY-F31-86-02}}\relax
\mciteBstWouldAddEndPuncttrue
\mciteSetBstMidEndSepPunct{\mcitedefaultmidpunct}
{\mcitedefaultendpunct}{\mcitedefaultseppunct}\relax
\EndOfBibitem
\end{mcitethebibliography}
 
\newpage
% LHCb collaboration author list
% Data extracted on August 11th, 2024 at 6:11pm for paper reference LHCb-PAPER-2024-033
\centerline
{\large\bf LHCb collaboration}
\begin
{flushleft}
\small
R.~Aaij$^{38}$\lhcborcid{0000-0003-0533-1952},
A.S.W.~Abdelmotteleb$^{57}$\lhcborcid{0000-0001-7905-0542},
C.~Abellan~Beteta$^{51}$,
F.~Abudin{\'e}n$^{57}$\lhcborcid{0000-0002-6737-3528},
T.~Ackernley$^{61}$\lhcborcid{0000-0002-5951-3498},
A. A. ~Adefisoye$^{69}$\lhcborcid{0000-0003-2448-1550},
B.~Adeva$^{47}$\lhcborcid{0000-0001-9756-3712},
M.~Adinolfi$^{55}$\lhcborcid{0000-0002-1326-1264},
P.~Adlarson$^{82}$\lhcborcid{0000-0001-6280-3851},
C.~Agapopoulou$^{14}$\lhcborcid{0000-0002-2368-0147}
% ...

{\footnotesize \it

% ...
$^{14}$Universit{\'e} Paris-Saclay, CNRS/IN2P3, IJCLab, Orsay, France\\
% ...
$^{38}$Nikhef National Institute for Subatomic Physics, Amsterdam, Netherlands\\
% ...
$^{47}$Instituto Galego de F{\'\i}sica de Altas Enerx{\'\i}as (IGFAE), Universidade de Santiago de Compostela, Santiago de Compostela, Spain\\
% ...
$^{51}$Physik-Institut, Universit{\"a}t Z{\"u}rich, Z{\"u}rich, Switzerland\\
% ...
$^{55}$H.H. Wills Physics Laboratory, University of Bristol, Bristol, United Kingdom\\
% ...
$^{57}$Department of Physics, University of Warwick, Coventry, United Kingdom\\
% ...
$^{60}$School of Physics and Astronomy, University of Glasgow, Glasgow, United Kingdom\\
$^{61}$Oliver Lodge Laboratory, University of Liverpool, Liverpool, United Kingdom\\
% ...
$^{69}$Syracuse University, Syracuse, NY, United States\\
% ...
$^{82}$Department of Physics and Astronomy, Uppsala University, Uppsala, Sweden, associated to $^{60}$\\
% ...
}
\end{flushleft}

\end{document}